\documentclass[aps,footinbib, notitlepage,superscriptaddress, longbibliography,eqsecnum]{revtex4-1}
\usepackage{graphicx}
 \usepackage{amssymb}
 \usepackage{amsmath}
 \usepackage{amsfonts}
\usepackage{accents}
\usepackage{overpic}
\usepackage{enumerate}
\usepackage{color}
\usepackage[dvipsnames]{xcolor}
\usepackage{xspace}
\usepackage[normalsize]{subfigure}
\usepackage{hyperref}
\usepackage{bm}
\usepackage{pifont}
\usepackage{empheq}
\usepackage[capitalize]{cleveref}

\crefname{section}{Sec.}{Secs.}

\newlength{\dhatheight}
\newcommand{\doublehat}[1]{%
    \settoheight{\dhatheight}{\ensuremath{\hat{#1}}}%
    \addtolength{\dhatheight}{-0.35ex}%
    \hat{\vphantom{\rule{1pt}{\dhatheight}}%
    \smash{\hat{#1}}}}

\DeclareFontFamily{OT1}{pzc}{}
\DeclareFontShape{OT1}{pzc}{m}{it}{<-> s * [1.10] pzcmi7t}{}
\DeclareMathAlphabet{\mathpzc}{OT1}{pzc}{m}{it}

\providecommand{\st}[1]{_{\text{#1}}}
\providecommand{\sfrac}[2]{#1/#2}

\providecommand{\pfrac}[2]{\left(\frac{#1}{#2}\right)}
\providecommand{\ut}[1]{^{\text{#1}}}

\def\onehalf{\frac{1}{2}}

\def\bra{\ensuremath{\langle}}
\def\ket{\ensuremath{\rangle}}
\def\eq{\st{eq}}

\def\const{\mathrm{const}}

\def\pd{\partial}

\def\im{\mathrm{i}}
\def\kv{\bv{k}}
\def\qv{\bv{q}}

\def\mv{\bv{m}}

\def\nv{\bv{n}}
\def\nd{{n_d}}
\def\nvp{\nv_\parallel}

\def\pv{\bv{p}}
\def\pnvp{\pv_{\nvp}}
\def\rv{\bv{r}}
\def\rvp{\bv{r}_\parallel}

\def\bvnp{\bv{0}_\parallel}

\def\Acal{\mathcal{A}}
\def\Fcal{\mathcal{F}}
\def\Gcal{\mathcal{G}}
\def\Hcal{\mathcal{H}}
\def\Ccal{\mathcal{C}}

\def\Kcal{\mathcal{K}}

\def\Ncal{\mathcal{N}}

\def\Ocal{\mathcal{O}}
\def\Pcal{\mathcal{P}}
\def\Qcal{\mathcal{Q}}
\def\Rcal{\mathcal{R}}
\def\Scal{\mathcal{S}}
\def\Tcal{\mathcal{T}}

\def\Vcal{\mathcal{V}}

\def\kcal{\mathpzc{g}}
\def\tphys{{\tilde t}}

\def\hyp13{{_1 F_3}}
\def\res{\st{res}}
\def\gc{\st{gc}}

\def\bcs{BCs\xspace}
\def\pbc{\ut{(p)}}

\def\Nbc{\ut{(N)}}

\def\izero{^{(0)}}

\def\tred{\ring{\tau}}
\def\tauInit{\tau_0}
\def\tauLG{\tau}

\def\th{\st{th}}

\def\amplPhit{\phi_t\izero}

\def\amplXip{\xi_+\izero}

\def\amplTrelax{\mathpzc{\tilde{t}}_{+}\izero}
\def\amplCorrel{\mathpzc{c}_b}
\def\d{\mathrm{d}}

\def\trelax{\tilde t_R}
\def\zdyn{\mathpzc{z}}
\def\reals{\mathbb{R}}
\def\dyn{\st{dyn}}
\def\stat{\st{stat}}

\def\eqGC{\st{eq,gc}}
\def\eqBlk{\st{b,eq}}
\def\rel{\st{rel}}
\def\flt{\st{flat}}
\def\initvar{v}
\def\LambdaRed{\chi}
\def\np{\nabla_\parallel}

\def\sump{\sideset{}{'}\sum}
\def\Lcaltildev{\tilde{\bm{\mathcal L}}(\bv{m})}
\def\Lcalv{\bm{\mathcal L}(\bv{m})}

\newcommand{\ave}[1]{\left\langle #1 \right\rangle}
\newcommand{\al}[1]{\begin{align} #1 \end{align}}

\newcommand{\bitem}{\begin{itemize}}
\newcommand{\eitem}{\end{itemize}}
\newcommand{\benum}{\begin{enumerate}}
\newcommand{\eenum}{\end{enumerate}}
\newcommand{\btab}[1]{\begin{tabular}{#1}}
\newcommand{\etab}{\end{tabular}}
\newcommand{\beq}{\begin{equation}}
\newcommand{\eeq}{\end{equation}}
\newcommand{\beqn}{\begin{equation*}}
\newcommand{\eeqn}{\end{equation*}}

\newcommand{\bv}[1]{\mathbf{#1}}

\graphicspath{{Figs/}{figs/}}

\begin{document}
\title{Dynamics of the critical Casimir force for a conserved order parameter\\ after a critical quench}
\author{Markus Gross}
\email{gross@is.mpg.de}
\affiliation{Max-Planck-Institut f\"{u}r Intelligente Systeme, Heisenbergstra{\ss}e 3, 70569 Stuttgart, Germany}
\affiliation{IV.\ Institut f\"{u}r Theoretische Physik, Universit\"{a}t Stuttgart, Pfaffenwaldring 57, 70569 Stuttgart, Germany}
 \author{Christian M. Rohwer}
\affiliation{Max-Planck-Institut f\"{u}r Intelligente Systeme, Heisenbergstra{\ss}e 3, 70569 Stuttgart, Germany}
\affiliation{IV.\ Institut f\"{u}r Theoretische Physik, Universit\"{a}t Stuttgart, Pfaffenwaldring 57, 70569 Stuttgart, Germany}
 \author{S. Dietrich}
 \affiliation{Max-Planck-Institut f\"{u}r Intelligente Systeme, Heisenbergstra{\ss}e 3, 70569 Stuttgart, Germany}
 \affiliation{IV.\ Institut f\"{u}r Theoretische Physik, Universit\"{a}t Stuttgart, Pfaffenwaldring 57, 70569 Stuttgart, Germany}
\date{\today}

\begin{abstract}
Fluctuation-induced forces occur generically when long-ranged correlations (e.g., in fluids) are confined by external bodies. In classical systems, such correlations require specific conditions, e.g., a medium close to a critical point. On the other hand, long-ranged correlations appear more commonly in certain non-equilibrium systems with conservation laws. Consequently, a variety of non-equilibrium fluctuation phenomena, including fluctuation-induced forces, have been discovered and explored recently. Here, we address a long-standing problem of non-equilibrium critical Casimir forces emerging after a quench to the critical point in a confined fluid with order-parameter-conserving dynamics and non-symmetry-breaking boundary conditions. The interplay of inherent (critical) fluctuations and dynamical non-local effects (due to density conservation) gives rise to striking features, including correlation functions and forces exhibiting oscillatory time-dependences. Complex transient regimes arise, depending on initial conditions and the geometry of the confinement. Our findings pave the way for exploring a wealth of non-equilibrium processes in critical fluids (e.g., fluctuation-mediated self-assembly or aggregation). In certain regimes, our results are applicable to active matter.
\end{abstract}


\maketitle

\section{Introduction}

Quenching a classical fluid or spin system from a homogeneous initial state into the multiphase region ($T<T_c$) induces nucleation of or spinodal decomposition into distinct phases \cite{onuki_phase_2002}. The subsequent dynamics of the domain evolution (i.e., coarsening and Ostwald ripening) has been extensively studied (see, e.g., Refs.\ \cite{bray_theory_1994,calabrese_ageing_2005} for reviews). 
A quench to the critical temperature $T_c$ is instead characterized not by the growth of well-ordered domains, but by a growing correlation length $\xi(t) \propto t^{1/\zdyn}$, where $t$ is time and $\zdyn$ is the dynamic critical exponent of the system.
The growth of $\xi$ reflects the fact that equilibrium evolves from smaller towards larger spatial scales.

On a coarse-grained level, the quench dynamics of a simple fluid is, in its most general form, described by the hydrodynamic equations of the so-called \emph{model H} \cite{hohenberg_theory_1977,onuki_phase_2002}. 
A simplified description, which retains the conserved nature of the order parameter (OP) but neglects heat and momentum transport, is provided by \emph{model B} \cite{hohenberg_theory_1977}, which, in other contexts is also known as Cahn-Hilliard \cite{cahn_free_1958,elliott_cahn-hilliard_1989} or Mullins-Herring equation \cite{mullins_theory_1957}.
For a one-component fluid, the OP $\phi$ is defined by $\phi\propto n-n_c$, where $n$ is the actual number density and $n_c$ is its critical value, while for a binary liquid mixture, $\phi \propto c_A-c_{A,c}$, where $c_A$ is the concentration of species A and $c_{A,c}$ is its critical value.
Quenches of fluid-like systems to critical or super-critical temperatures have been studied previously for various dynamical models \cite{majumdar_growth_1995, godreche_non-equilibrium_2004, sire_autocorrelation_2004, calabrese_ageing_2005, baumann_kinetics_2007, rothlein_symmetry-based_2006, henkel_non-equilibrium_2010, albano_study_2011} and particularly extensively in the context of interfacial roughening (see Refs.\ \cite{barabasi_fractal_1995,krug_origins_1997} and references therein). 
However, these studies considered mostly the behavior of correlation functions. Here, we focus instead on the dynamics of the (non-equilibrium) critical Casimir force (CCF) as the confined system relaxes towards equilibrium after the quench.

In considering post-quench dynamics, it is important to distinguish distinct sources of long-ranged correlations. On the one hand, systems may exhibit inherent correlations, e.g., in the vicinity of critical points. On the other hand, it is well-recognized that driving certain systems out of equilibrium in the presence of conservation laws (e.g., conserved particle number, momentum, etc.) can give rise to purely non-equilibrium correlations \cite{spohn1983,dorfmankirkpatricksengers1994} which vanish in thermal equilibrium.

In turn, the confinement of long-ranged correlations (irrespective of their source) by external objects (e.g., plates), generally gives rise to fluctuation-induced forces \cite{kardargolestanian1999}. In classical equilibrium systems, the prototypical example is the well-established notion of an equilibrium CCF \cite{fisher_wall_1978, krech_free_1992, krech_casimir_1994, brankov_theory_2000}, which arises due to confinement of long-ranged correlations in near-critical fluid media. Regarding non-equilibrium situations, the aforementioned conservation laws and the associated long-ranged correlations have been demonstrated to give rise to purely non-equilibrium Casimir-like forces in a variety of settings, including hydrodynamic systems with density gradients \cite{aminovkardarkafri2015} or temperature gradients \cite{kirkpatricksengers2013,kirkpatrick2015prl,kirkpatrick2016pre},  sheared systems \cite{Gompper2017shear, KirkpatrickSengers2018shear}, far-from-critical fluids undergoing quenches of their temperature or activity \cite{rohwer_transient_2017,rohwer2018forces} as well as shear flow~\cite{rohwer_correlations_2019}, and stochastically driven systems \cite{Mohammadi17}. 

Fluctuation-induced forces in (far from critical) hydrodynamic systems have been shown to be vanishingly small in thermal equilibrium \cite{Monahan16}. It thus is interesting to consider the interplay of quench dynamics and inherent correlations, which has been studied in the setting of nonequilibrium, time-dependent generalizations of the CCF (see, e.g., Refs.~\cite{gambassi_critical_2006,gambassi2008EPJB,deangopinathan2009JStatMech,dean_out--equilibrium_2010}). We emphasize that these studies considered model A dynamics, for which the order parameter is not conserved. 
For conserved dynamics, solving the full problem in the presence of surfaces becomes significantly more difficult; thus far, discussions of the quench dynamics for non-symmetry breaking \bcs are limited to semi-infinite geometries \cite{diehl_boundary_1992,wichmann_dynamic_1995}. In Ref.\ \cite{gross_surface-induced_2018}, the noise-free dynamics of the CCF after a quench of a fluid film in the presence of surface adsorption has been investigated within mean-field theory. We also note that particular care must be taken in applying field-theoretic stress tensors when computing forces out of equilibrium \cite{deangopinathan2009JStatMech, dean_out--equilibrium_2010, gross_surface-induced_2018, kruger_stresses_2018}. 

In the present study, we consider an instantaneous quench of a confined fluid to the critical point for non-symmetry-breaking boundary conditions (\bcs) within model B, which describes the relaxation of a conserved OP under the influence of thermal noise. 
In order to facilitate an analytical study we focus on the Gaussian limit. The Gaussian limit correctly describes the universal features of a critical fluid within the Ising universality class in $d>4$ spatial dimensions. It furthermore provides the leading contribution to the actual critical behavior in $d<4$ dimensions within a systematic expansion of the full theory around $d=4$ \cite{le_bellac_quantum_1991, amit_field_2005}. We emphasize that the model considered here is also applicable to the macroscopic description of active matter in certain parameter regimes \cite{tjhung_cluster_2018,caballero_bulk_2018}, and thus is of timely relevance.

The paper is organized as follows. Section \ref{sec_sysmod} describes the system and model under consideration. Next, post-quench correlation functions are computed in the bulk (\cref{sec_bulk_correl}) and in confinement with Neumann and periodic boundary conditions in \cref{sec_correl_confined}. In \cref{sec_CCF} we compute critical Casimir forces for film and cuboidal box geometries. Finally, a summary and outlook are provided in \cref{sec_sum}.

\section{System and model}
\label{sec_sysmod}

\begin{figure}[t]\centering
    \includegraphics[width=0.26\linewidth]{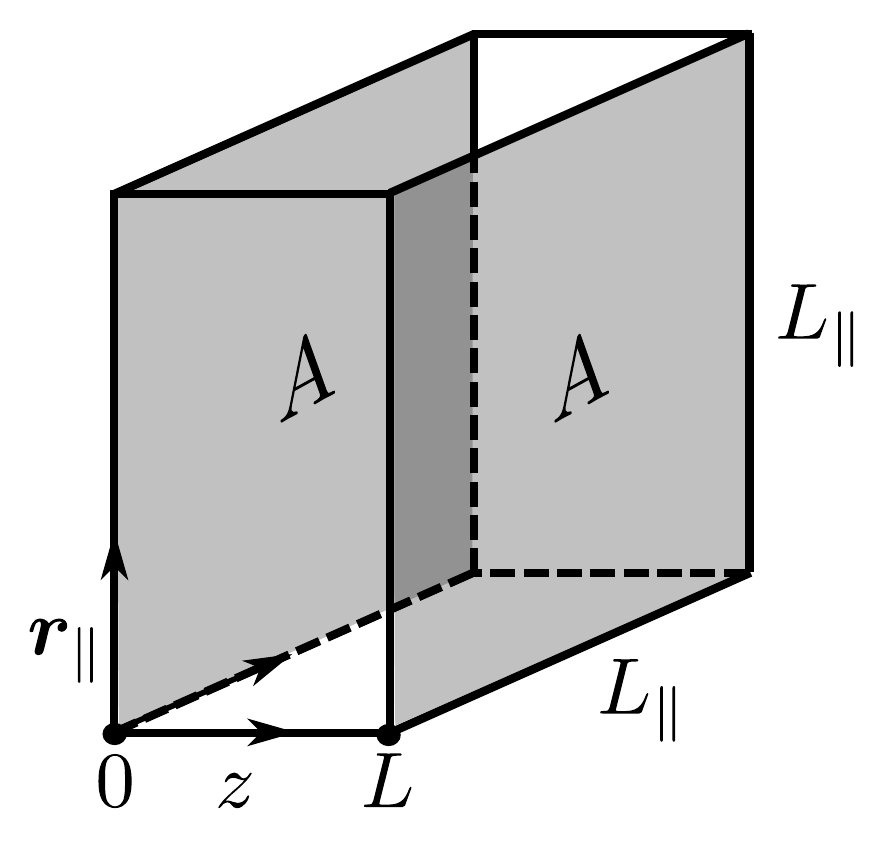}
    \caption{
    The geometry under consideration. The fluid is a slab of thickness $L$ in the $z$-direction. It is confined by two parallel surfaces of area $A=L_\parallel^{d-1}$ lying in the transverse planes at $z=0$ and $z=L$, respectively; and the vector $\rv_\parallel$ has $d-1$ components. We generally consider $d$ spatial dimensions. The shaded surfaces carry periodic or Neumann \bcs, while the remaining (lateral) surfaces of the box exhibit periodic \bcs. The limit $A\to\infty$ corresponds to a \emph{thin film}.
    }
    \label{fig_boxsketch}
\end{figure}

\subsection{Quench protocol, geometry, and OP conservation}
Initially, the fluid resides in a homogeneous high-temperature (i.e., far-from-critical) phase, and the OP is taken to have a vanishing mean value and short-ranged correlations, characterized by the strength $\initvar k_B T$ of their \emph{v}ariance:
\beq \bra \phi(\rv,t=0) \ket = 0,\qquad \bra \phi(\rv,t=0) \phi(\rv',t=0)\ket =  \initvar k_B T \delta(\rv-\rv').
\label{eq_IC_correl}\eeq 
For generic $\initvar>0$ these are called \emph{thermal} initial conditions (ICs). 
For theoretical purposes, it is useful to study also the extreme limit of vanishing initial variance: 
\beq \bra \phi(\rv,t=0) \ket = 0,\qquad \bra \phi(\rv,t=0) \phi(\rv',t=0)\ket =  0,
\label{eq_IC_correl_flat}\eeq 
which are called \emph{flat} ICs.
At time $t=0$, the system is instantaneously quenched to a near-critical reduced temperature
\beq \tred \equiv \frac{T-T_{c}}{T_{c}} > 0.
\label{eq_T_red}\eeq 
It proceeds to evolve towards thermal equilibrium, which is reached in the late-time limit $t\to\infty$. We emphasize that the initial conditions \cref{eq_IC_correl,eq_IC_correl_flat} are non-equilibrium initial conditions with respect to the post-quench dynamics (in particular, in \cref{eq_IC_correl}  $T$ is the post-quench temperature).
In a generic fluid, flat ICs can only be realized at zero temperature. In this case, a quench would correspond to an instantaneous heating.

We consider a $d$-dimensional cuboid \emph{box} geometry with volume $V= L A$. It is characterized by an \emph{aspect ratio} 
\beq \varrho \equiv \frac{L}{L_\parallel} = \frac{L}{A^{\frac{1}{d-1}}},
\label{eq_aspectratio}\eeq 
where $L$ and $L_\parallel$ denote the transverse and lateral extension of the system, respectively, and $A=L_\parallel^{d-1}$ is the transverse area (see \cref{fig_boxsketch}).
An extreme case, facilitating analytical calculations, is the \emph{thin film} limit $\varrho\to 0$. 
At the system boundaries at $z=0$ and $z=L$, we impose either periodic or Neumann \bcs, the latter being given by $\pd_z\phi(\rv,t)|_{z\in\{0,L\}}=0$. In the lateral directions, we generally apply periodic \bcs. These \bcs ensure that the total OP inside the box,
\beq \Phi(t) \equiv \int_V \d^d r\, \phi(\rv,t) = \const,
\label{eq_total_OP}\eeq 
is conserved in time. While \cref{eq_IC_correl} implies $\Phi(t)=0$, and thus $\bra\phi(\rv,t)\ket=0$ for all $t$, we will occasionally treat the general case $\Phi=\const$, restricting it to zero where necessary.

\subsection{General scaling considerations}
\label{sec_scaling}

We first discuss the expected scaling behavior of the OP correlation function, focusing on equal-time correlations,  which are the most relevant case for the present study. More general considerations can be found in Refs.\ \cite{calabrese_ageing_2005, henkel_non-equilibrium_2010}.
Here and in the following, the physical time is denoted by $\tphys$, which is to be distinguished from a rescaled time $t$ introduced in \cref{eq_time_resc} below.
At late times after the quench, the influence of the initial condition [\cref{eq_IC_correl}] diminishes. 
In fact, it can be shown that, at the corresponding fixed-point of the renormalization group flow, which determines the universal behavior \cite{janssen_new_1989, calabrese_ageing_2005}, the correlation strength vanishes, i.e., $\initvar=0$ [see \cref{eq_IC_correl}].
In this scaling regime, the equal-time OP correlation function $\Ccal(\rv, \rv',\tphys\,)\equiv \bra \phi(\rv,\tphys\,) \phi(\rv',\tphys\,)\ket$ fulfills the finite-size scaling relation \cite{barber_finite-size_1983, gambassi_critical_2006}
\beq \Ccal(\rv, \rv',\tphys,\tred, L, \initvar=0) = \amplCorrel \pfrac{L}{\amplXip}^{2-d-\eta} \mathfrak{C} \left( \frac{\rv}{L}, \frac{\rv'}{L},  \pfrac{L}{\amplXip}^{-\zdyn} \frac{\tphys}{\amplTrelax} ,  \pfrac{L}{\amplXip}^{1/\nu} \tred, \varrho \right)
\label{eq_scalform_C}\eeq
with a dimensionless scaling function $\mathfrak{C}$.
Here, the non-universal amplitude $\amplXip$, which carries the dimension of length, is defined in terms of the critical behavior of the bulk correlation length $\xi$ above $T_c$, i.e., $\xi(\tred\to 0^+) = \amplXip \tred^{-\nu}$; the non-universal amplitude $\amplTrelax$ describes the critical divergence of the characteristic relaxation time $\trelax= \amplTrelax \tred^{-\nu \zdyn} = \amplTrelax (\xi/\amplXip)^z$ of the system; $\eta$ and $\nu$ are standard bulk critical exponents; the quantity $\amplCorrel$ is a non-universal correlation function amplitude (see Ref.\ \cite{gambassi_critical_2006} for explicit values).
With these expressions the scaling function in \cref{eq_scalform_C} takes the form $\mathfrak{C}\left(\rv/L, \rv'/L, (L/\xi)^{-z} \tphys/{\trelax}, (L/\xi)^{1/\nu}, \varrho\right)$.
Within model B, the dynamic critical exponent  $\zdyn$ fulfills the relation \cite{hohenberg_theory_1977, tauber_critical_2014} 
\beq \zdyn= 4-\eta.
\label{eq_zdyn}\eeq 
In a uniform \emph{bulk} system, the associated bulk correlation function $\Ccal_b$ is isotropic and thus depends only on $r\equiv |\rv|$. 
The bulk analogue of \cref{eq_scalform_C} is given by \cite{tauber_critical_2014}
\beq \Ccal_b(\rv,\tphys,\tred, \initvar=0) = \amplCorrel \pfrac{r}{\amplXip}^{2-d-\eta} \mathfrak{C}_b \left( \pfrac{r}{\amplXip}^{-\zdyn} \frac{\tphys}{\amplTrelax}, \pfrac{r}{\amplXip}^{1/\nu} \tred \right),
\label{eq_scalform_Cb}\eeq 
where $\mathfrak{C}_b$ is normalized such that $\amplCorrel$ is the same amplitude as for the confined case above.
At early times, the influence of a nonzero initial OP variance $\initvar\neq 0$ is not negligible; instead one expects the following scaling behavior of the bulk correlation function (see also Ref.\ \cite{sire_autocorrelation_2004}):
\begin{multline} \Ccal_b(\rv,\tphys,\tred,L,\initvar) = \initvar \frac{\amplCorrel}{(\amplXip)^{2}} \pfrac{r}{\amplXip}^{-d-\eta} \mathfrak{C}_0 \left( \pfrac{r}{\amplXip}^{-\zdyn} \frac{\tphys}{\amplTrelax}, \pfrac{r}{\amplXip}^{1/\nu} \tred \right) \\ + \amplCorrel \pfrac{r}{\amplXip}^{2-d-\eta} \mathfrak{C}_b \left( \pfrac{r}{\amplXip}^{-\zdyn} \frac{\tphys}{\amplTrelax}, \pfrac{r}{\amplXip}^{1/\nu} \tred \right),
\label{eq_scalform_Cb_init}\end{multline} 
with a further scaling function $\mathfrak{C}_0$. 
Note that $\initvar$ has the dimension $[\initvar]\sim [L]^2$ \cite{calabrese_ageing_2005}. The prefactors of $\mathfrak{C}_0$ are motivated by an explicit calculation (see \cref{eq_Stherm} below).
In passing, we remark that the scaling properties expressed in \cref{eq_scalform_C,eq_scalform_Cb} can be conveniently summarized in terms of the homogeneity relation
\cite{barber_finite-size_1983, pelissetto_critical_2002}
\beq \Ccal(\rv, \rv',\tphys, \tred, \varrho, L, \initvar=0) = \amplCorrel\, b^{2-d-\eta} \, \hat\Ccal\left(\frac{\rv}{\amplXip} b^{-1}, \frac{\rv'}{\amplXip} b^{-1}, \frac{\tphys}{\amplTrelax} b^{-\zdyn}, \tred b^{1/\nu}, \varrho, \frac{L}{\amplXip}b^{-1}, \initvar=0\right),
\label{eq_homrel_C}\eeq 
where $b$ is an arbitrary scaling factor and $\hat\Ccal$ is a dimensionless function. The existence of the latter follows on dimensional grounds upon expressing all dimensional quantities in terms of the fundamental non-universal bulk amplitudes $\amplXip$, $\amplTrelax$, and the amplitude $\amplPhit$ of the bulk OP ($\amplCorrel$ is a function of these, see Ref.\ \cite{gambassi_critical_2006}).

The (thermally averaged) CCF $\bra \Kcal\ket$ can be defined as the difference between the pressure of the fluid in the film, $\bra \Pcal_f\ket$, and in the surrounding bulk medium, $\bra \Pcal_b \ket$, \footnote{We explicitly indicate the thermal average $\bra\ldots\ket$ because \cref{eq_CCF_def} can be formulated also for the instantaneous CCF $\Kcal$ and the pressures $\Pcal_{f,b}$.}
\beq \bra\Kcal\ket = \bra\Pcal_f\ket -\bra \Pcal_b \ket.
\label{eq_CCF_def}\eeq 
In thermal \emph{equilibrium}, the averaged film pressure $\bra\Pcal_f\ket$ can be obtained from the derivative of the film free energy $\Fcal_f$ via $\bra\Pcal_f\ket = -\d \Fcal_f / \d L$.
Alternatively, the film pressure can be obtained by evaluating a stress tensor at the boundaries (see, e.g., Refs.\ \cite{krech_casimir_1994,gross_critical_2016} and references therein). Based on the notion of a generalized force \cite{dean_out--equilibrium_2010}, the stress tensor approach can be naturally extended to \emph{nonequilibrium} situations \cite{kruger_stresses_2018,rohwer_transient_2017,rohwer2018forces,gross_surface-induced_2018}. The present study follows this approach (see \cref{sec_CCF}). 
Both in and out of equilibrium, the bulk pressure can be obtained from a scaling limit: 
\beq \bra\Pcal_b\ket = \lim_{L\to\infty} \bra\Pcal_f\ket,
\label{eq_Pbulk_Pfilm_lim}\eeq 
which is taken by keeping the relevant thermodynamic control parameters fixed (see Refs.\  \cite{gross_critical_2016, gross_statistical_2017, gross_surface-induced_2018} for further details concerning ensemble differences).
The thermally averaged \emph{dynamical} CCF $\bra\Kcal\ket$, in units of $k_B T$, is expected to exhibit the following scaling behavior:
\beq \bra\Kcal(\tphys,\tred,\varrho, L, \initvar)\ket = L^{-d} \left[ L^{-2}\initvar\,\Xi_0\left(  \pfrac{L}{\amplXip}^{-\zdyn} \frac{\tphys}{\amplTrelax} ,  \pfrac{L}{\amplXip}^{1/\nu} \tred, \varrho \right) + \Xi \left(  \pfrac{L}{\amplXip}^{-\zdyn} \frac{\tphys}{\amplTrelax} ,  \pfrac{L}{\amplXip}^{1/\nu} \tred, \varrho \right)\right].
\label{eq_scalform_CCF}\eeq 
Since the effect of initial conditions gives rise to purely transient behaviors, the scaling function $\Xi_0\big(\theta = (L/\xi)^{-z} \tphys/\trelax,\allowbreak x=(L/\xi)^{1/\nu}, \varrho\big)$ vanishes for $\theta\to 0$ and $\theta\to \infty$. In contrast, the approach to equilibrium correlations is characterized by $\Xi(\theta\to\infty, x, \varrho)$, which ultimately attains a nonzero value corresponding to the equilibrium CCF. The scaling factor $L^{-2-d}$ in front of $\initvar\, \Xi_0$ follows from dimensional considerations [compare \cref{eq_scalform_Cb_init}].
Note that, at fixed arguments of the scaling functions $\Xi_0$ and $\Xi$, the influence of the thermal IC ($\initvar\neq 0$) becomes negligible for large $L$.

\subsection{Dynamical model}

Model B describes the conserved dynamics of an OP field $\phi(\rv,\tphys\,)$ evolving according to 
\al{ \pd_\tphys \phi &= \gamma \nabla^2 \mu(\phi) + \bar\eta ,
\label{eq_modelB_gen}}
where $\gamma$ represents a mobility coefficient with dimension $[\gamma] = [L^4/\tphys\,]$, which, within the Gaussian approximation, can be expressed as \footnote{\Cref{eq_mobility_ampl} follows analogously to model A, which has been discussed in Ref.\ \cite{gambassi_critical_2006}.} 
\beq \gamma = \frac{(\amplXip)^z}{\amplTrelax} ,\qquad \zdyn=4,
\label{eq_mobility_ampl}\eeq 
in terms of the amplitudes introduced in \cref{sec_scaling}.
Furthermore,
\beq \mu(\phi)  \equiv \frac{\delta \Fcal_b(\tauLG, g; [\phi])}{\delta \phi}
\label{eq_chempot_gen}
\eeq 
represents the bulk chemical potential obtained from the bare bulk free energy functional
\beq \Fcal_b(\tauLG, g; [\phi]) \equiv \int_V \d^d r\, \Hcal_b(\phi(\rv), \nabla\phi(\rv), \tauLG, g),
\label{eq_freeEn}\eeq
which plays the role of a Hamiltonian. 
We take the bulk Hamiltonian density $\Hcal_b$ to be of the Landau-Ginzburg form, i.e.,
\beq \Hcal_b(\phi,\nabla\phi, \tauLG, g) = \frac 1 2 (\nabla \phi)^2 + \frac{\tauLG}{2} \phi^2 + \frac{g}{4!} \phi^4.
\label{eq_Hamiltonian}\eeq 
In the present study we restrict ourselves to \emph{linear} dynamics, i.e., a \emph{Gaussian} form of $\Hcal_b$, and thus henceforth the quartic coupling constant is set to $g=0$.
The coupling constant $\tau$ in \cref{eq_Hamiltonian}, henceforth also called ``temperature parameter'', can be expressed as \cite{gross_critical_2016}
\al{
\tauLG =(\amplXip)^{-2}\tred
\label{eq_tauLGDef}
}
in terms of the reduced temperature introduced in \cref{eq_T_red}.
In \cref{eq_modelB_gen}, $\bar\eta$  represents a Gaussian white noise correlated as
$\bra \bar\eta(\rv, \tphys) \bar\eta(\rv', \tphys\,')\ket = -2 k_B T \gamma\, \nabla^2 \delta(\rv-\rv')\delta(\tphys-\tphys\,')
$
\ with $\bra \bar\eta(\rv, \tphys)\ket=0$, which ensures that, in equilibrium, the OP is distributed according to \cite{tauber_critical_2014}
\beq P\st{eq}(\tauLG,g; [\phi]) \sim e^{-\Fcal_b(\tauLG,g; [\phi])/(k_B T)}.
\label{eq_eq_dist}\eeq 

Equation \eqref{eq_modelB_gen} can be written as a continuity equation, $\pd_\tphys \phi= - \nabla\cdot \bm J$, where the current
\al{
\bm J \equiv   - \gamma\nabla \mu(\phi) +  \bm { \Ncal}\,
\label{eq_flux}}
comprises a deterministic and a stochastic contribution, with $\nabla\cdot\bm{\Ncal} = \bar\eta$, such that $\ave{\Ncal_\alpha(\rv, \tphys) \Ncal_\beta(\rv', \tphys\,')}=2 k_B T \gamma\, \delta(\rv-\rv')\delta(\tphys-\tphys\,') \delta_{\alpha \beta}$. As stated in the Introduction, we consider systems having either a thin film or a cuboid box geometry, and we impose periodic or Neumann \bcs [see \cref{eq_eigenspec} below] at the confining surfaces at $z=0,L$. 
Importantly, while Neumann \bcs automatically eliminate fluxes across the individual surfaces, i.e., $J_z= \gamma \left[\partial_z^3 \phi(z) - \tau  \partial_z \phi(z)\right] + \Ncal_z=0$ for $z =0$ and $z=L$, the total OP is in fact also conserved with periodic \bcs, because the net flux out of the system is zero.

In order to simplify the notation, we remove the temperature $k_B T$ from the description by defining a new OP field $\phi/(k_B T)^{1/2}$. 
Moreover, we introduce a rescaled time
\beq t = \gamma \tphys,
\label{eq_time_resc}\eeq
having dimension $[t] \sim [L]^4$.
This allows us to write \cref{eq_modelB_gen} as
\beq \pd_t \phi = -\nabla^4 \phi + \tauLG \nabla^2 \phi + \eta,
\label{eq_modelB_resc}\eeq 
where the noise $\eta \equiv \bar\eta/\gamma$ is correlated as
\beq \bra \eta(\rv, t) \eta(\rv', t')\ket = -2 \, \nabla^2 \delta(\rv-\rv')\delta(t-t'),
\label{eq_noise_correl}\eeq
which readily follows from the correlations of $\bar\eta$, taking into account the transformation law of the $\delta$ function.
We do not rescale coordinates by the length $L$, because it is instructive to keep all lengths explicit during the calculations.
While we mostly keep the quench temperature parameter $\tauLG\geq 0$ in \cref{eq_modelB_resc} arbitrary in our discussion, in order to facilitate analytical calculations, we will present our main results for $\tauLG=0$, i.e., for a quench to the critical point.

\subsection{Construction of the OP}

Here, the solution of the OP for model B in a box geometry is constructed. It is also shown how the corresponding results in the case of a thin film can be obtained via an appropriate substitution. To this end we consider the following set of orthonormal eigenfunctions $\sigma_n(z)$ and eigenvalues $\lambda_n^2$ of the operator $-\pd_z^2$:
\begin{subequations}
\begin{align}
\sigma_n\pbc(z) &= \frac{1}{\sqrt{L}} \exp \left(\im \lambda_n\pbc z\right) ,\qquad \lambda_n\pbc = \frac{2\pi n}{L}, \qquad n=0,\pm 1, \pm 2, \ldots, \qquad  \text{periodic \bcs}, \label{eq_eigenf_pbc} \\
\sigma_n\Nbc(z) &= \sqrt{\frac{2-\delta_{n,0}}{L}} \cos\left(\lambda_n\Nbc z\right), \qquad \lambda_n\Nbc = \frac{\pi n}{L}, \qquad n=0,1,2,\ldots,  \qquad \text{Neumann \bcs}, \label{eq_eigenf_Nbc}
\end{align}\label{eq_eigenspec}
\end{subequations}
which fulfill (for both p and N)
\beq \int_0^L \d z\, \sigma_m(z) \sigma_n^*(z) = \delta_{m,n}.
\label{eq_eigenf_ortho}
\eeq 
Since the system has periodic \bcs in the $d-1$ lateral directions perpendicular to the $z$-direction, we expand the OP and the noise as 
\beq \phi(\rv,t) = \frac{1}{\sqrt{A}} \sum_{\nvp,\nd} e^{\im \pnvp\cdot \rvp} \sigma_\nd(z) a_\nd(\pnvp, t)
\label{eq_OP_exp_gen}\eeq 
and
\beq \eta(\rv,t) = \frac{1}{\sqrt{A}} \sum_{\nvp,\nd} e^{\im \pnvp\cdot \rvp} \sigma_\nd(z) \zeta_\nd(\pnvp, t).
\label{eq_noise_exp_gen}\eeq 
For clarity, we denote the $z$ direction as the $d$th coordinate, while the remaining directions ($1,\ldots,d-1$) refer to the lateral coordinates $\rvp\in \reals^{d-1}$:
\al{
\rv = \{\rv_\parallel,z\} \equiv \big\{(r_1,\ldots,r_{d-1}),r_d\big\}\in\reals^d.
\label{eq_rvsplit}
}
The lateral wavevector $\pnvp$ has the components
\beq  p_{n_\alpha} \equiv p_\alpha  = \frac{2\pi n_\alpha}{L_\parallel}, \qquad \alpha=1,\ldots,d-1,
\label{eq_eigenval_lateral}\eeq 
where $n_\alpha= 0,\pm 1, \pm 2,\ldots$ and $\nvp \equiv \{n_1,\ldots, n_{d-1}\}$. 
In order to avoid a clumsy notation, in the following we drop the subscripts on $(\pnvp,\nd)$, and write $(\pv,n)$. 
It is useful to note that, for the eigenfunctions considered here, one has
\beq \int_0^L \d z\, \sigma_n(z) = \sqrt{L} \, \delta_{n,0}.
\label{eq_eigen_volint}\eeq 
The noise modes $\zeta_n$ have the expansion
\beq \zeta_n(\pv,t) = \frac{1}{\sqrt{A}} \int_A \d^{d-1} r_\parallel \int_0^L \d z\, e^{-\im \pv\cdot \rvp } \sigma^*_n(z) \eta(\{\rvp, z\}, t),
\label{eq_noise_mode_exp}\eeq 
and are correlated as [see \cref{eq_noise_correl}]
\beq \bra \zeta_n(\pv,t) \zeta_{n'}^*(\pv',t') \ket = 2 (\pv^2+ \lambda_n^2) \delta(t-t') \delta_{\pv,\pv'} \delta_{n,n'}, \quad \bra \zeta_n(\pv,t)\ket = 0.  
\label{eq_noise_mode_correl}\eeq 
Analogously to \cref{eq_noise_mode_exp}, the OP modes are given by
\beq a_n(\pv,t) = \frac{1}{\sqrt{A}} \int_A \d^{d-1} r_\parallel \int_0^L \d z\, e^{-\im \pv\cdot \rvp } \sigma^*_n(z) \phi(\{\rvp, z\}, t).
\label{eq_a_n_mode_exp}\eeq 
The notation in and after \cref{eq_OP_exp_gen} is chosen to highlight the special nature of the lateral directions and extensions, which become infinite for a thin film.
The latter (limiting) case is obtained by the replacement \beq \frac{1}{\sqrt{A}}\sum_{\pv} \to \frac{1}{\sqrt{(2\pi)^{d-1}}} \int \d^{d-1} p
\label{eq_continuum_repl_modes}\eeq 
in \cref{eq_OP_exp_gen,eq_noise_exp_gen}.

Inserting \cref{eq_OP_exp_gen,eq_noise_exp_gen} into \cref{eq_modelB_resc} and using \cref{eq_eigenf_ortho}
yields the equation for the OP modes $a_n$:
\beq \pd_t a_n(\pv,t) = -\Lambda_n(\pv) a_n(\pv,t) + \zeta_n(\pv,t)
\label{eq_modelB_mode}\eeq
with
\beq \Lambda_n(\pv,\tauLG) \equiv (\pv^2 + \lambda_n^2)^2 + \tauLG (\pv^2 + \lambda_n^2).
\label{eq_Lambda}\eeq
We occasionally omit the argument $\tauLG$ in order to keep the notation simple. Unless otherwise stated, all results apply to arbitrary $\tau\geq 0$.
The solution to \cref{eq_modelB_mode} is
\beq a_n(\pv,t) = e^{-\Lambda_n(\pv) t} a_n(\pv,0) + \int_0^t \d s\ e^{-\Lambda_n(\pv) (t-s)} \zeta_n(\pv,s)
\label{eq_modelB_modesol}\eeq 
in terms of the initial condition $a_n(\pv,t=0)$.
From \cref{eq_OP_exp_gen,eq_eigen_volint} it readily follows that the zero mode $a_{n=0}(\pv=0, t)$ is related to the total integrated OP $\Phi(t)$:
\beq \Phi(t) = \int_V \d^d r\, \phi(\rv,t) = \sqrt{A L}\, a_{n=0}(\pv=0, t).
\label{eq_zeromode_totalOP}\eeq 
In the case of an infinite transverse area ($A\to\infty$), one has $\int_V \d^d r\, \phi(\rv,t) = \sqrt{L}\, a_{n=0}(\pv=0,t)$ instead of \cref{eq_zeromode_totalOP}.
Since $\Lambda_{n=0}(\pv=0)=0$ and $\zeta_{n=0}(\pv=0,t)=0$ [as implied by the correlations in \cref{eq_noise_mode_correl}], the zero mode in fact remains constant in time for all \bcs considered here (see \cref{eq_zeromode_totalOP}):
\beq a_{n=0}(\pv=\bv0, t) = a_0(\bv0, 0) = \frac{\Phi}{\sqrt{AL}} = \const, 
\label{eq_zeromode_sol} \eeq 
which is consistent with the global OP conservation stated in \cref{eq_total_OP}.

The two-time correlation function in mode space follows from \cref{eq_noise_mode_correl,eq_modelB_modesol}:
\beq \bra a_n(\pv,t) a_{n'}^*(\pv',t')\ket = \Big[ \bra |a_n(\pv,0)|^2\ket e^{-\Lambda_n(\pv)(t+t')}   + \frac{1}{\LambdaRed_n(\pv,\tauLG)}\, \left( e^{-\Lambda_n(\pv)|t-t'|} - e^{-\Lambda_n(\pv)(t+t')} \right)  \Big] \delta_{\pv,\pv'} \delta_{n,n'},
\label{eq_modecorrel_gen}\eeq 
where we used $\bra a_n(\pv,0) \zeta_n(\pv,t)\ket =0$ for $t>0$ and the fact that the initial correlations are given by \cref{eq_IC_correl}, i.e., $\bra a_n(\pv,0) a_{n'}^*(\pv',0)\ket \propto \delta_{\pv,\pv'} \delta_{n,n'}$.
We have introduced the shorthand notation 
\beq \LambdaRed_n(\pv,\tauLG) \equiv \frac{\Lambda_n(\pv,\tauLG)}{\pv^2 + \lambda_n^2} = \pv^2 + \lambda_n^2 + \tauLG.
\label{eq_Lambda_red}\eeq 
Accordingly, the equal-time correlator is given by
\beq \bra a_n(\pv,t) a_{n'}^*(\pv',t)\ket = 
\Big[ \bra |a_n(\pv,0)|^2\ket e^{-2\Lambda_n(\pv)t}   + \frac{1}{\LambdaRed_n(\pv,\tauLG)} \left(1-e^{-2\Lambda_n(\pv) t}\right)\,  \Big] \delta_{\pv,\pv'} \delta_{n,n'}.
\label{eq_modecorrel_eqt}
\eeq 
At any finite time, for $\pv=\bv0$, $n=0$ this expression reduces to the time-independent correlations $\bra |a_{n=0}(\pv=\bv0,t)|^2\ket = \Phi^2/AL$ of the zero mode as implied by \cref{eq_zeromode_sol}. 
In \cref{eq_modecorrel_eqt}, the limit $t\to\infty$ (for general $n$ and $\pv$) yields $ \bra |a_n(\pv,t\to\infty)|^2\ket = \sfrac{1}{\LambdaRed_n(\pv,\tau)}$. 
For $\pv=0$, $n=0$, this expression reduces to $1/\tauLG$, which is inconsistent with \cref{eq_zeromode_sol}. This indicates that the limit $t\to\infty$ does not commute with that of a vanishing wave vector \footnote{The non-interchangeability of the two limits essentially arises from the second exponential in \cref{eq_modecorrel_eqt}: for arbitrary $\pv$, $n$ and $t\to\infty$, one has $e^{-2\Lambda_n(\pv) t}\to 0$, while, for $\pv\to 0$, $n\to 0$ and finite $t$, one has $e^{-2\Lambda_n(\pv) t}\to 1$.}. 
In fact,
\beq \bra |a_n(\pv)|^2\ket\eqGC = \frac{1}{\LambdaRed_n(\pv,\tauLG)}
\label{eq_eqmodecorrel_eqt_gc}\eeq 
describes the static equilibrium correlations of the modes (including the zero mode) in the \emph{grand canonical} ensemble (see also Eq.~(84) in Ref.\ \cite{gross_statistical_2017}), in which $\Phi(t)$ is allowed to fluctuate and has a variance which diverges at criticality, i.e., for $\tauLG\to 0$, constituting the so-called ``zero mode problem'' (see, e.g., Ref.\ \cite{gruneberg_thermodynamic_2008}). 
The present model, instead, realizes the equilibrium of the \emph{canonical} ensemble, within which the zero mode is constant in time and the equilibrium mode correlations are given by
\begin{subequations}
  \begin{empheq}[left ={ \bra |a_n(\pv)|^2\ket\eq \equiv \bra |a_n(\pv,t\to\infty)|^2\ket =  \empheqlbrace }]{align}
    &  |a_{n=0}(\pv=\bv0, 0)|^2  = \frac{\Phi^2}{AL}, \qquad & n=0, \pv=\bv0 \label{eq_eqmodecorrel_eqt_zero}, \\
    & \frac{1}{\LambdaRed_n(\pv,\tauLG)}, & \text{otherwise}. \label{eq_eqmodecorrel_eqt_other}
    \end{empheq}\label{eq_eqmodecorrel_eqt}
\end{subequations}

Thermal ICs [\cref{eq_IC_correl}] are expressed in mode space by
\beq \bra |a_n(\pv,0)|^2\ket = \initvar.
\label{eq_IC_correl_mode}\eeq 
For uncorrelated (i.e., ``flat'') ICs [\cref{eq_IC_correl_flat}], instead, one has $\initvar=0$, which, within Gaussian theory, implies 
\beq \phi(\rv,t=0)=0,\qquad a_n(\pv,0)=0,
\label{eq_IC_flat}\eeq 
such that only the second term in the square brackets in \cref{eq_modecorrel_eqt} remains.
We note that (grand canonical) equilibrium fluctuations within the Gaussian Landau-Ginzburg model [see \cref{eq_eq_dist}] at a reduced temperature $\tred_0$ [corresponding to a coupling constant $\tauInit = (\amplXip)^{-2} \tred_0$, see \cref{eq_tauLGDef}] are described by the standard Ornstein-Zernike form 
\beq \bra |a_n(\pv)|^2\ket\eqGC = \frac{1}{\pv^2 + \lambda_n^2 + \tauInit}.
\label{eq_static_OZ_correl}\eeq 
Thermal ICs [\cref{eq_IC_correl_mode}] can thus be considered as an approximation to equilibrium OP correlations at high temperatures $\tauInit=1/\initvar \to \infty$. 
At finite $\tauInit$, we expect the approximation of \cref{eq_static_OZ_correl} by \cref{eq_IC_correl_mode} to be reliable also for sufficiently large times, because in this case the exponential factor in \cref{eq_modecorrel_eqt} suppresses the modes with large $\pv$ or $\lambda_n$. 
Upon introducing the abbreviations
\begin{subequations}
\begin{align}
\Scal_{\text{dyn},n}(\pv,t) &\equiv  \frac{e^{-2\Lambda_n(\pv,\tauLG) t}}{\LambdaRed_n(\pv,\tauLG)},\qquad \Scal_{\text{stat},n}(\pv) \equiv \Scal_{\text{dyn},n}(\pv,t=0) = \frac{1}{\LambdaRed_n(\pv,\tauLG)}, \label{eq_fund_dyncorrel_1} \\ \intertext{for the \textit{dyn}amic and \textit{stat}ic contributions, respectively, as well as}
\Scal_{\text{rel},n}(\pv,t) &\equiv \frac{1}{\tauInit} e^{-2\Lambda_n(\pv,\tauLG) t}
\label{eq_fund_dyncorrel_2}
\end{align} \label{eq_fund_dyncorrel}
\end{subequations}
\hspace{-0.12cm}for the \textit{rel}axing contribution, the correlation function in \cref{eq_modecorrel_eqt} can be expressed as 
\begin{subequations}
\begin{align}
\bra a_n(\pv,t) a_{n'}^*(\pv',t)\ket\flt &= \left[ \Scal_{\text{stat},n}(\pv,t) - \Scal_{\text{dyn},n}(\pv,t) \right] \delta_{\pv,\pv'} \delta_{n,n'}, \label{eq_modecorrel_S_flat} \\
\intertext{and}
\bra a_n(\pv,t) a_{n'}^*(\pv',t)\ket\th &= \left[ \Scal_{\text{rel},n}(\pv,t) + \Scal_{\text{stat},n}(\pv,t) - \Scal_{\text{dyn},n}(\pv,t) \right] \delta_{\pv,\pv'} \delta_{n,n'}, \label{eq_modecorrel_S_th}
\end{align} \label{eq_modecorrel_S}
\end{subequations}
for flat and \emph{th}ermal ICs, respectively.

By writing the solution in \cref{eq_modelB_modesol} as 
\beq a_n(\pv,t) = \int_{0^-}^t \d s\, \Gcal_n(\pv,t-s) \left[\delta(s) a_n(\pv,0) + \zeta_n(\pv,s) \right],\qquad \Gcal_n(\pv,t) \equiv \tauInit \Scal_{\text{rel},n}(\pv,t/2),
\label{eq_Srel_Green}\eeq 
we infer that $\Scal\rel$ essentially corresponds to the Green function $\Gcal$ for the fourth-order diffusion equation [\cref{eq_modelB_resc}].
Furthermore, in equilibrium, time-translation invariance implies $\bra|a_n(\pv,t=0)|^2\ket = 1/\chi_n(\pv,\tauLG)$, such that \cref{eq_modecorrel_gen} renders
\beq \bra a_n(\pv,t) a_n^*(\pv,0)\ket\eq = \Scal_{\text{dyn},n}(\pv,t/2).
\label{eq_Sdyn_twotime_eq}\eeq 
Accordingly, $\Scal\dyn$ essentially represents the equilibrium two-time correlation function of the model.

\section{Bulk correlation functions}
\label{sec_bulk_correl}

The present approach to obtain the CCF requires knowledge of the bulk correlation function $\Ccal_b$,  which we discuss in this section.
The bulk counterparts of the correlators in mode space, as given by \cref{eq_modecorrel_gen,eq_modecorrel_eqt}, are obtained upon replacing the discrete eigenvalue $\lambda_n$ (corresponding to the direction of the confinement) by a continuous wave number $k$ and by attaining the continuum in line with \cref{eq_continuum_repl_modes}.
The mode $a_n(\pv)$ will be denoted in the bulk as $a(\qv=\{\pv,k\})$, while $\Lambda_n(\pv)$ defined in \cref{eq_Lambda} turns into
$\Lambda(\qv=\{\pv,k\}) = (\pv^2+k^2)^2 + \tauLG (\pv^2+k^2) = \qv^4 + \tauLG \qv^2$, where $\qv\equiv \{\pv,k\}$ represents the $d$-dimensional vector having $p_{\alpha=1,\ldots,d-1}$ as its first $d-1$ entries. Analogous mappings apply also to $\chi_n$ [\cref{eq_Lambda_red}] and $\Scal_n$ [\cref{eq_fund_dyncorrel}].

Accordingly, the two-time bulk OP correlation function is given by the Fourier transform of \cref{eq_modecorrel_gen}, i.e., 
\beq \Ccal_b(\rv, t,t') \equiv \Ccal_b(r,t,t') \equiv  \left\bra \phi(\{\rvp, z\}, t) \phi(\{\rvp'=0, z'=0\}, t') \right\ket  
= \int \frac{\d^{d-1} p}{(2\pi)^{d-1}} \int \frac{\d k}{2\pi} e^{\im \pv\cdot\rvp + \im k z} \bra a(\{\pv,k\},t) a^*(\{\pv,k\},t')\ket ,
\label{eq_blkOP_correl_gen}\eeq 
where we made use of translational invariance in all $d$ directions as well as of isotropy, due to which $\Ccal_b$ is solely a function of $r\equiv |\rv|$. 
Accordingly, the equal-time bulk correlation function is given by
\beq \Ccal_b(\rv, t) \equiv \Ccal_b(r,t) 
= \left\bra \phi(\rv, t) \phi(\rv'=\bv0, t) \right\ket  
= \int \frac{\d^{d} q}{(2\pi)^{d}} e^{\im \qv\cdot\rv} \bra |a(\qv,t)|^2\ket .
\label{eq_blkOP_correl_eqt}
\eeq 
In a bulk system, global OP conservation [\cref{eq_zeromode_totalOP}] is formally expressed as 
\beq \int \d^d r\, \phi(\rv,t) = a(\qv=\bv0,t) = a(\bv0,0)=\Phi=\const,
\label{eq_blkOP_totalint}\eeq 
where we have used \cref{eq_zeromode_sol}.
It is useful to remark that this implies
\beq \int \d^d r\, \Ccal_b(\rv,t) = \left\bra \phi(\bv0, t) \int \d^d r\, \phi(\rv,t) \right\ket = 0.
\label{eq_blkOP_correl_int}\eeq 

\subsection{Preliminaries}

\begin{figure}[t]\centering
    \subfigure[]{\includegraphics[width=0.39\linewidth]{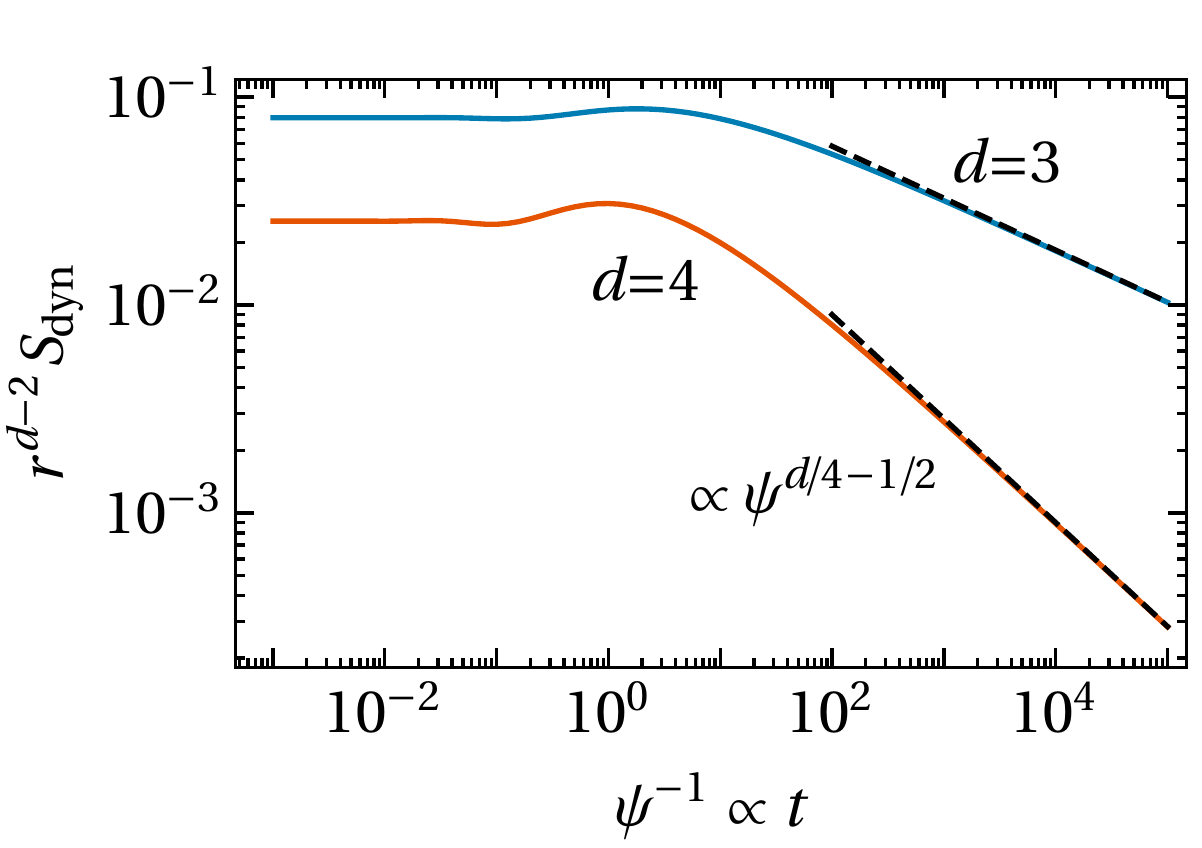} }\qquad
    \subfigure[]{\includegraphics[width=0.39\linewidth]{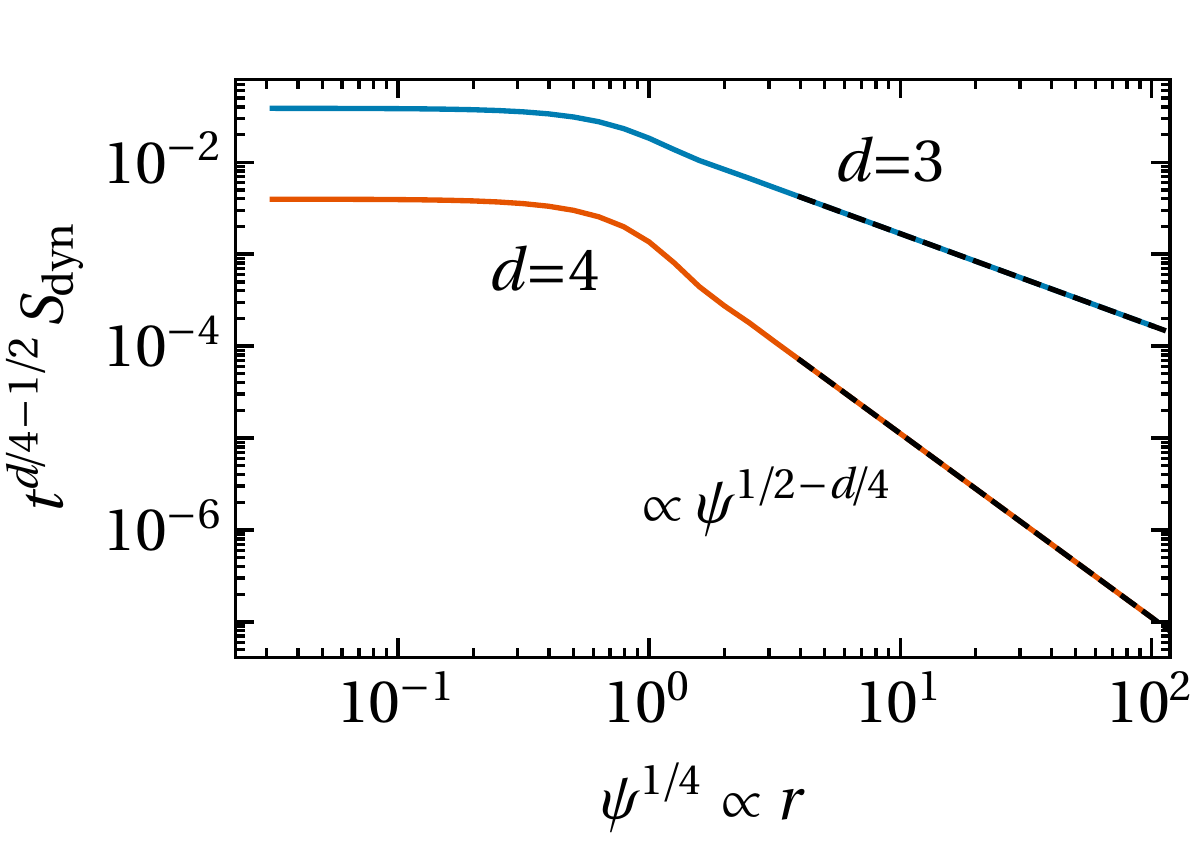} } 
    \caption{Auxiliary function $\Scal\dyn(r,t)$ [\cref{eq_Sdyn_def}] as function of (a) time $t$ and (b) distance $r$ for spatial dimensions $d=3$ and $d=4$. $\Scal\dyn(r,t/2)$ represents the equilibrium two-time correlation function in the bulk [see \cref{eq_Sdyn_twotime_eq}]. $\Scal\dyn$ is multiplied by appropriate prefactors which render it a dimensionless function of $\psi=r^4/(512 t)$ with $t=\gamma \tphys$. The actual time and distance dependences are expressed in terms of the dimensionless scaling variable $\psi$ [\cref{eq_scalvar_rt}]. The dashed lines represent the asymptotic behaviors applying to (a) late times [\cref{eq_Sdyn_limt} and (b) large distances [\cref{eq_Sstat}].}
    \label{fig_Sdyn}
\end{figure}

The equal-time bulk OP correlator given in \cref{eq_blkOP_correl_eqt,eq_modecorrel_eqt} can be determined explicitly for general $d$ at bulk criticality ($\tauLG=0$) \cite{baumann_kinetics_2007}.
According to \cref{eq_modecorrel_S} it suffices to discuss the real-space expressions of the auxiliary quantities defined in \cref{eq_fund_dyncorrel}, the bulk counterparts of which will correspondingly be denoted by $\Scal(\qv,t)$.
Turning first to the contribution $\Scal\dyn$ [see \cref{eq_fund_dyncorrel_1}], which can be interpreted as the equilibrium two-time bulk correlation function [see \cref{eq_Sdyn_twotime_eq}], one obtains
\beq\begin{split} 
\Scal\dyn(\rv, t)  &\equiv  \int \frac{\d^d q}{(2\pi)^d} \frac{e^{-2 \qv^4 t}}{\qv^2} e^{\im \qv\cdot\rv} 
= \frac{2^{1-d}\, }{\pi^{(1+d)/2} \Gamma[(d-1)/2]} \int_0^\infty \d q\, q^{d-3}  e^{-2 q^4 t} \int_{0}^\pi \d \theta\, (\sin \theta)^{d-2} \,  e^{\im q r \cos\theta}  \\
&= \frac{1 }{(2\pi)^{d/2}} r^{1-d/2} \int_0^\infty \d q\, q^{d/2-2} e^{-2 q^4 t}  J_{d/2-1}(q r) \\
&= \frac{2^{d/2-1} \pi^{1/2-d/2} }{ d\, \Gamma((2+d)/4)\, r^{d-2}} \psi^{d/4-1/2} \Bigg[ \frac{d\, \Gamma((d-2)/4)}{8\,\Gamma(d/4)}\, {}_1 F_3\left(\frac d 4 - \frac 1 2; \frac 1 2, \frac 1 2 + \frac d 4, \frac d 4; \psi \right) \\ 
&\qquad - \sqrt{\psi}\, {}_1 F_3\left(\frac d 4; \frac 3 2, \frac 1 2 + \frac d 4, 1+ \frac d 4; \psi \right)\Bigg], 
\end{split} \label{eq_Sdyn_def}\eeq 
where $J_\nu$ is a Bessel function of the first kind, ${}_1 F_3$ is a hypergeometric function \cite{olver_nist_2010}, and 
\beq \psi \equiv \frac{r^4}{512 t}
\label{eq_scalvar_rt}\eeq 
is a dimensionless scaling variable [recall \cref{eq_time_resc}], which is defined such that the crossover between the early- and late-time (or, correspondingly, the large- and small-distance) regime typically occurs for $\psi\sim\Ocal(1)$.
As in \cref{eq_blkOP_correl_eqt}, besides its dependence on $t$, $\Scal\dyn(\rv,t) = \Scal\dyn(r,t)$ is in fact a function of $r$ only.
The associated static contribution $\Scal\stat$ [\cref{eq_fund_dyncorrel_1}] follows from evaluating the first equation in \cref{eq_Sdyn_def} at $t=0$: 
\beq \Scal\stat(r) \equiv \Scal\dyn(r,t=0) = \frac{\Gamma(d/2-1)}{4\pi^{d/2} r^{d-2}}  \simeq  \Scal\dyn(r\to \infty, t).  
\label{eq_Sstat}\eeq 
As indicated, \Cref{eq_Sstat} also provides the leading asymptotic behavior of $\Scal\dyn$ [\cref{eq_Sdyn_def}] for $r\to \infty$. 
Accordingly, $\Scal\dyn$ is generally finite for $t\to 0$ and $r\neq 0$.
The asymptotic behaviors of $\Scal\dyn$ for $t\to\infty$ (at fixed $r\neq 0$) or, equivalently, for small $r$ (at fixed $t>0$) are given by 
\beq \Scal\dyn(r \neq 0, t\to\infty) \simeq \Scal\dyn(r\to 0, t>0) \simeq  \frac{2^{5/2-7d/4} \pi^{1/2-d/2}}{(d-2)\Gamma(d/4)} t^{1/2-d/4}. 
\label{eq_Sdyn_limt}\eeq 
We note that $\Scal\dyn$ is also finite for $r=0$, provided $t>0$.
The divergence of the static correlation function $\Scal\stat(r)$ [\cref{eq_Sstat}] for $r\to 0$ corresponds to the divergence of $\Scal\dyn(r\to 0, t>0)$ upon approaching $t=0$.
The behavior of $\Scal\dyn$ is illustrated in \cref{fig_Sdyn}.

\begin{figure}[t]\centering
    \subfigure[]{\includegraphics[width=0.38\linewidth]{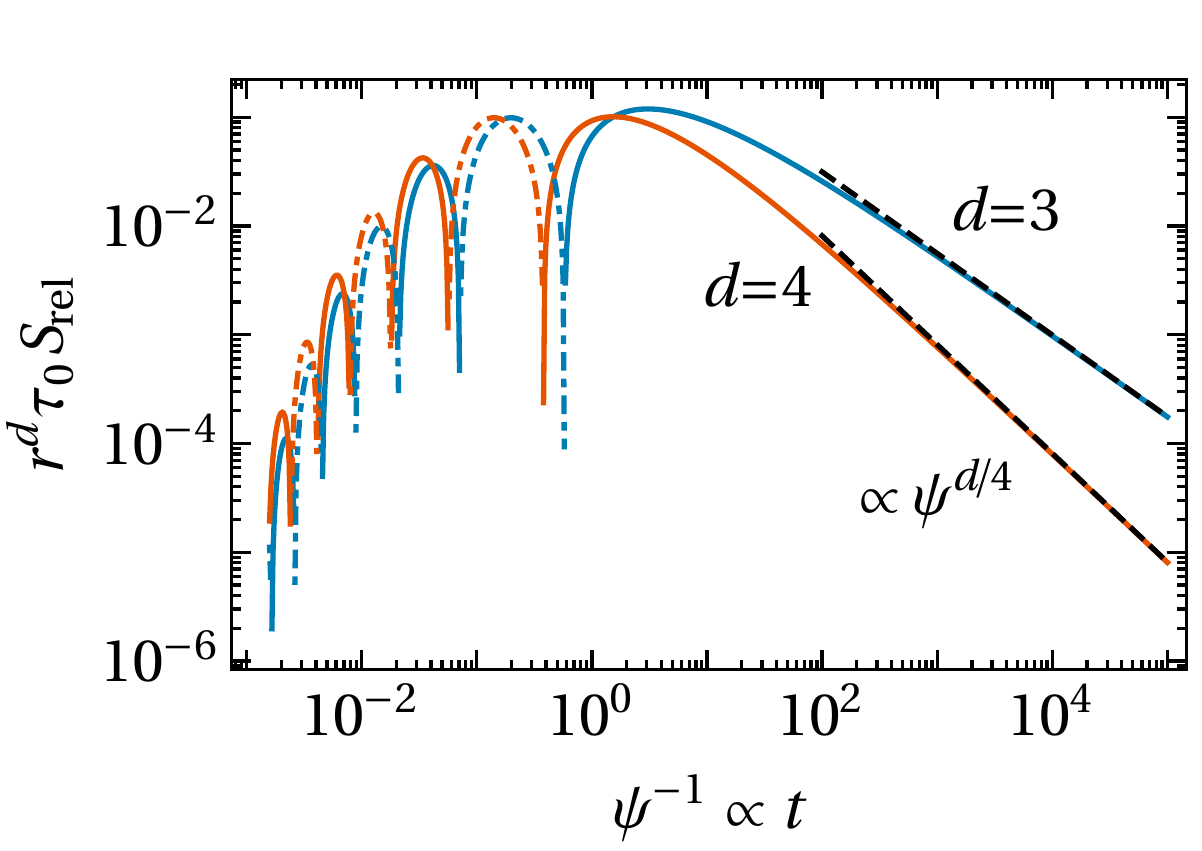} } \qquad 
    \subfigure[]{\includegraphics[width=0.386\linewidth]{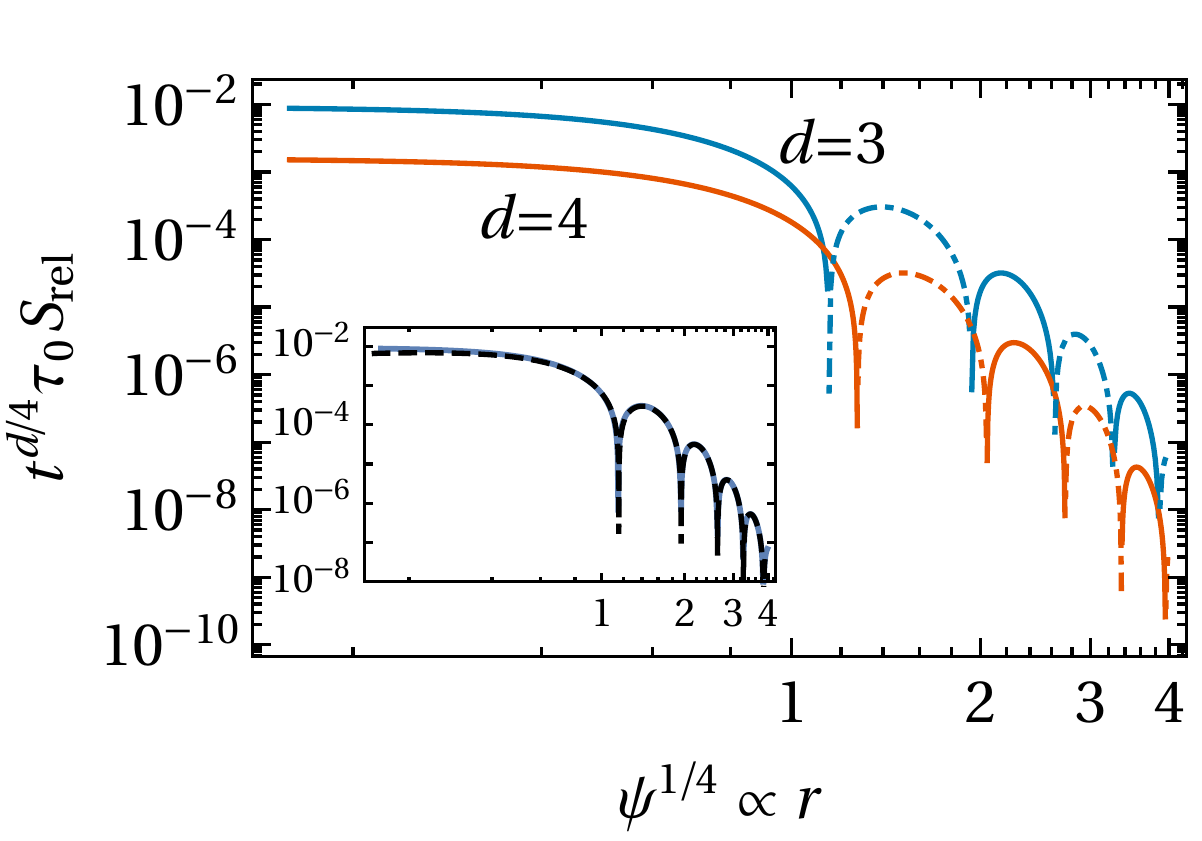} } \qquad
    \caption{Auxiliary function $\Scal\rel(r,t)$ [\cref{eq_Srel_def}] as function of (a) time $t$ and (b) distance $r$ for spatial dimensions $d=3$ and $d=4$. $\Scal\rel(r,t/2)$ is proportional to the Green function for the linear model B [see \cref{eq_Srel_Green}]. $\Scal\rel$ is multiplied by appropriate prefactors which render it a dimensionless function of $\psi=r^4/(512t)$ with $t=\gamma\tphys$. The actual time and distance dependences are expressed in terms of the dimensionless scaling variable $\psi$ [\cref{eq_scalvar_rt}]. The dashed-dotted parts correspond to $-\Scal\rel$. The dashed lines represent the asymptotic behaviors applying to (a) late times [\cref{eq_Srel_late_time} and (b) large distances [\cref{eq_Srel_largeR_asympt}]. For better visibility, the latter asymptote is compared with $\Scal\rel$ for $d=3$ (solid curve) separately in the inset. }
    \label{fig_Srel}
\end{figure}

We now turn to the auxiliary function $\Scal\rel$ [\cref{eq_fund_dyncorrel_2}], which can be interpreted as the bulk Green function for the linearized model B [see \cref{eq_Srel_Green}].
At bulk criticality ($\tauLG=0$), its real-space expression is given by 
\beq\begin{split} \Scal\rel(\rv, t)  &\equiv \frac{1}{\tauInit} \int \frac{\d^d q}{(2\pi)^d} e^{-2 \qv^4 t} e^{\im \qv\cdot\rv} \\
&= \frac{2^{d/2} \pi^{1/2-d/2} }{ \Gamma[(2+d)/4] \tauInit r^d } \psi^{d/4} \Big[\, {}_0 F_2\left( \frac{1}{2}, \frac{1}{2}+\frac{d}{4}; \psi \right) - \frac{2\sqrt{\psi}\Gamma[(2+d)/4]}{\Gamma(1+d/4)}\, {}_0 F_2\left( \frac 3 2, 1 + \frac d 4; \psi \right) \Big]. 
\end{split}\label{eq_Srel_def}\eeq 
This result can be readily obtained from \cref{eq_Sdyn_def} using the relation $\tauInit\, \Scal\rel = - \nabla^2\Scal\dyn$ (with $\tauInit=1/\initvar$), which follows directly from the Fourier representations of $\Scal\dyn$ and $\Scal\rel$. 
Due to translational invariance one has $\Scal\rel(\rv,t)=\Scal\rel(r,t)$.
For $t\to\infty$ (and any $r\geq 0$) as well as for $r\to 0$ (at fixed $t>0$), $\Scal\rel$ behaves asymptotically as
\beq \Scal\rel(r\neq 0, t\to\infty)  \simeq \Scal\rel(r\to 0, t>0) \simeq     \frac{2^{-7d/4} \pi^{(1-d)/2}}{\tauInit \Gamma[1/2+ d/4]} t^{-d/4} . 
\label{eq_Srel_late_time}\eeq 
In order to determine the leading behavior of $\Scal\rel(r,t)$ in the limit $r\to \infty$ for $t>0$ or, equivalently, in the limit $t\to 0$ for $r>0$, the subdominant asymptotics of the generalized hypergeometric function ${}_0 F_2$ is required (see \S 5.11.2 in Ref.\ \cite{luke_special_1969}). One obtains \footnote{Further studies of the asymptotic behavior of the Green function of general parabolic PDEs can be found, e.g., in Refs.\ \cite{boyd_fourier_2014, li_asymptotic_1993, evgrafov_asymptotic_1970, barbatis_higher_2012}.}
\beq \Scal\rel(r\to\infty, t>0) \simeq \Scal\rel(r>0, t\to 0) \simeq \frac{2^{1+d/2} }{\tauInit \pi^{d/2} \sqrt{3}\, r^d} \psi^{d/6} \exp\left(-\frac{3\psi^{1/3}}{2}\right) \cos[(d\pi - 9\sqrt{3} \psi^{1/3})/6].
\label{eq_Srel_largeR_asympt}\eeq 
This shows that $\Scal\rel$ vanishes in an exponentially damped oscillatory fashion at large distances or short times, respectively.
From the Fourier representation in \cref{eq_Srel_def} one infers that the divergence for $t\to 0$ of the expression in \cref{eq_Srel_late_time} turns, for $t=0$, into $\Scal\rel(\rv, t=0) = \tauInit^{-1} \delta^{(d)}(\rv)$.
\Cref{fig_Srel} summarizes the behavior of $\Scal\rel$.

The above expressions for $\Scal\dyn$ and $\Scal\stat$ are valid for $d\neq 2$ \footnote{Although we assume $d\geq 2$ in \cref{eq_Sdyn_def}, an explicit calculation confirms the final result to hold also for $d=1$. Note furthermore that \cref{eq_Sstat} develops a pole $\sim 1/(2-d)$ for $d\to 2$}. In two dimensions, the static correlation function, obtained as the fundamental solution of the Laplace equation, is given by the dimensionless expression $\Scal\stat = -\ln (r/\ell)/(2\pi)$ with some regularization length scale $\ell$ \cite{le_bellac_quantum_1991}. The two-dimensional case requires a careful treatment of the logarithmic divergences at short and large wavelengths and is not considered here further.

\subsection{Equal-time bulk correlation functions}
\label{sec_bulk_correl_onetime}

\subsubsection{Flat initial conditions}

\begin{figure}[t]\centering
    \subfigure[]{\includegraphics[width=0.38\linewidth]{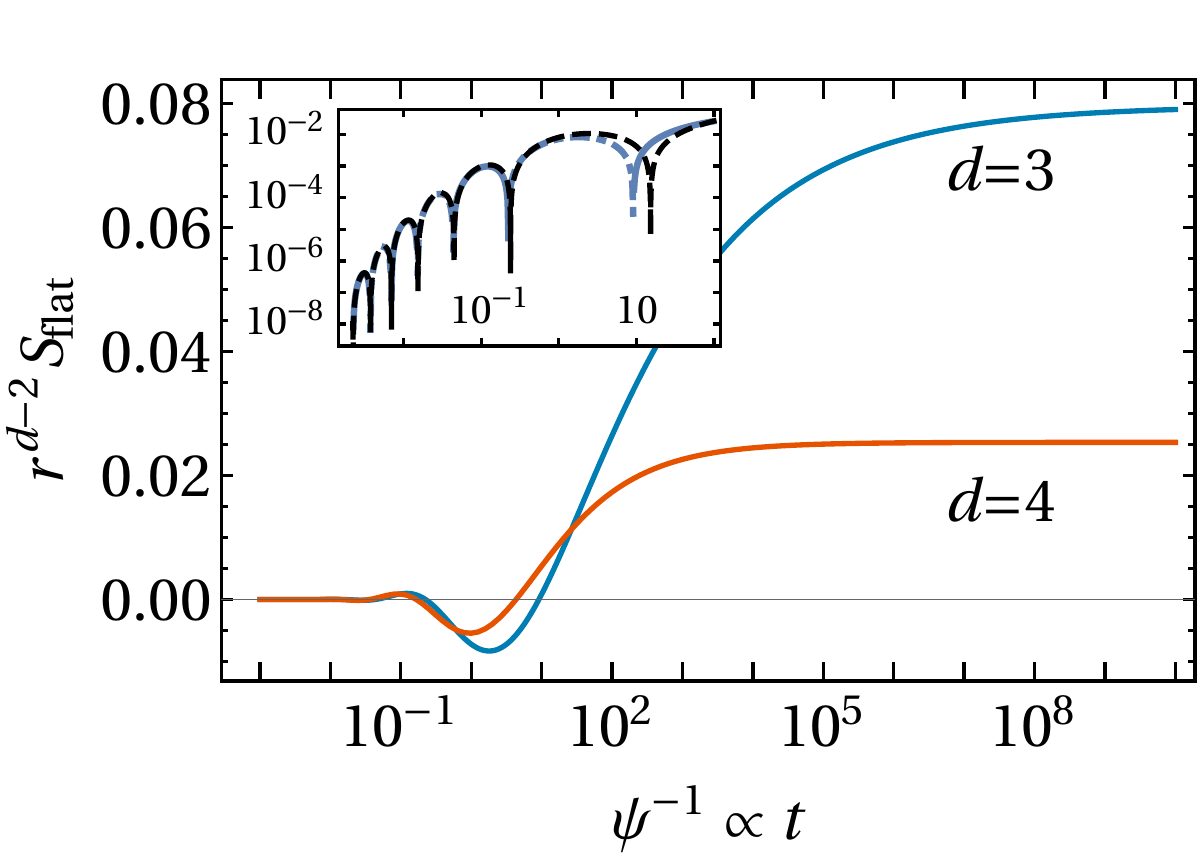} } \qquad 
    \subfigure[]{\includegraphics[width=0.385\linewidth]{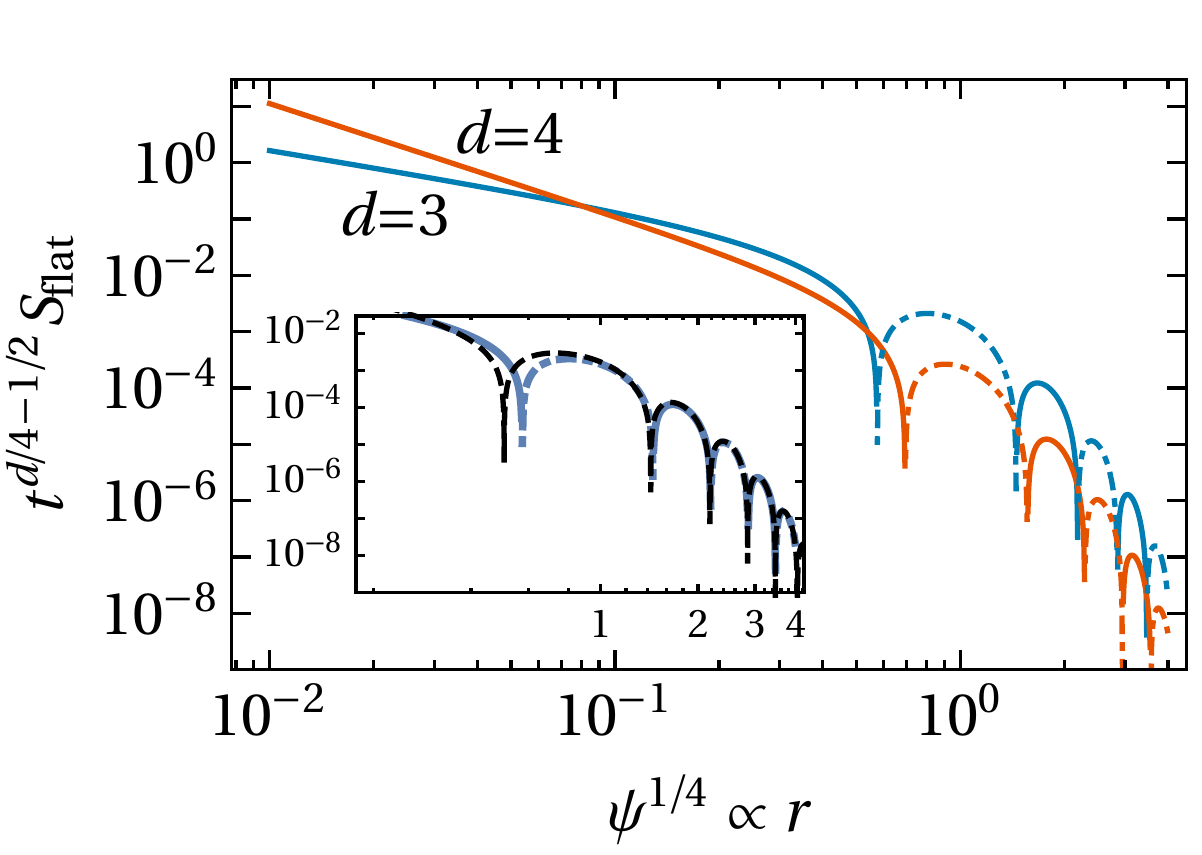} } \qquad
    \caption{Bulk correlation function $\Scal\flt(r,t)$ for flat ICs [\cref{eq_Sflat}], as function of (a) time $t$ and (b) distance $r$ for spatial dimensions $d=3$ and $d=4$. $\Scal\flt$ is multiplied by appropriate prefactors which render it a dimensionless function of $\psi=r^4/(512 t)$ with $t=\gamma \tphys$. The actual time and distance dependences are expressed in terms of the dimensionless scaling variable $\psi$ [\cref{eq_scalvar_rt}]. The dashed-dotted parts correspond to $-\Scal\flt$. In the insets, $\Scal\flt$ (for $d=3$) is compared with the asymptotic expression in \cref{eq_Sflat_asympt} (black dashed curve), which captures the exponentially damped oscillatory behavior of $\Scal\flt$ at early times or large distances, respectively.}
    \label{fig_Sflat}
\end{figure}

For flat ICs [\cref{eq_IC_correl_flat}] the equal-time OP correlator in the bulk $\Ccal_b$ [\cref{eq_blkOP_correl_eqt}] follows from \cref{eq_modecorrel_S_flat}:
\beq \Ccal_b(r,t) = \Scal\flt(r,t) \equiv \Scal\stat(r) - \Scal\dyn(r,t)  .
\label{eq_Sflat}\eeq 
Using \cref{eq_Sdyn_def,eq_Sstat} it can be shown that $\Scal\flt$ fulfills the scaling behavior expressed in \cref{eq_scalform_Cb} (with $\amplCorrel \sim (\amplXip)^{2-d}$ up to numerical constants and within the Gaussian approximation, consistent with Ref.\ \cite{gambassi_critical_2006}).
In particular, $r^{d-2}\Scal\flt$ is solely a function of the scaling variable $\psi$ [\cref{eq_scalvar_rt}], implying that the asymptotic behaviors of $\Scal\flt$ for $r\to\infty$ and $t\to 0$ are closely related. 
The leading asymptotic behavior of $\Scal\flt$ is given by
\beq \Scal\flt(r\to \infty, t) \simeq \Scal\flt(r>0, t\to 0) \simeq - \frac{1  }{ 2^{3-d/2} 3^{1/2} \pi^{d/2}\, r^{d-2} } \psi^{(d-4)/6} e^{-3\psi^{1/3}/2} \sin\left[\frac{(1-d)\pi}{6} + \frac{3\sqrt{3}}{2} \psi^{1/3}\right],
\label{eq_Sflat_asympt}\eeq 
which follows from the asymptotic behavior of the generalized hypergeometric function ${}_1 F_3$ in next-to-leading order (see \S 5.11.2 in Ref.\ \cite{luke_special_1969}).
According to \cref{eq_Sdyn_limt}, $\Scal\dyn$ approaches $\Scal\stat$ algebraically in $t$ for large $t$, such that 
\beq \Scal\flt(r,t\to \infty)\simeq \Scal\stat(r).
\label{eq_Sflat_eqlim}\eeq 
This reflects that, in the bulk, equilibrium is established for $t\to\infty$.
The behavior of $\Scal\flt$ for $r\to 0$ is singular. The intermediate result in \cref{eq_Sdyn_def} can be written as
\beq \Scal\flt(r,t) = \frac{1 }{(2\pi)^{d/2}} \, r^{2-d} \int_0^\infty \d u \left(1- e^{-2 u^4 t/r^4}\right) u^{d/2-2} J_{d/2-1}(u).
\label{eq_Sflat_int}\eeq
Next, performing the limit $t\to 0$ for $r>0$ fixed, yields $\Scal\flt(r,t=0)=0$, which is also expected from \cref{eq_Sstat}.
However, for any $t>0$ we can find a sufficiently small $r$ such that the exponential term in the integral in \cref{eq_Sflat_int} becomes negligible. 
This renders the asymptotic behavior
\beq \Scal\flt(r\to 0, t>0) \simeq   \frac{1}{4\pi^{d/2} r^{d-2}} \Gamma(d/2-1) = \Scal\stat(r),
\label{eq_Sflat_smallR}\eeq 
which implies that, for any finite $t>0$, $\Scal\flt$ diverges $\propto r^{2-d}$ at small $r$. (Determining the auto-correlation function $\Ccal_b(\rv=\bv0, t, t')$ thus requires a suitable regularization in general, and is not considered here further.)
The behavior of $\Scal\flt$ is illustrated in \cref{fig_Sflat}. 
As shown in the insets, the asymptotic expression in \cref{eq_Sflat_asympt} accurately captures the exponentially damped oscillatory behavior of $\Scal\flt$ at early times or, correspondingly, at large distances.

As anticipated in the Introduction, the qualitative behavior of a correlation function $\Ccal$ is often characterized in terms of a time-dependent correlation length $\xi(t)$ \cite{amit_field_2005}. 
In the present case, however, due to \cref{eq_blkOP_correl_int}, the standard definition of $\xi(t)$ as the second moment of $\Ccal_b$,
$\xi^2(t) = \frac{\int_V \d^d r\, r^2 \Ccal_b(\rv, t) }{\int_V \d^d r\, \Ccal_b(\rv,t)}$, where $V$ is an arbitrary but sufficiently large volume, is ill-defined.
This definition of $\xi^2(t)$ is problematic even if one restricts the integral to a lower-dimensional plane: since, for model B, $\Ccal_b(r,t)$ is typically decaying in an oscillatory way, one could obtain $\xi^2<0$.
Instead, a more suitable measure of the growth of the correlation volume is provided by the first zero-crossing $\xi_\times$  of $\Ccal_b$. 
Applying this criterion to $\Scal\flt$, the asymptote in \cref{eq_Sflat_asympt} renders 
\beq
\xi_\times \propto t^{1/4},
\eeq 
which is consistent with the dynamic scaling hypothesis for model B in the Gaussian limit \cite{hohenberg_theory_1977}. 
The same scaling emerges also for an effective $\xi$ which is defined on the basis of the asymptotic exponential decay reported in \cref{eq_Sflat_asympt}.

\subsubsection{Thermal initial conditions}

In the case of thermal ICs [\cref{eq_IC_correl}], the bulk correlation function is given according to \cref{eq_modecorrel_S_th} by 
\beq \Ccal_b(r,t) = \Scal\th(r,t) = \Scal\rel(r,t) + \Scal\flt(r,t).
\label{eq_Stherm}\eeq 
From \cref{eq_Sdyn_def,eq_Sstat,eq_Srel_def} it follows that $\Scal\th$ fulfills the scaling behavior expressed in \cref{eq_scalform_Cb_init} (with $\amplCorrel \sim (\amplXip)^{2-d}$ up to numerical constants and within the Gaussian approximation).
Since, at late times, $\Scal\flt$ approaches $\Scal\stat$ [see \cref{eq_Sflat_eqlim}] while $\Scal\rel$ vanishes [see \cref{eq_Srel_late_time}], the late-time behavior of $\Scal\th$ is always dominated by $\Scal\flt$. 
Furthermore, owing to \cref{eq_fund_dyncorrel_2}, one has $\Scal\rel\propto 1/\tauInit$, such that, in the limit of a large initial temperature, i.e.,  $\tauInit\to \infty$ [see \cref{eq_static_OZ_correl}], one obtains $\Scal\th \simeq \Scal\flt$.
We recall that, in fact, thermal ICs accurately describe an initial equilibrium state only in this limit $\tauInit\to \infty$ [see the discussion around \cref{eq_static_OZ_correl}].

\section{Correlation functions in confinement}
\label{sec_correl_confined}

In the following, we focus on the \emph{equal-time} correlation function $\Ccal(\rvp,z,z',t)\equiv \Ccal(\rvp-\rvp',z,z',t,t'=t)$ of a finite $d$-dimensional system having a cuboidal box geometry.
Since we assume periodic \bcs in the subspace containing $\rvp$ (see \cref{fig_boxsketch}), $\Ccal$ is translationally invariant in $\rvp$. This is important for determining critical Casimir forces.
The thin film limit is readily obtained from the results for a box, as discussed below.
We recall that, at finite times $t$, the expression in \cref{eq_modecorrel_eqt} for the mode correlations $\bra |a_n(\pv,t) |^2 \ket$ applies to all modes including the zero mode $a_0(\bv0,t)$. [This is not the case in the limit $t\to\infty$, see the discussion around \cref{eq_eqmodecorrel_eqt_gc}.]
Accordingly, we obtain
\beq 
\Ccal(\rvp, z, z', t) = \frac{1}{A} \sum_{\nvp,n_d}  e^{\im \pv_{\nvp}\cdot\rvp  } \sigma_{n_d}(z)\sigma_{n_d}^*(z') \bra |a_{n_d}(\pv_{\nvp},t)|^2\ket ,
\label{eq_OPcorrel_eqt}
\eeq
where the eigenfunctions $\sigma_n$ are given in \cref{eq_eigenspec} and the notation is explained in \cref{eq_rvsplit,eq_eigenval_lateral}.
Applying the Poisson resummation formula [see \cref{eq_Poisson_periodic,eq_Poisson_Neumann} in \cref{app_Poisson}] renders 
\beq 
\Ccal(\rvp, z, z', t) =\Ncal \sum_{\{m_\alpha=-\infty\}}^{\{\infty\}}  \int \frac{\d^d q}{(2\pi)^d} e^{\im
\qv_\parallel\cdot \rvp + \qv \cdot \Lcaltildev } \hat\sigma_{q_d}(z) \hat\sigma_{q_d}^*(z') \bra |a(\qv,t)|^2\ket,
\label{eq_OPcorrel_box}
\eeq
where we use the notation $\sum_{\{m_\alpha=-\infty\}}^{\{\infty\}}\equiv  \sum_{m_1=-\infty}^\infty \cdots \sum_{m_d=-\infty}^\infty$ and define
\begin{subequations}
\begin{align}
\Ncal\pbc = 1,\qquad \hat\sigma_{q_d}\pbc(z) = e^{\im q_z z}, \\
\Ncal\Nbc = 2,\qquad  \hat\sigma_{q_d}\Nbc(z) = \cos(q_d z), 
\end{align}
\end{subequations}
for periodic and Neumann \bcs, respectively. Furthermore, the vector $\qv$ is decomposed into lateral and transverse components as $\qv=\{\qv_\parallel,q_d\}$, while $\Lcaltildev$ stands for the $d$-dimensional vector 
\beq \Lcaltildev\equiv \{m_1 L_\parallel, \ldots, m_{d-1} L_\parallel, \Ncal m_d L\},
\label{eq_mL_notation}\eeq 
which, for $\Ncal=1$ (periodic \bcs), we shall simply write as $\Lcalv$.
The notation $a(\qv,t)$ indicates the continuum case, as explained in \cref{sec_bulk_correl}.

For periodic \bcs, \cref{eq_OPcorrel_box} reduces to 
\beq 
\Ccal\pbc(\rvp, z-z', t,L) \equiv \Ccal\pbc(\rvp, z, z', t,L)  = \sum_{\{m_\alpha=-\infty\}}^{\{\infty\}}  \Ccal_b(\{\rvp, z-z'\} + \Lcalv, t) ,
\label{eq_OPcorrel_box_pbc}\eeq 
where $\Ccal_b$ denotes the (equal-time) bulk correlation function [see \cref{eq_blkOP_correl_eqt}].
For Neumann \bcs, instead, using 
\beq \cos(q z)\cos(q z') = \onehalf \left[ \cos(q(z-z')) + \cos(q(z+z')) \right] ,
\eeq 
\cref{eq_OPcorrel_box} turns into (see also Ref.\ \cite{diehl_dynamic_2009})
\beq \begin{split} 
\Ccal\Nbc(\rvp, z, z', t,L) &= \onehalf \sum_{\{m_\alpha=-\infty\}}^{\{\infty\}} \Big[ \Ccal_b(\{\rvp, z-z'\} + \Lcaltildev, t) + \Ccal_b(\{\rvp, -(z-z')\} + \Lcaltildev, t) \\ &\qquad + \Ccal_b(\{\rvp, z+z'\} + \Lcaltildev, t) + \Ccal_b(\{\rvp, -(z+z')\} + \Lcaltildev, t) \Big] \\
&= \sum_{\{m_\alpha=-\infty\}}^{\{\infty\}}  \left [\Ccal_b(\{\rvp, z-z'\} + \Lcaltildev, t)  + \Ccal_b(\{\rvp, z+z'\} + \Lcaltildev, t) \right] \\
&= \Ccal\pbc(\rvp,z-z',t,2L) + \Ccal\pbc(\rvp,z+z',t,2L),
\end{split}\label{eq_OPcorrel_box_Nbc}\eeq  
where we have employed \cref{eq_OPcorrel_box_pbc} in order to obtain the final result, which involves the correlation function for periodic \bcs in a box of thickness $2L$ \footnote{The second equality in \cref{eq_OPcorrel_box_Nbc} can be proven by using the Fourier expression for $\Ccal_b$ [\cref{eq_blkOP_correl_eqt}] and the fact that $\bra |a(-\qv,t)|^2\ket = \bra |a(\qv,t)|^2\ket$.}.
Owing to the periodicity $\Ccal\pbc(\rvp,z=2L,t,2L)=\Ccal\pbc(\rvp,z=0,t,2L)$, so that, in the special case $z=z'=0$ or $z=z'=L$, \cref{eq_OPcorrel_box_Nbc} reduces to
\beq 
\Ccal\Nbc(\rvp, z=z'\in\{0,L\}, t,L) = 2\sum_{\{m_\alpha=-\infty\}}^{\{\infty\}}  \Ccal_b(\{\rvp, 0\} + \Lcaltildev, t)  = 2\, \Ccal\pbc(\rvp,0,t,2L).
\label{eq_OPcorrel_box_Nbc_zcoinc}\eeq 
This equation applies also to derivatives of $\Ccal\Nbc$ with respect to $\rv_{\parallel,\alpha}$, $z$, or $z'$, noting that $\pd_z^n\Ccal\pbc(\rvp,z,t,2L)=0$ for $z\in \{0,2L\}$ and $n$ odd.

In \emph{equilibrium}, obtained for $t\to\infty$, the zero mode must be treated carefully due to the non-interchangeability of the limits $t\to\infty$ and $\{\nvp,n_d\}\to \bm 0$ in the expression for the correlation function in \cref{eq_OPcorrel_box} [see \cref{eq_eqmodecorrel_eqt} and the associated discussion].
This is reflected by the fact that $\bra|a_{n_d}(\pnvp)|^2\ket$ given in \cref{eq_eqmodecorrel_eqt} is a discontinuous function of $\{\nvp,n_d\}$.
Before applying the Poisson resummation formula in the equilibrium case, one must therefore separate the zero mode from the mode sum in \cref{eq_OPcorrel_box}, i.e.,
\beq\begin{split} 
\Ccal\eq(\rvp, z, z',L) &= \frac{1}{A} \sump_{\nvp,n_d}  e^{\im \pnvp\cdot\rvp  } \sigma_{n_d}(z)\sigma_{n_d}^*(z') \bra |a_{n_d}(\pnvp)|^2\ket\eq + \pfrac{\Phi}{AL}^2 \\
&=  \Ccal\eqGC(\rvp, z, z') - \frac{1}{AL\, \tauLG}, 
\end{split}\label{eq_OPcorrel_box_eq}\eeq
where $\sump$ stands for the sum excluding the single mode $n_1=\ldots = n_d=0$, $\Phi$ is the total OP [see \cref{eq_total_OP}], and we have used $\LambdaRed_{0}(\bv0,\tauLG)=\tauLG$ [see \cref{eq_Lambda_red}].
Furthermore, we have introduced here the \emph{grand canonical} static equilibrium correlation function
\beq \Ccal\eqGC(\rvp, z, z',L) = \frac{1}{A} \sum_{\nvp,n_d}  e^{\im \pnvp\cdot\rvp  } \sigma_{n_d}(z)\sigma_{n_d}^*(z') \frac{1}{\LambdaRed_{n_d}(\pnvp,\tauLG)} + \pfrac{\Phi}{AL}^2,
\label{eq_OPcorrel_box_eq_gc}\eeq 
which does not involve a constraint on the zero mode [see also \cref{eq_eqmodecorrel_eqt_gc}].
The last term in \cref{eq_OPcorrel_box_eq_gc} is due to the fact that we consider the correlations of the actual OP $\phi$ and not of its fluctuation $\phi-\Phi/(AL)$. (In the present study, we focus on $\Phi=0$.)
Since \cref{eq_OPcorrel_box_eq} holds for all \bcs considered here, upon applying the Poisson resummation formula to \cref{eq_OPcorrel_box_eq_gc}, one obtains the same expression for $\Ccal\eqGC$ as reported in \cref{eq_OPcorrel_box_pbc,eq_OPcorrel_box_Nbc} but with $\Ccal_b(\rv,t)$ replaced by $\Ccal_{b,\text{eq}}(\rv)$.

\Cref{eq_OPcorrel_box_eq} coincides with the expression for the correlation function in the canonical ensemble, derived within equilibrium field theory in Ref.\ \cite{gross_statistical_2017}.
The term $-1/(AL\tauLG)$ represents the constraint-induced correction stemming from the non-fluctuating character of $\Phi$ [see \cref{eq_zeromode_sol}]. This term diverges at criticality ($\tauLG\to 0$) and cancels the corresponding divergence of the grand canonical correlation function $\Ccal\eqGC$ [\cref{eq_OPcorrel_box_eq_gc}], such that the canonical one, $\Ccal\eq$ [\cref{eq_OPcorrel_box_eq}], stays finite for $\tauLG\to 0$ \footnote{The divergent behavior of the correlation function is a known artifact of Gaussian field theory in the grand canonical ensemble (see, e.g., Refs.\ \cite{chen_order-parameter_1997, gruneberg_thermodynamic_2008, gross_statistical_2017}).}.

In the case of a thin film, where the transverse area $A=L_\parallel^{d-1} \to\infty$ is infinitely extended, the $(d-1)$-dimensional sum over $\pv_n$ in \cref{eq_OPcorrel_eqt} has to be replaced according to \footnote{The replacement rule in \cref{eq_continuum_repl_C} applies to a correlation function and is therefore different from the one in \cref{eq_continuum_repl_modes}, which applies to the OP field.}
\beq \frac{1}{A} \sum_{\nv} f(\pv_{\nv}) \to \int \frac{\d^{d-1} p}{(2\pi)^{d-1}} f(\pv).
\label{eq_continuum_repl_C}\eeq 
Consequently, the associated Poisson representation of the correlation function in \cref{eq_OPcorrel_box} involves only the sum over $m_z$, such that \cref{eq_OPcorrel_box_pbc,eq_OPcorrel_box_Nbc,eq_OPcorrel_box_Nbc_zcoinc} reduce to
\begin{subequations}
\begin{align}
\Ccal\pbc(\rvp, z-z', t,L) &= \sum_{m=-\infty}^{\infty} \Ccal_b(\{\rvp, z-z' + m L\} , t) , \label{eq_OPcorrel_thinfilm_per} \\
\Ccal\Nbc(\rvp,z,z',t,L) &= \sum_{m=-\infty}^{\infty}  \left [\Ccal_b(\{\rvp, z-z' + 2 m L\} , t)  + \Ccal_b(\{\rvp, z+z' + 2 m L\} , t) \right], \label{eq_OPcorrel_thinfilm_Neu}\\
\Ccal\Nbc(\rvp, z=z'\in\{0,L\}, t,L) &= 2\sum_{m=-\infty}^{\infty} \Ccal_b(\{\rvp, 2m L\} , t)  = 2\, \Ccal\pbc(\rvp,0,t,2L). \label{eq_OPcorrel_thinfilm_Neu_zcoinc}
\end{align}\label{eq_OPcorrel_thinfilm}
\end{subequations}
\hspace{-0.12cm}Further analytic expressions for these correlation functions are provided in \cref{sec_film_correl}.
Importantly, the constraint-induced correction $-1/(AL\tauLG)$ vanishes in \cref{eq_OPcorrel_box_eq} both in the thin film limit and in the bulk limit, implying that the corresponding correlation function is identical in the canonical and grand canonical ensembles: 
\beq \Ccal\eq(\rvp,z,z') = \Ccal\eqGC(\rvp,z,z'),\qquad \text{($A\to\infty$ or $V\to\infty$)}.
\label{eq_OPcorrel_thinfilm_eq}
\eeq 
We note, however, that this property is specific to non-symmetry-breaking \bcs, while, for symmetry-breaking \bcs, in a thin film \cite{gross_critical_2016} ensemble differences can be relevant.

\section{Critical Casimir force}
\label{sec_CCF}

\begin{figure}[t]\centering
    \includegraphics[width=0.39\linewidth]{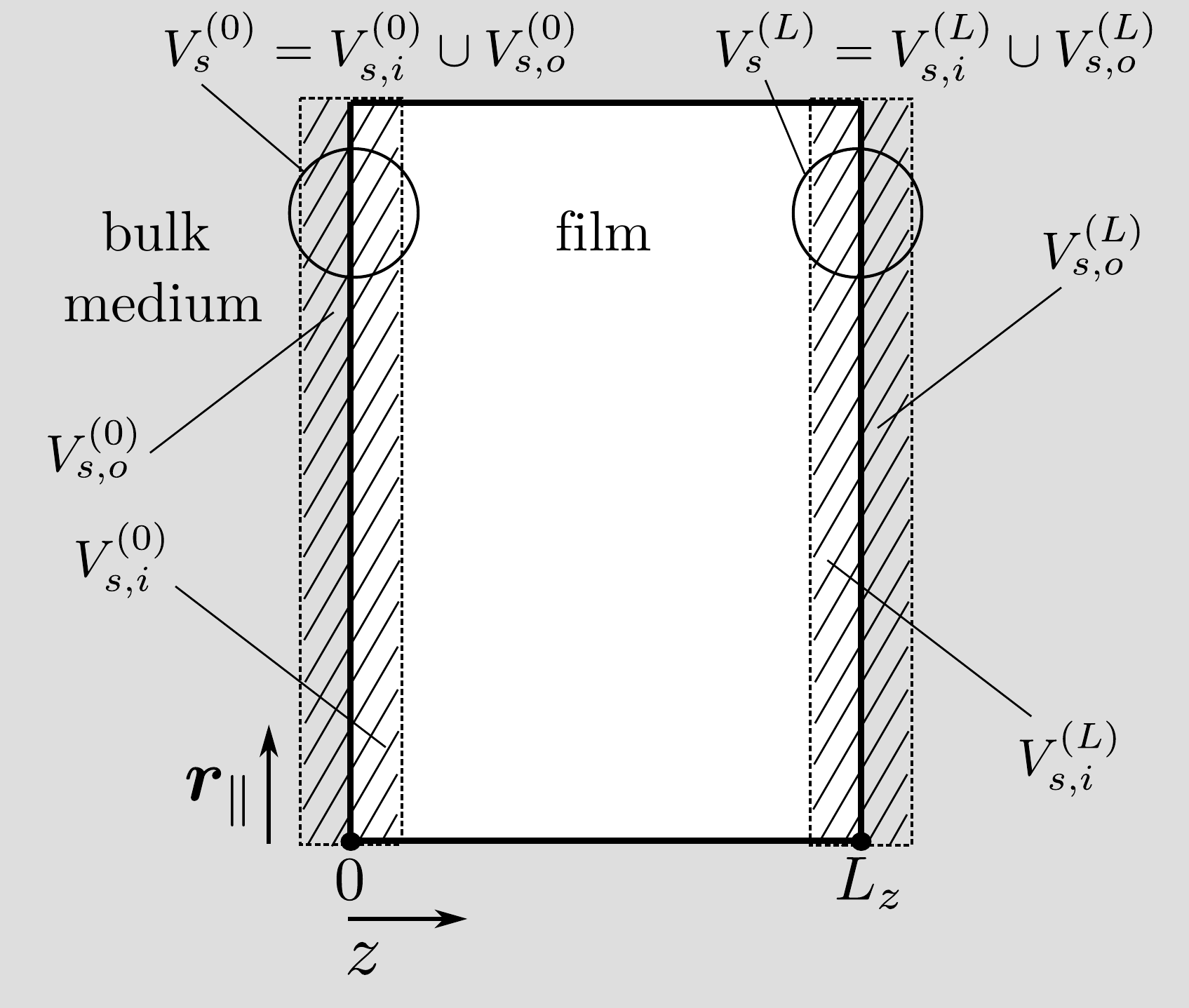}
    \caption{Schematic view of the system under consideration, projected onto one of the lateral dimensions. 
    The hatched areas represent the integration volumes $V_s^{(0,L)}$ around the surfaces at $z=0$ and $z=L$, entering into \cref{eq_CCF_stress}, while the shaded area illustrates the bulk medium at the outside. The thickness of $V_s$ is taken to be infinitesimal, which facilitates the use of no-flux \bcs [or, correspondingly, the symmetry of the in-going and out-going fluxes in the case of periodic \bcs; see the discussion after \cref{eq_CCF_genforce}]. $V_{s,\{i,o\}}^{(0,L)}$ denotes the volumes of the inner/outer fluid at the wall at $z=0$ and $z=L$, respectively; $V_s^{(0,L)}=V_{s,i}^{(0,L)}\cup V_{s,o}^{(0,L)}$.
    }
    \label{fig_sketch_CCF}
\end{figure}

Following Refs.\ \cite{dean_out--equilibrium_2010, rohwer_transient_2017, rohwer2018forces, kruger_stresses_2018, gross_surface-induced_2018}, we define the CCF in terms of the generalized force exerted by the OP field on an inclusion (e.g., a surface) in the system.
For a given instantaneous (i.e., spatially varying, but fixed) field configuration $\phi(\rv)$, the generalized force acting in direction $i$ on a single surface described by an energy density $\Hcal_s(l_i, \phi(\rv))$ localized at $r_i=l_i$ is defined by \cite{dean_out--equilibrium_2010,kruger_stresses_2018, gross_surface-induced_2018}
\beq
K_i  \equiv - \frac{\pd \Fcal}{\pd  l_i} =  -\int_V \d^d r \frac{\partial \Hcal(\rv,\phi(\rv),\nabla\phi(\rv))}{\pd l_i}.
\label{eq_CCF_potential}
\eeq 
Here, $\Fcal = \int_V \d^d r\, \Hcal$ denotes the total energy, with $\Hcal=\Hcal_b+\Hcal_s$ and the bulk energy density $\Hcal_b$ [see \cref{eq_Hamiltonian}].
$V$ is the system volume independent of $l_i$.
In the case of the bounding surfaces of the system, located at $z=L$ and $z=0$, respectively, the generalized force $K_i$ in direction $i=z$ can be expressed as \cite{kruger_stresses_2018, gross_surface-induced_2018}
\begin{subequations}
\begin{align}
K_i^{(L)}   &= - \int_{\pd V_s^{(L)}} \d^{d-1} s_j\, \bar \Tcal_{ij} + \int_{V_s^{(L)}} \d^d r\, (\nabla_i \mu)\phi\, ,\\
K_i^{(0)}   &=  + \int_{\pd V_s^{(0)}} \d^{d-1} s_j\, \bar \Tcal_{ij} - \int_{V_s^{(0)}} \d^d r\, (\nabla_i \mu)\phi\, ,
\end{align}\label{eq_CCF_stress}
\end{subequations}
\hspace{-0.12cm}where $V_s^{(0)}$ and $V_s^{(L)}$ are cuboid-shaped volumes of infinitesimal thickness enclosing the surfaces at $z=L$ and $z=0$, respectively (see \cref{fig_sketch_CCF}),
\beq \bar \Tcal_{ij} \equiv \Tcal_{ij} + \mu\phi \delta_{ij}
\label{eq_stressten_dyn}\eeq 
defines the \emph{dynamical} stress tensor,  
\beq \Tcal_{ij} = \frac{\pd \Hcal}{\pd\nabla_i\phi} \nabla_j \phi - \delta_{ij} \Hcal
\label{eq_stressten_gc}\eeq 
is the standard (grand canonical) stress tensor \cite{krech_casimir_1994}, and 
\beq \mu =  \frac{\delta \Fcal}{\delta\phi} =\frac{\pd\Hcal}{\pd\phi} - \nabla_j\left(\frac{\pd\Hcal}{\pd\nabla_j\phi}\right) 
\label{eq_chempot}\eeq 
is the chemical potential. 
The different signs in \cref{eq_CCF_stress} stem from the symmetrization of the derivative with respect to $l_z$ in \cref{eq_CCF_potential}, which is necessary in order to obtain equal but opposite forces on the boundaries of the system (see Ref.\ \cite{gross_surface-induced_2018}).
The above expressions apply for any generic $\Hcal_b$ which is at most quadratic in gradients of $\phi$, but can contain arbitrary powers of $\phi$.

The generalized force $K_z$ (per area $A$), averaged over the left and right boundary of the system, renders the dynamic CCF $\Kcal$:
\beq \Kcal = \frac{1}{2A}\left( K_z^{(0)} + K_z^{(L)} \right).
\label{eq_CCF_genforce}\eeq 
Upon thermal averaging, the first term on the r.h.s.\ of \cref{eq_CCF_stress} provides identical contributions to $\Kcal$ at the two boundaries.
If $\phi$ as well as the noise [see \cref{eq_flux}] obey \emph{no-flux} \bcs at the surfaces, the second term on the r.h.s.\ of \cref{eq_CCF_stress} vanishes due to the infinitesimal thickness of $V_s$. In fact, in this case, the symmetrized definition in \cref{eq_CCF_genforce} is not necessary. 
For periodic \bcs, instead, non-vanishing fluxes across the surfaces imply that the last term in \cref{eq_CCF_stress} does not vanish at any single boundary.
However, we have $\nabla_i \mu|_{z=0} = \nabla_i\mu|_{z=L}$, such that the last term on the r.h.s.\ of \cref{eq_CCF_stress} cancels in the sum in \cref{eq_CCF_genforce}.
We note that, in thermal equilibrium, the second term on the r.h.s.\ of \cref{eq_CCF_stress} is zero on average, i.e., $\langle (\nabla_i \mu)\phi \rangle = 0$ [see \cref{eq_SchwingerDyson} below and Ref.~\cite{dean_out--equilibrium_2010}].

Taking into account the direction of the surface normals (which point outward from the respective volumes $V_s$), the mean CCF follows from \cref{eq_CCF_stress,eq_CCF_genforce} as
\beq\begin{split} \bra \Kcal(t) \ket   &= \frac{1}{A} \int_{\pd V^{(L,0)}_{s,i}} \d^{d-1} s^{(L,0)}_z(\rvp,z)\, \bra \bar\Tcal_{zz}(\rvp, z, t) \ket - \frac{1}{A} \int_{\pd V^{(L,0)}_{s,o}} \d^{d-1} s^{(L,0)}_z(\rvp,z)\, \bra \bar\Tcal_{zz,b}(\rvp, z, t) \ket \\
&\equiv  \bra \Pcal_f(t)\ket - \bra \Pcal_b(t)\ket  ,
\end{split}
\label{eq_CCF_Tzz_gen}\eeq 
where $\pd V^{(L,0)}_{s,\{i,o\}}$ denotes the surface of the inner/outer fluid at $z=L$ and $z=0$, respectively (see \cref{fig_sketch_CCF})
[with the corresponding area element $\d^{d-1} s_z(\rvp,z)$], and where $\bar \Tcal_{zz,b}$, accordingly, represents the bulk stress.
As set out in \cref{eq_CCF_def}, the last line in \cref{eq_CCF_Tzz_gen} defines the averaged film and bulk pressures. 
We shall use the notion of a \emph{film pressure} both in the case of a thin film and that of a box geometry.

For a Gaussian Hamiltonian density of the Landau-Ginzburg form [\cref{eq_Hamiltonian}], the chemical potential [\cref{eq_chempot}] reduces to the bulk value $\mu=-\nabla^2 \phi + \tauLG\phi$ and the dynamic stress tensor in \cref{eq_stressten_dyn} takes the form
\beq \bar \Tcal_{zz} = \onehalf (\pd_z \phi)^2  - \onehalf \sum_{\alpha=1}^{d-1} (\pd_\alpha \phi)^2 + \onehalf \tauLG \phi^2 - \phi \nabla^2\phi .
\label{eq_Tzz_dyn}\eeq 
In the following, we focus on a \emph{critical quench}, i.e., $\tauLG=0$; as before, we also assume $\Phi=0$.
Accordingly, the mean CCF in \cref{eq_CCF_Tzz_gen} is given by 
\beq
\begin{split}
\bra \Kcal(t) \ket &= \bra \Pcal_f(t) \ket - \bra\Pcal_b(t) \ket \\
&= \Bigg[\onehalf \bra (\pd_z \phi(\rvp, z, t) )^2\ket - \bra \phi(\rvp, z, t) \pd_z^2 \phi(\rvp, z, t)\ket - \onehalf \sum_{\alpha=1}^{d-1} \bra (\pd_\alpha \phi(\rvp, z, t))^2\ket \\
&\quad - \Big\bra \phi(\rvp, z, t) \sum_{\alpha=1}^{d-1} \pd_\alpha^2 \phi(\rvp, z, t) \Big\ket \Bigg]_{{\rvp=\bvnp}\atop{z \in \{0,L\}}}  - \text{(bulk)} ,
\end{split}\label{eq_CCF_Tzz}
\eeq
where `(bulk)' denotes the corresponding bulk contribution, explicit expressions of which will be provided below. 
In \cref{eq_CCF_Tzz}, translational invariance along the lateral directions allows one to set $\rvp=\bv0$. 
The correlations of the OP derivatives can be cast into derivatives of the OP correlation function, as will be shown below.

Here we emphasize that \cref{eq_CCF_Tzz} strictly applies only to \emph{finite} times. 
In fact, upon taking the limit $t\to\infty$ in order to calculate equilibrium quantities, it is necessary to regularize the zero-mode divergence occurring for $\tauLG\to 0$ [see \cref{eq_OPcorrel_box_eq,eq_OPcorrel_thinfilm_eq} and the associated discussion].
Accordingly, in the case $t\to\infty$, a nonzero $\tauLG>0$ must be kept within the intermediate calculations and the limit $\tau\to 0$ is carried out only at the end. 
An exception is a \emph{thin film} ($\varrho= 0$) where, within Gaussian approximation, the zero-mode divergence does not play a role because the corresponding problematic term in \cref{eq_OPcorrel_box_eq} vanishes  \cite{gross_statistical_2017} \footnote{This can be different at higher orders in perturbation theory (see Refs.\ \cite{diehl_fluctuation-induced_2006,gruneberg_thermodynamic_2008}).}. For a thin film, one may thus set $\tauLG=0$ from the outset if one takes the limit $t\to\infty$.

We proceed by analyzing the dynamical and equilibrium CCF for various geometries and \bcs based on the stress tensor formalism.

\subsection{Thin film with periodic \bcs}
\label{sec_CCF_pbc_film}

We first consider a thin film with periodic \bcs at its confining surfaces ($z=0,L$), which is the simplest geometry to study CCFs.
Using \cref{eq_OPcorrel_box} [reduced to the special case of a thin film via \cref{eq_continuum_repl_C}], we determine the following auto-correlators:
\begin{subequations}
\begin{align}
\bra\Rcal\pbc(t)\ket &\equiv \bra \pd_z \phi(\bvnp, z, t) \pd_z \phi(\bvnp, z, t)\ket\big|_{z\in\{0,L\}} =   \frac{1}{L} \int \frac{\d^{d-1} p}{(2\pi)^{d-1}} \sum_{n} (\lambda_n\pbc)^2 \bra |a_n(\pv,t)|^2\ket \nonumber \\ 
&= -\bra \phi(\bvnp, z, t) \pd_z^2 \phi(\bvnp, z, t)\ket\big|_{z\in\{0,L\}}  = -\pd_z^2\Ccal\pbc(\bvnp, z, t)\big|_{z=0}, \\
\bra\Qcal\pbc(t)\ket &\equiv \Big\bra \phi(\rvp, z, t) 
\np^2\phi(\rvp, z, t) \Big\ket\Big|_{{\rvp=\bvnp}\atop{z\in\{0,L\}}} 
= -\frac{1}{L} \int \frac{\d^{d-1} p}{(2\pi)^{d-1}} \sum_{n} \pv^2 \bra |a_n(\pv,t)|^2\ket \nonumber \\
&= - \sum_{a = 1}^{d-1} \bra (\pd_\alpha \phi(\rvp,z,r))^2\ket\Big|_{{\rvp=\bvnp}\atop{z\in\{0,L\}}} = \np^2 \Ccal\pbc(\rvp,0,t)\big|_{\rvp=\bvnp} ,
\end{align}\label{eq_CCF_pbc_correlators}
\end{subequations}
\hspace{-0.12cm}where we have introduced the operator
\al{
\np^2 \equiv \sum_{\alpha=1}^{d-1}  \pd_\alpha^2.
\label{eq_npdef}
}
The averaged instantaneous film pressure $\bra\Pcal_f\ket$ follows according to \cref{eq_CCF_Tzz} as
\beq \bra\Pcal_f\pbc(t)\ket = \frac{3}{2}\bra \Rcal\pbc(t) \ket - \frac{1}{2}\bra \Qcal\pbc(t) \ket .
\label{eq_Pfilm_pbc}\eeq 
Upon applying the Poisson resummation formula via \cref{eq_OPcorrel_thinfilm_per} (see also \cref{app_Poisson}), one obtains 
\begin{subequations}
\begin{align}
\bra \Rcal\pbc(t) \ket &=  -\sum_{m=-\infty}^\infty \pd_z^2 \Ccal_b(\{\bvnp, z\}, t)\big|_{z=Lm},
\label{eq_CCF_pbc_Poisson_1} \\
\bra \Qcal\pbc(t) \ket &=  \sum_{m=-\infty}^\infty 
\np^2\Ccal_b(\{\rvp, z\}, t)\big|_{z=Lm,\, \rvp=\bvnp},
\label{eq_CCF_pbc_Poisson_2}\end{align} \label{eq_CCF_pbc_Poisson}
\end{subequations}
in terms of the equal-time bulk correlation function $\Ccal_b$ defined in \cref{eq_blkOP_correl_eqt}.

In the following we consider flat as well as thermal ICs, for which the bulk correlator $\Ccal_b$ is provided in \cref{eq_Sflat,eq_Stherm}, respectively.
The derivatives required in \cref{eq_CCF_pbc_Poisson} are readily obtained from the analytic expressions for $\Ccal_b$ provided in \cref{sec_bulk_correl}.
Since $\Ccal_b$ depends only on $r= |\rv|$, one has
\beq \pd_\alpha^2 \Ccal_b(r,t) = \pd_\alpha \left[\frac{r_\alpha}{r}\, \pd_r \Ccal_b(r,t)\right]  
=  \frac{r_\alpha^2}{r^2} \pd_r^2 \Ccal_b(r,t)  + \left( \frac{1}{r} - \frac{r_\alpha^2}{r^3} \right)  \pd_r\Ccal_b(r,t) .
\label{eq_Cb_gen_deriv}\eeq 
Consequently \cref{eq_CCF_pbc_Poisson_2} can be written as
$\bra \Qcal\pbc(t) \ket =  \sum_{m=-\infty}^\infty \frac{d-1}{z}\pd_z \Ccal_b(\{\bvnp, z\}, t)\big|_{z=Lm}$.
Explicit expressions for the derivatives of $\Scal\dyn$ and $\Scal\rel$ are rather lengthy, and are not stated here; they can be readily obtained from \cref{eq_Sdyn_def,eq_Srel_def} using standard properties of hypergeometric functions \cite{olver_nist_2010}.
However, it is useful to report the following asymptotic behaviors ($\beta = 1,\ldots,d$): 
\begin{subequations}
\begin{align}
\pd_r^2 \Scal\dyn(r,t\to \infty) &\simeq \pd_r^2 \Scal\dyn(r, t)\big|_{z\to 0} \simeq \pd_\beta^2 \Scal\dyn(r,t\to \infty)\big|_{\rvp=\bvnp} \simeq \pd_\beta^2 \Scal\dyn(r,t)\big|_{\rvp=\bvnp,\, z\to 0} \nonumber \\ 
&\simeq r^{-1} \pd_r \Scal\dyn(r,t\to \infty) \simeq r^{-1}\pd_r \Scal\dyn(r, t)\big|_{r\to 0}  \nonumber \\
&\simeq - \frac{\Gamma(d/4)}{2^{2+5d/4} \pi^{d/2} \Gamma(1+d/2) t^{d/4}},
\label{eq_Cb_Dz2_lim_tinf} \\
\pd_r^2 \Scal\rel(r\neq 0, t\to \infty) &\simeq r^{-1} \pd_r \Scal\rel(r\neq 0,t\to\infty) \simeq - \frac{\pi^{1/2-d/2} }{2^{5/2+7d/4} \Gamma(1+d/4)\, t^{1/2+d/4}}, 
\label{eq_Cb_Dz2_rel_lim_tinf} \\
\pd_r^2 \Scal\rel(r\neq 0, t\to 0) &\sim r^{-1}\pd_r \Scal\rel(r\neq 0, t\to 0) \sim \pd_r^2 \Scal\rel(r, t)\big|_{r \to\infty } \sim r^{-1}\pd_r \Scal\rel(r, t\to 0)\big|_{r\to\infty} \to 0 , \label{eq_Srel_lim_t0} \\
\pd_r^2 \Scal\flt(r,t\to 0) &\simeq \pd_r^2 \Scal\flt(r,t)\big|_{r\to\infty} \simeq r^{-1}\pd_r \Scal\flt(r,t\to 0) \simeq r^{-1}\pd_r \Scal\flt(r,t)\big|_{r\to\infty} \to 0.
\label{eq_Sflat_lim_t0}
\end{align}\label{eq_Cb_dyn_Dz2}
\end{subequations}
\hspace{-0.15cm}In fact, according to \cref{eq_Sflat_asympt}, $\pd_r^2\Scal\dyn$ approaches $\pd_r^2\Scal\stat$ exponentially for large $r$ or small $t$.
Furthermore, the leading asymptotic behavior extending \cref{eq_Srel_lim_t0} to large but finite $r$ or small but nonzero $t$ can be straightforwardly obtained from \cref{eq_Srel_largeR_asympt}.

Owing to \cref{eq_Srel_lim_t0,eq_Sflat_lim_t0}, the derivatives of $\Ccal_b$ appearing in \cref{eq_CCF_pbc_Poisson} vanish for $|z|\to\infty$.
Therefore in \cref{eq_CCF_pbc_Poisson}, in the \emph{bulk} limit $L\to\infty$, only terms for $m=0$ survive. These render, via \cref{eq_Pbulk_Pfilm_lim}, the bulk pressure $\bra\Pcal_b(t)\ket$. The bulk pressure can equivalently be obtained by directly evaluating the r.h.s.\ of \cref{eq_Pfilm_pbc} with the bulk correlation function [\cref{eq_blkOP_correl_eqt}].

Using the fact that $\Ccal_b(z,t) = \Ccal_b(-z,t)$, the time-dependent CCF [\cref{eq_CCF_Tzz}] at bulk criticality ($\tauLG=0$) for a slab with periodic \bcs eventually follows as
\beq \bra \Kcal\pbc(t) \ket = -\sum_{m=1}^\infty \left[ 3 \pd_z^2 \Ccal_b(z,t) + \frac{d-1}{z}\pd_z \Ccal_b(z,t) \right]_{z=Lm}. 
\label{eq_CCF_pbc_final}\eeq
Since $\Ccal_b$ fulfils the scaling behavior in \cref{eq_scalform_Cb_init} with $\zdyn=4$ [and recalling \cref{eq_mobility_ampl,eq_time_resc}], one readily demonstrates that $\bra \Kcal\pbc(t) \ket$ indeed has the scaling form anticipated in \cref{eq_scalform_CCF} (with $\tred=0$).
In particular, $L^d \bra \Kcal\pbc \ket$ is a function of the dimensionless time scaling variable $t/L^\zdyn$.

\subsubsection{Equilibrium CCF}

The contributions from $\Scal\dyn$ and $\Scal\rel$ vanish in the long-time limit, owing to \cref{eq_Cb_Dz2_rel_lim_tinf,eq_Cb_Dz2_lim_tinf}. 
Both for flat and thermal ICs [see \cref{eq_Sflat,eq_Stherm}], the equilibrium CCF $\bra \Kcal\pbc\ket\eq$ at bulk criticality thus follows by inserting $\Ccal_b=\Scal\stat$ [\cref{eq_Sstat}] into \cref{eq_CCF_pbc_final}, rendering \footnote{For a thin film in $d\geq 2$ dimensions, the limit $t\to\infty$ can be exchanged with the sum over $m$. This is not permitted in the case of a box geometry, as will be discussed in \cref{sec_CCF_dyn_box_pbc}.}
\beq \bra \Kcal\pbc\ket\eq = \bra\Kcal\pbc(t\to \infty)\ket = - \sum_{m=1}^\infty \left[ 3 \pd_z^2\Scal\stat(z) + \frac{d-1}{z} \pd_z\Scal\stat(z) \right]  = L^{-d}  \pi^{-d/2} \Gamma(d/2) (1-d) \zeta(d),
\label{eq_CCF_pbc_eqampl}\eeq 
where, in the last step, we have used \cref{eq_Sstat} and introduced the Riemann zeta-function $\zeta(s)=\sum_{n=1}^\infty n^{-s}$~\cite{olver_nist_2010}. 
In spatial dimensions $d=3$ and $4$, one obtains
\beq \bra \Kcal\pbc\ket\eq  = \begin{cases} \displaystyle
                           -\frac{1}{L^3} \frac{\zeta(3)}{ \pi}, \qquad & d=3,\\[12pt] \displaystyle
                           -\frac{1}{L^4} \frac{\pi^2}{30 } , & d=4,
                          \end{cases}
\label{eq_CCF_pbc_eqampl_dim}\eeq 
respectively.
The same expression as in \cref{eq_CCF_pbc_eqampl} is obtained from a calculation of $\bra \Kcal\pbc\ket\eq$ based on the residual finite-size free energy (see Ref.\ \cite{krech_free_1992}).
Note that, concerning the CCF, the ensemble difference is immaterial for a thin film with periodic \bcs [see also \cref{eq_OPcorrel_thinfilm_eq}].

\begin{figure}[t]\centering
    \subfigure[]{\includegraphics[width=0.42\linewidth]{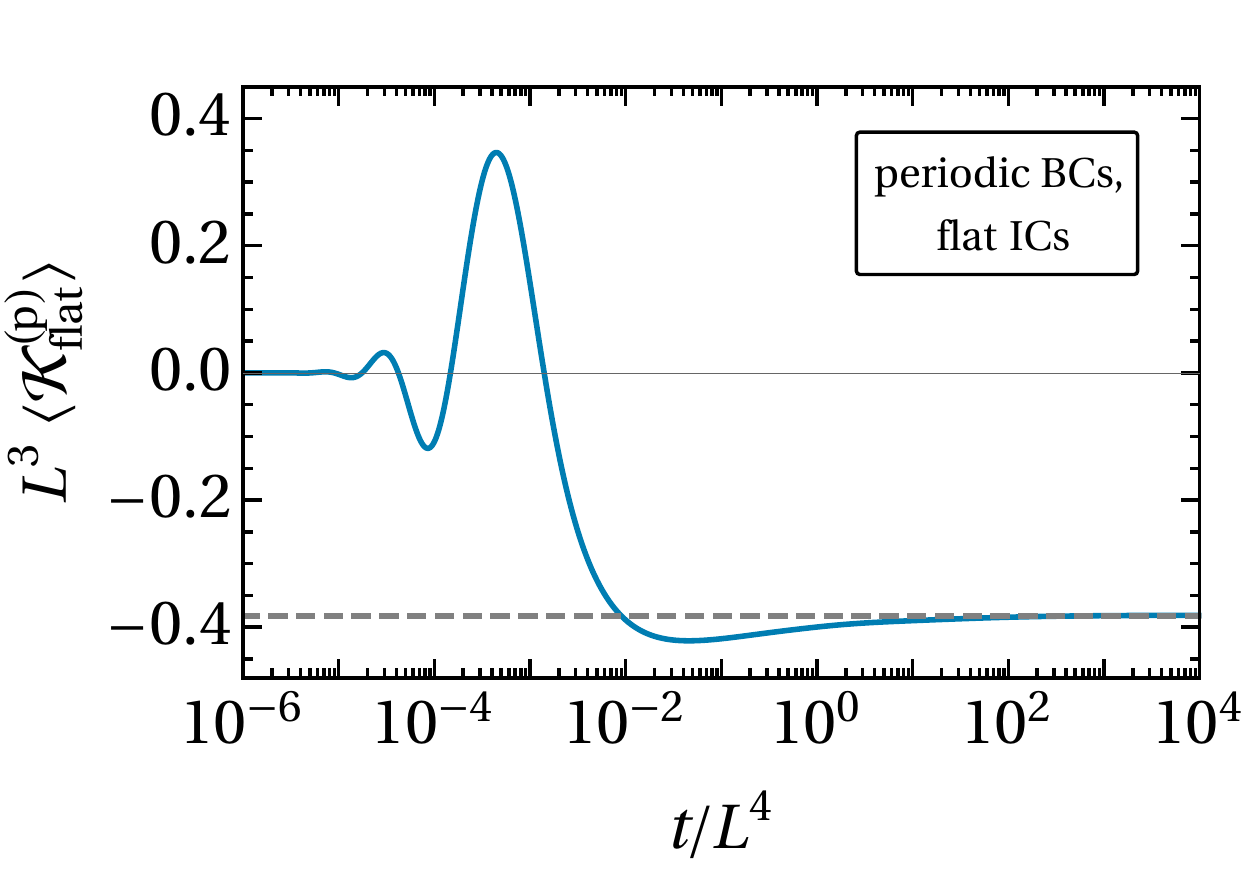} \label{fig_CCF_pbc_flatIC} } \qquad 
    \subfigure[]{\includegraphics[width=0.42\linewidth]{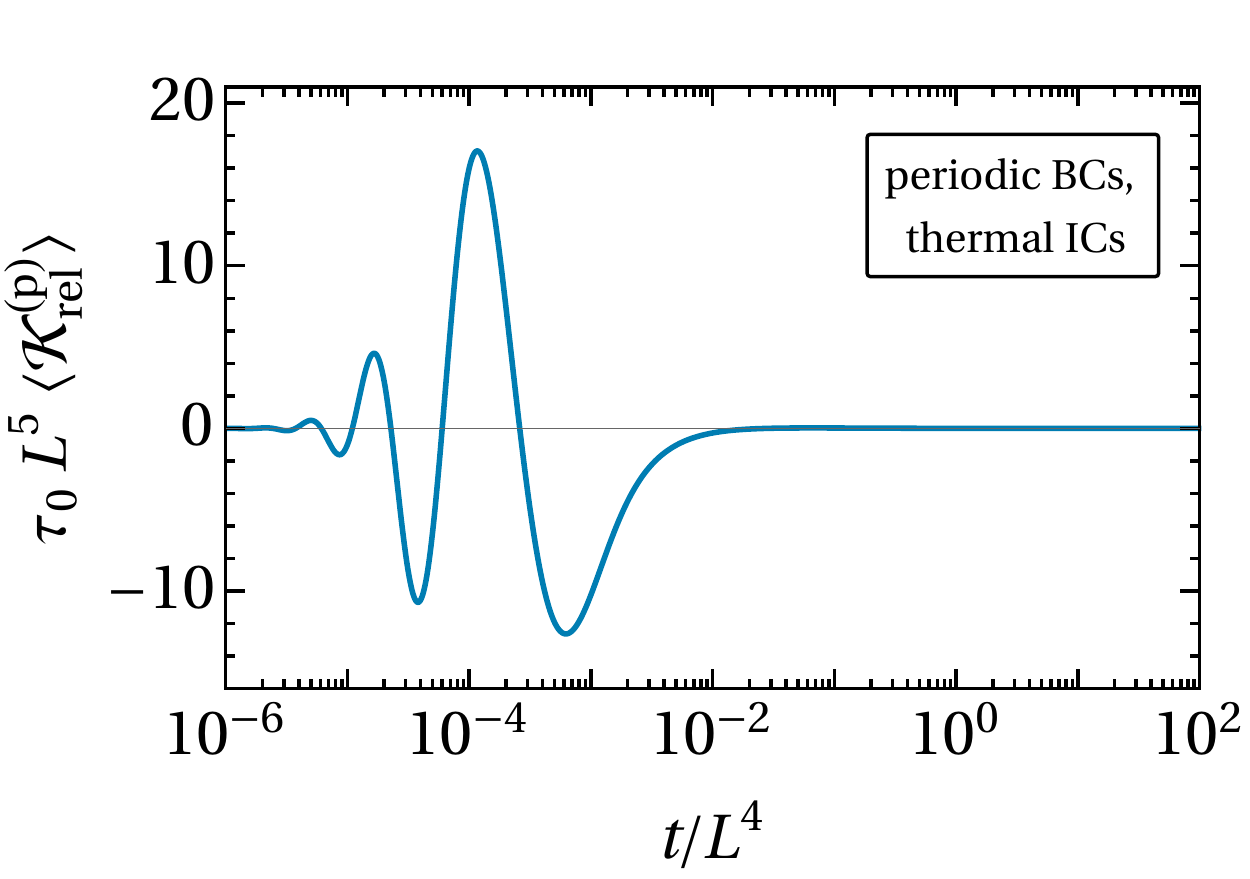} \label{fig_CCF_pbc_relax}}
    \caption{Time evolution of the dynamic CCF $\bra\Kcal\pbc(t)\ket$ [per $k_B T_c$, \cref{eq_CCF_pbc_final}] in a thin film for periodic \bcs and $d=3$ dimensions. 
    At time $t=0$, the system is instantaneously quenched to the bulk critical temperature ($\tauLG=0$). 
    Panel (a) shows the CCF for flat ICs [\cref{eq_IC_correl_flat}] and the dashed line indicates the value of the equilibrium CCF at $T_c$ [\cref{eq_CCF_pbc_eqampl}].
    Panel (b) shows the contribution $\tauInit \bra\Kcal\pbc\rel(t)\ket$ [\cref{eq_CCF_relax_contrib}] to the CCF $\bra\Kcal\pbc\th(t)\ket = \bra\Kcal\pbc\flt(t)\ket + \bra\Kcal\pbc\rel(t)\ket$ for thermal ICs [\cref{eq_IC_correl}], where $\tauInit=1/\initvar$ is a measure of the initial temperature [see \cref{eq_static_OZ_correl}]. In both panels, CCF and time are scaled according to \cref{eq_scalform_CCF}, such that the universal behavior of the CCF scaling function is exhibited by the plot explicitly.
    }
    \label{fig_CCF_slab_per}
\end{figure}

\subsubsection{Dynamic CCF for flat initial conditions}
\label{sec_CCF_dyn_film_pbc}

The CCF for flat ICs, which is obtained by inserting $\Ccal_b=\Scal\flt$ [\cref{eq_Sflat}]  into \cref{eq_CCF_pbc_final}, is denoted by $\bra\Kcal\pbc\flt(t)\ket$.
At finite times, a closed analytical expression for $\bra\Kcal\pbc\flt(t)\ket$ is not available, and the CCF thus has to be determined numerically.  
Due to the rapid exponential decay of $\Scal\flt$ for large values of $r$ [see \cref{eq_Sflat_asympt}], it suffices to retain only the first few terms of the sum in \cref{eq_CCF_pbc_final} in order to obtain an accurate estimate.
The CCF $\bra\Kcal\pbc\flt(t)\ket$ obtained in this way is illustrated in \cref{fig_CCF_slab_per}(a) for $d=3$ spatial dimensions.
From \cref{eq_Sflat_lim_t0} it follows that the CCF vanishes initially: 
\beq \bra \Kcal\flt\pbc(t=0) \ket = 0.
\label{eq_CCF_film_asympt_smallt}\eeq
At short times $t/L^\zdyn\lesssim \Ocal(1)$, the CCF grows in an oscillatory fashion.
At late times $t/L^\zdyn \gtrsim \Ocal(1)$ the CCF approaches its equilibrium value [\cref{eq_CCF_pbc_eqampl}] algebraically in time from below (see \cref{app_CCF_film_asympt} for a derivation of this behavior):
\beq \bra\Kcal\flt\pbc(t)\ket - \bra\Kcal\pbc\ket\eq \propto -t^{1/4-d/4}.
\label{eq_CCF_asymp_larget}\eeq

\subsubsection{Dynamic CCF for thermal initial conditions}

According to \cref{eq_Stherm}, the dynamic CCF for thermal ICs can be written as 
\beq \bra \Kcal\pbc\th(t)\ket = \bra\Kcal\pbc\rel(t)\ket + \bra \Kcal\pbc\flt(t)\ket,
\label{eq_CCF_pbc_thIC}\eeq 
where $\bra\Kcal\pbc\flt\ket$ is defined in the preceding subsection, while 
\beq \bra\Kcal\pbc\rel(t) \ket \equiv  -\sum_{m=1}^\infty \left[ 3 \pd_z^2 \Scal\rel(z,t) + \frac{2}{z}\pd_z \Scal\rel(z,t) \right]_{z=Lm}
\label{eq_CCF_relax_contrib}\eeq 
captures the contribution stemming from $\Scal\rel$ [\cref{eq_Srel_def}].
According to \cref{eq_Cb_Dz2_rel_lim_tinf,eq_Srel_lim_t0}, $\bra\Kcal\pbc\rel\ket$ vanishes both initially and at late times, implying that $\bra\Kcal\pbc\rel\ket$ essentially modifies only the transient behavior of the dynamic CCF.
Since $\bra\Kcal\pbc\rel\ket \propto 1/\tauInit$, for large initial temperatures [see \cref{eq_static_OZ_correl}] the influence of the thermal IC is diminished, such that $\bra \Kcal\pbc\th(t)\ket\big|_{\tauInit\to\infty} \simeq \bra \Kcal\pbc\flt(t)\ket$. 
\Cref{fig_CCF_slab_per}(b) displays $\tauInit \bra\Kcal\pbc\rel(t) \ket$, which is independent of the initial temperature $\tauInit$, as function of time.

\subsection{Thin film with Neumann \bcs}
\label{sec_CCF_Nbc_film}

In order to obtain the dynamic CCF in a thin film with Neumann \bcs, we proceed as in \cref{sec_CCF_pbc_film}. Accordingly, using \cref{eq_OPcorrel_box,eq_continuum_repl_C}, we obtain the following correlation functions: 
\begin{subequations}
\begin{align}
\bra \Rcal\Nbc(t) \ket &\equiv - \bra \phi(\bvnp, z, t) \pd_z^2 \phi(\bvnp, z, t)\ket\big|_{z \in \{0,L\}} =   \frac{1}{L} \int \frac{\d^{d-1} p}{(2\pi)^{d-1}}  \sum_{n=0}^\infty (\lambda_n\Nbc)^2 (2-\delta_{n,0}) \bra |a_n(\pv,t)|^2\ket \nonumber \\
 &= -\pd_z^2\Ccal\Nbc(\rvp=\bvnp, z, z', t)\big|_{z=z'\in \{0,L\}},
 \\
\bra \Qcal\Nbc(t) \ket &\equiv \big\bra \phi(\rvp, z, t) \sum_{\alpha=1}^{d-1} \pd_\alpha^2  \phi(\rvp, z, t) \big\ket\Big|_{{\rvp=\bvnp}\atop{z\in \{0,L\}}} 
= - \sum_{\alpha=1}^{d-1} \bra (\pd_\alpha \phi(\rvp, z, r))^2\ket\Big|_{{\rvp=\bvnp}\atop{z\in \{0,L\}}} \nonumber \\
&= -\frac{1}{L} \int \frac{\d^{d-1} p}{(2\pi)^{d-1}} \sum_{n=0}^\infty \pv^2 (2-\delta_{n,0}) \bra |a_n(\pv,t)|^2\ket = \np^2 \Ccal\Nbc(\rvp, z, z, t)\big|_{z\in \{0,L\}}  ,
\end{align}\label{eq_CCF_Nbc_correlators}
\end{subequations}
\hspace{-0.12cm}in terms of which \cref{eq_CCF_Tzz} renders the film pressure \footnote{Since $\phi$ is constructed as a sum of Neumann eigenfunctions [see \cref{eq_OP_exp_gen}], \cref{eq_eigenf_Nbc} implies $\pd_z \phi(\bv0, z=0, t)=0$, such that the correlator involving $\pd_z\phi$ in \cref{eq_CCF_Tzz} does not contribute to $\bra\Pcal_f\Nbc\ket$. This is the reason for the prefactors in \cref{eq_Pfilm_Nbc,eq_Pfilm_pbc} to be different.} \beq \bra\Pcal_f\Nbc(t)\ket = \bra \Rcal\Nbc(t) \ket - \frac{1}{2} \bra \Qcal\Nbc(t) \ket. 
\label{eq_Pfilm_Nbc}\eeq 
Upon invoking the Poisson summation formula via \cref{eq_OPcorrel_thinfilm_Neu} (see also \cref{app_Poisson}), the correlators in \cref{eq_CCF_Nbc_correlators} can be expressed as 
\begin{subequations}
\begin{align}
\bra \Rcal\Nbc(t) \ket &= -2\sum_{m=-\infty}^\infty \pd_z^2 \Ccal_b(\{\rvp=\bvnp, z\}, t)\big|_{z=2Lm} = 2\bra \Rcal\pbc(t) \ket\big|_{L\to 2L},
\label{eq_CCF_Nbc_Poisson_1} \\
\bra \Qcal\Nbc(t) \ket &= 2\sum_{m=-\infty}^\infty \np^2\Ccal_b(\{\rvp, z\}, t)\big|_{z=2Lm,\, \rvp=\bvnp} = 2 \sum_{m=-\infty}^\infty \frac{d-1}{z}\pd_z \Ccal_b(\{\bvnp, z\}, t)\big|_{z=2Lm} = 2 \bra \Qcal\pbc(t) \ket\big|_{L\to 2L},
\label{eq_CCF_Nbc_Poisson_2}\end{align} \label{eq_CCF_Nbc_Poisson}
\end{subequations}
\hspace{-0.12cm}where $\Ccal_b$ is the equal-time bulk correlation function [\cref{eq_blkOP_correl_eqt}] and where we have used \cref{eq_Cb_gen_deriv} as well as \cref{eq_CCF_pbc_Poisson}.
On the r.h.s., the expressions for periodic \bcs given in \cref{eq_CCF_pbc_Poisson} are to be evaluated for $2L$ instead of $L$. 

As before, the bulk pressure $\bra\Pcal_b\Nbc(t)\ket$ is provided by the terms pertaining to $m=0$ in \cref{eq_CCF_Nbc_Poisson}. 
Consequently, the dynamic CCF for a film with Neumann \bcs follows from \cref{eq_Pfilm_Nbc,eq_CCF_Nbc_Poisson} as
\beq \bra\Kcal\Nbc(t)\ket = -2\sum_{m=1}^\infty \left[ 2 \pd_z^2 \Ccal_b(z,t) + \frac{d-1}{z}\pd_z \Ccal_b(z,t) \right]_{z=2Lm} = 
-2^{-d+1}\sum_{m=1}^\infty \left[ 2 \pd_z^2 \Ccal_b(z,t/2^\zdyn) + \frac{d-1}{z}\pd_z \Ccal_b(z,t/2^\zdyn) \right]_{z=Lm}. 
\label{eq_CCF_Nbc_final}
\eeq 
In the last equation, we made use of the scaling behavior expressed in \cref{eq_scalform_CCF} (with $\tred=0$), according to which a change in the film thickness $L$ is equivalent to a change of the amplitude together with a rescaling of the time which appears in the dynamic CCF.

\subsubsection{Equilibrium CCF}

Due to \cref{eq_Cb_Dz2_lim_tinf,eq_Cb_Dz2_rel_lim_tinf}, in the long-time limit, only $\Scal\stat$ [\cref{eq_Sstat}] contributes to the dynamic CCF, independent of the type of IC [see \cref{eq_Sflat,eq_Stherm}].
Hence, upon inserting $\Scal\stat$ into \cref{eq_CCF_Nbc_final}, we obtain the equilibrium CCF for Neumann \bcs at bulk criticality ($\tauLG=0$):
\beq \bra \Kcal\Nbc\ket\eq = \bra\Kcal\Nbc(t\to \infty)\ket = 2^{-d} \bra \Kcal\pbc\ket\eq ,
\label{eq_CCF_Nbc_eq}\eeq 
which can be expressed in terms of the equilibrium CCF for periodic \bcs reported in \cref{eq_CCF_pbc_eqampl} (for the same value of the film thickness $L$), consistent with Refs.\ \cite{krech_free_1992,gross_statistical_2017}.
As is the case for periodic \bcs, in the thin film geometry with Neumann \bcs the CCF is the same in the canonical and the grand canonical ensembles, respectively.

\subsubsection{Dynamic CCF}

\begin{figure}[t]\centering
	\subfigure[]{\includegraphics[width=0.42\linewidth]{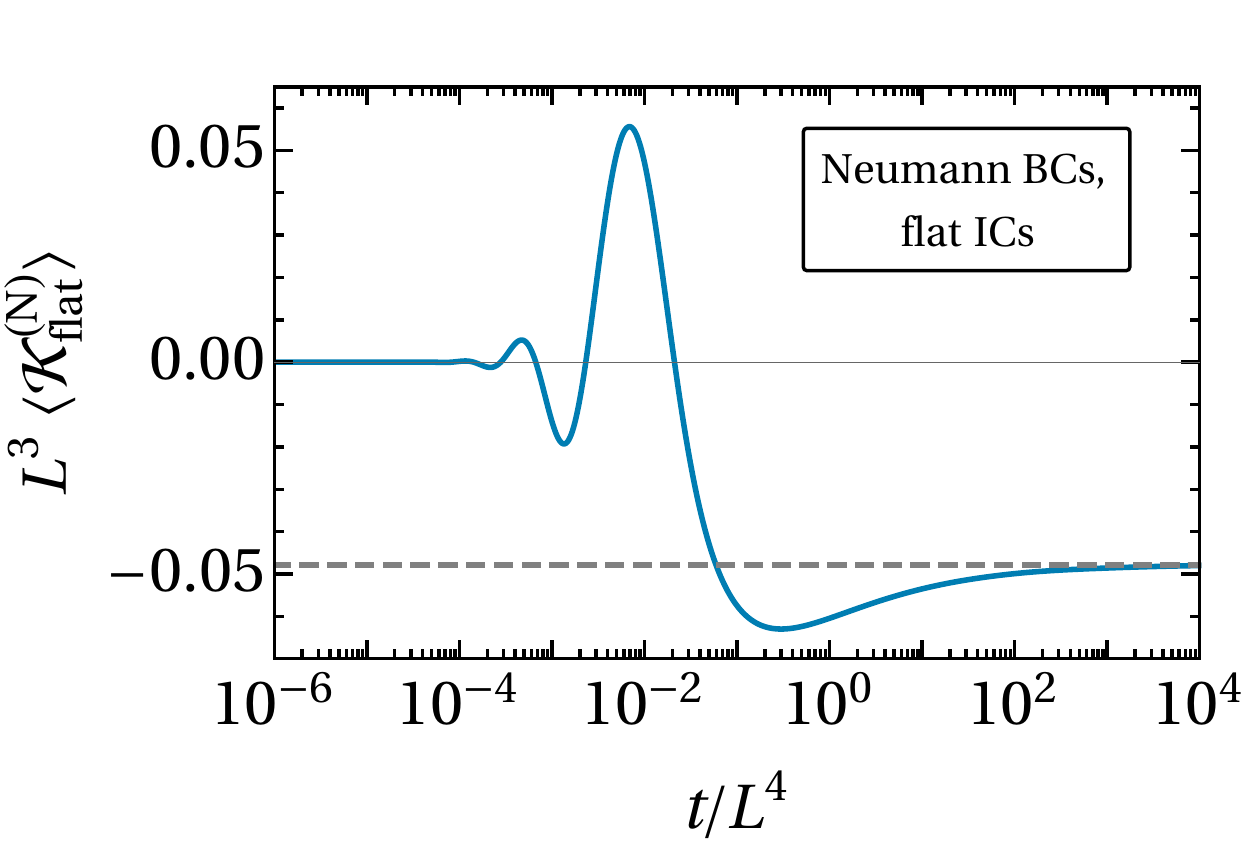} \label{fig_CCF_Nbc_flatIC} }\qquad 
    \subfigure[]{\includegraphics[width=0.425\linewidth]{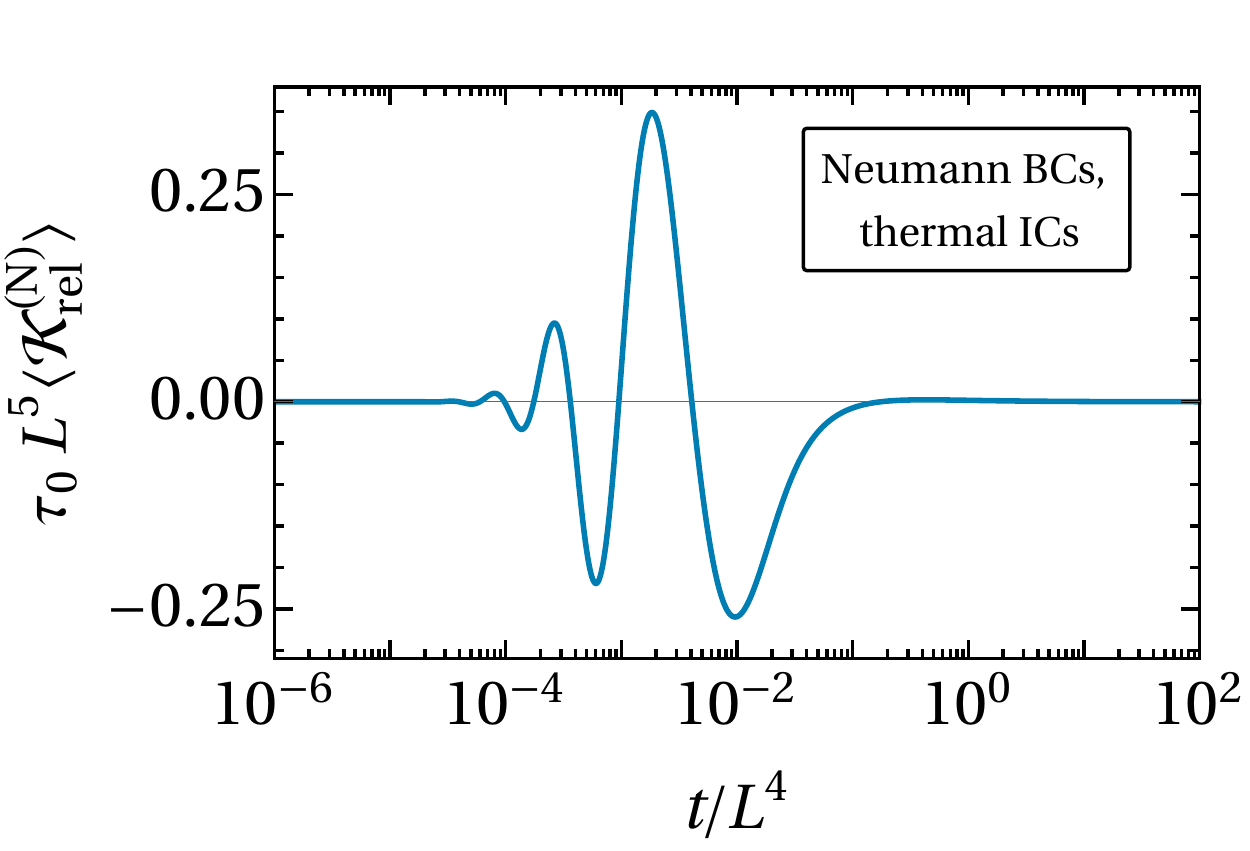} \label{fig_CCF_Nbc_relax_contrib}}
    \caption{Time evolution of the dynamic CCF $\bra\Kcal\Nbc(t)\ket$ [per $k_B T_c$, \cref{eq_CCF_Nbc_final}] in a thin film  for Neumann \bcs and $d=3$ dimensions. 
    At time $t=0$, the system is instantaneously quenched to the bulk critical temperature ($\tauLG=0$). 
    Panel (a) shows the CCF for flat ICs and the dashed line indicates the value of the equilibrium CCF at $T_c$ [\cref{eq_CCF_Nbc_eq}].
    Panel (b) shows the contribution $\tauInit \bra\Kcal\Nbc\rel(t)\ket$ to the CCF $\bra\Kcal\Nbc\th(t)\ket = \bra\Kcal\Nbc\flt(t)\ket + \bra\Kcal\Nbc\rel(t)\ket$ for thermal ICs [see \cref{eq_IC_correl}], where $\tauInit=1/\initvar$ is a measure of the initial temperature [see \cref{eq_static_OZ_correl}]. In both panels, CCF and time are scaled according to \cref{eq_scalform_CCF}, such that the universal behavior of the CCF scaling function is exhibited by the plot explicitly.
    }
    \label{fig_CCF_slab_Neu}
\end{figure}

Inserting Eq.\ \eqref{eq_Sflat} or \eqref{eq_Stherm} for $\Ccal_b$ into \cref{eq_CCF_Nbc_final} renders the CCF for flat and thermal ICs, respectively.
The numerically determined dynamic CCF $\bra \Kcal\Nbc(t)\ket$ for flat ICs is illustrated in \cref{fig_CCF_slab_Neu}(a).
Analogously to periodic \bcs [see \cref{sec_CCF_pbc_film}], the dynamic CCF vanishes initially, $\bra \Kcal\Nbc(t\to 0)\ket\to 0$, and approaches its equilibrium value [\cref{eq_CCF_Nbc_eq}] at late times in an oscillatory fashion. The contribution $\bra\Kcal\rel\Nbc(t)\ket$ to the CCF for thermal ICs, obtained by inserting $\Ccal_b=\Scal\rel$ into \cref{eq_CCF_Nbc_final}, is shown in \cref{fig_CCF_slab_Neu}(b).

\subsection{Cubical box with periodic \bcs}
\label{sec_CCF_pbc_box}

In the case of a finite cuboidal system with periodic \bcs at all surfaces, the film pressure follows from \cref{eq_CCF_Tzz} [analogously to \cref{eq_Pfilm_pbc}] as
\beq \bra\Pcal_f\pbc(t)\ket = \frac{3}{2}\bra \Rcal\pbc(t) \ket - \frac{1}{2}\bra \Qcal\pbc(t) \ket ,
\label{eq_Pfilm_box_pbc}\eeq 
with the correlators 
\begin{subequations}
\begin{align}
\bra\Rcal\pbc(t)\ket &\equiv \bra \pd_z \phi(\bvnp, z, t) \pd_z \phi(\bvnp, z, t)\ket\big|_{z\in\{0,L\}} =   \frac{1}{AL} \sum_{\pv,n} (\lambda_n\pbc)^2 \bra |a_n(\pv,t)|^2\ket = -\bra \phi(\bvnp, z, t) \pd_z^2 \phi(\bvnp, z, t)\ket\big|_{z\in\{0,L\}} \nonumber \\
&= -\pd_z^2 \Ccal\pbc(\bvnp, z, t)\big|_{z\in \{0,L\}} = -\sum_{\{m_\alpha=-\infty\}}^{\{\infty\}} \pd_z^2 \Ccal_b(r, t)\big|_{\rv = \Lcalv}\, ,
\label{eq_CCF_pbc_correlators_1}
 \\
\bra\Qcal\pbc(t)\ket &\equiv \Big\bra \phi(\rvp, z, t) 
\np^2 \phi(\rvp, z, t) \Big\ket\Big|_{{\rvp=\bv0}\atop{z\in\{0,L\}}} 
= -\frac{1}{AL} \sum_{\pv,n} \pv^2 \bra |a_n(\pv,t)|^2\ket 
= - \sum_{\alpha=1}^{d-1} \bra (\pd_\alpha \phi(\rvp,z,r))^2\ket\Big|_{{\rvp=\bvnp}\atop{z\in\{0,L\}}} \nonumber \\
&= \np^2\Ccal\pbc(\rvp, z, t)\big|_{{\rvp=\bvnp}\atop{z\in\{0,L\}}} =  \sum_{\{m_\alpha=-\infty\}}^{\{\infty\}} \np^2\Ccal_b(r, t)\big|_{\rv = \Lcalv},
\label{eq_CCF_pbc_correlators_2} 
\end{align}\label{eq_CCF_pbc_box_correlators}
\end{subequations}
\hspace{-0.15cm}where we have applied the Poisson resummation formula via \cref{eq_OPcorrel_box_pbc}. [For further details regarding the notation, see \cref{eq_OPcorrel_box,eq_mL_notation}.] 
With the aid of \cref{eq_Cb_gen_deriv}, the required derivatives are obtained as 
\begin{subequations}
\begin{align}
 \pd_z^2 \Ccal_b(r,t) &= \left( \frac{1}{r} - \frac{z^2}{r^3} \right) \pd_r \Ccal_b + \frac{z^2}{r^2} \pd_r^2 \Ccal_b(r,t),  \\
 \np^2\Ccal_b &=  \left(\frac{d-2}{r} + \frac{z^2}{r^3} \right) \pd_r \Ccal_b  + \left(1- \frac{z^2}{r^2}\right) \pd_r^2 \Ccal_b. 
\end{align}
\label{eq_Cb_derivs}\end{subequations}
In the Poisson representation in \cref{eq_CCF_pbc_box_correlators}, the terms pertaining to $\mv=\bv0$ (i.e., $m_1=\ldots = m_d=0$) provide the \emph{bulk} contribution to the CCF.
Accordingly, defining the indicator function 
\beq \Theta(\bv{m}) \equiv
\begin{cases} 0 \quad  &\text{if }\bv{m}=\bv0, \\
			  1 	& \text{otherwise,}
\end{cases}			  
\label{eq_Theta_m}\eeq 
the CCF follows from \cref{eq_Pfilm_box_pbc} as
\beq \bra \Kcal\pbc(t)\ket = - \sum_{m_1=-\infty}^\infty \cdots \sum_{m_d=-\infty}^\infty \Theta(\bv{m}) \left[ \left(\frac{1}{2} + \frac{z^2}{r^2}\right) \pd_r^2 \Ccal_b + \left(\frac{1+d}{2r} - \frac{z^2}{r^3}\right)\pd_r\Ccal_b \right]_{\rv =  \Lcalv}. 
\label{eq_CCF_pbc_box}\eeq

\subsubsection{Equilibrium CCF}
\label{sec_CCF_eq_box}

We now demonstrate that \cref{eq_CCF_pbc_box} leads to the \emph{canonical} equilibrium CCF which has previously been obtained in Ref.\ \cite{gross_statistical_2017} based on statistical field theory.
We emphasize that evaluating \cref{eq_CCF_pbc_box} with the correlation function obtained at bulk criticality $\tauLG=0$, i.e., $\Ccal_b=\Scal\stat$ [see \cref{eq_Sstat}], does \emph{not} result in the correct equilibrium CCF at $\tauLG=0$ (see \cref{app_static_CCF}).
Instead, for the purpose of regularizating the zero-mode divergence [see the discussion following \cref{eq_OPcorrel_box_eq}] it is necessary to consider a nonzero $\tauLG$ and perform the limit $\tauLG\to 0$ only at the end of the calculation. 
For nonzero $\tauLG$, \cref{eq_Tzz_dyn} renders, analogously to \cref{eq_Pfilm_box_pbc}, the canonical equilibrium film pressure 
\beq\begin{split} \bra\Pcal_f\pbc\ket\eq &= \frac{3}{2}\bra \Rcal\pbc \ket\eq - \frac{1}{2}\bra \Qcal\pbc \ket\eq +  \onehalf \tauLG \Ccal\pbc\eq(\rvp=\bv0_\parallel, z=0),  \\
&= \frac{3}{2}\bra \Rcal\pbc \ket\eq - \frac{1}{2}\bra \Qcal\pbc \ket\eq + \onehalf \tauLG \Ccal\pbc\eqGC(\rvp=\bv0_\parallel,z=0)  -  \onehalf L^{-d} \varrho^{d-1} ,
\end{split}\label{eq_Pfilm_pbc_can_eq}\eeq 
where we have used \cref{eq_OPcorrel_box_eq} in order to replace the canonical by the grand canonical static correlation function (for $\Phi=0$).
In order to reformulate \cref{eq_Pfilm_pbc_can_eq}, we invoke the Schwinger-Dyson equation (see, e.g., Refs.\ \cite{le_bellac_quantum_1991, dean_out--equilibrium_2010}), 
\beq \left\bra \phi(\rv) \frac{\delta \Fcal}{\delta \phi(\rv')} \right\ket = \delta(\rv-\rv')  ,
\label{eq_SchwingerDyson}
\eeq 
which, in the Gaussian case and for $\tauLG\neq0$, implies the following identity for the static bulk correlation function (see also Ref.\ \cite{mikheev_sum_1991}): 
\beq -\nabla^2 \Ccal\eqBlk(\rv) + \tauLG \Ccal\eqBlk(\rv) = \delta(\rv).
\label{eq_WardIdent}\eeq 
Inserting this relation into \cref{eq_CCF_pbc_correlators_2}, with $\nabla^2 = \nabla_\parallel^2 + \pd_z^2$, renders
\beq \bra\Qcal\pbc \ket\eq = \bra\Rcal\pbc \ket\eq + \tauLG \underbrace{\sum_{\{m_\alpha=-\infty\}}^{\{\infty\}} \Ccal\eqBlk(\Lcalv)}_{\Ccal\pbc\eqGC(\bv0_\parallel,0)} - \delta(\bv0),
\label{eq_Qcal_pbc_Schwinger}\eeq 
where we have used \cref{eq_OPcorrel_box_pbc,eq_OPcorrel_box_eq_gc} in order to identify the static finite-size correlation function $\Ccal\pbc\eqGC$ (which pertains to the grand canonical ensemble).
Accordingly, we obtain from \cref{eq_Pfilm_pbc_can_eq} the canonical equilibrium film pressure 
\beq \bra\Pcal_f\pbc \ket\eq = \bra\Rcal\pbc \ket\eq  - \onehalf \delta(\bv0) -  \onehalf L^{-d} \varrho^{d-1} .
\label{eq_Pfilm_box_pbc_eq}\eeq
The singular term $\delta(\bv0)$ acquires a well-defined meaning by regularizing the theory on a lattice. Specifically, upon expressing \cref{eq_WardIdent} in Fourier space, one obtains the regularized form $\delta(0) = \int_{-\pi/\ell}^{\pi/\ell} \d k/(2\pi) = 1/\ell$, where $\ell$ denotes the lattice constant.   
However, since $\delta(\bv0)$ is actually independent of $L$ (and therefore represents a bulk term irrelevant for the CCF), it is not necessary to actually perform this regularization here.
The last term on the r.h.s.\ of \cref{eq_Pfilm_box_pbc_eq} stems from the absence of the zero-mode fluctuations [see \cref{eq_total_OP}] in the canonical ensemble \cite{gross_statistical_2017}. Its specific form is closely related to the contribution $\mu\phi$ in the dynamical stress tensor [\cref{eq_stressten_dyn}], such that, without it, one would obtain the wrong sign for the last term in \cref{eq_Pfilm_box_pbc_eq}.
The zero-mode contribution $L^{-d}\varrho^{d-1}/2$ in \cref{eq_OPcorrel_box_eq} plays no role for the correlators $\bra\Rcal\pbc\ket\eq$ and $\bra\Qcal\pbc\ket\eq$, because they are derivatives of the correlation function. Hence, they take the same form in the canonical and the grand canonical ensemble.
Furthermore, in the thin film limit $\varrho\to 0$, \cref{eq_Pfilm_box_pbc_eq} implies that $\bra\Pcal_f\ket\eq = \bra\Pcal_f\ket\eqGC$ and hence 
\beq \bra\Kcal\pbc\ket\eq = \bra\Kcal\pbc\ket\eqGC, \qquad (\text{thin film}, \varrho\to 0),
\label{eq_CCF_gc_eq_film}\eeq
consistent with \cref{eq_CCF_pbc_eqampl}.

One is left with determining the expression of $\bra \Rcal\pbc \ket\eq$ in \cref{eq_Pfilm_box_pbc_eq} for arbitrary $\tauLG\geq 0$. Inserting \cref{eq_eqmodecorrel_eqt} into the first line of \cref{eq_CCF_pbc_correlators_1}, one obtains
\beq\begin{split} 
\bra\Rcal\pbc \ket\eq &= \frac{1}{AL} \sum_{\bv{n_\parallel},n_d} \frac{(\lambda_{n_d}\pbc)^2}{(\lambda_{n_d}\pbc)^2 + \pv_\bv{n_\parallel}^2 + \tauLG} = \frac{1}{AL} \sum_{\bv{n_\parallel},n_d} (\lambda_{n_d}\pbc)^2 \int_0^\infty \d s\, e^{-s \left[(\lambda_{n_d}\pbc)^2 + \pv_\bv{n_\parallel}^2 + \tauLG \right]} \\
&= \frac{1}{2A} \frac{\d}{\d L} \sum_{\bv{n_\parallel},n_d} \int_0^\infty \frac{\d s}{s}  \exp\left\{-s \left [ \tauLG + \pfrac{2\pi n_d }{L}^2 + \sum_{\alpha=1}^{d-1} \pfrac{2\pi n_\alpha }{L_\parallel}^2 \right]\right\} \\
&= \frac{1}{2A} \frac{\d}{\d L} \sum_{\bv{n_\parallel},n_d} \int_0^\infty \frac{\d s}{s}  \exp\left\{- s \left[ \frac{L^2 \tauLG }{4\pi^2}  + n_d^2  + \varrho^2 \sum_{\alpha=1}^{d-1} n_\alpha^2  \right]\right\} \\
&= \frac{1}{2A} \frac{\d}{\d L} \int_0^\infty \frac{\d s}{s}\, \exp\left(-\frac{L^2\tauLG s}{4\pi^2}\right) \vartheta(s) [\vartheta(\varrho^2 s)]^{d-1},
\end{split}\label{eq_Pfilm_pbc_box_eq}\eeq 
where  
\beq \vartheta(y) = \sum_{n=-\infty}^\infty e^{-y n^2}
\label{eq_Jacobi_theta}\eeq
is the Jacobi theta function \cite{olver_nist_2010}.
Note that the total derivative $\d/\d L$ also acts on $\varrho$, which is a function of $L$ [see \cref{eq_aspectratio}].
The corresponding bulk contribution follows from \cref{eq_Pfilm_pbc_box_eq} by replacing the sum over the eigenspectrum $\bv{n}$ by integrals according to \cref{eq_continuum_repl_C}, which essentially amounts to replacing $\vartheta(y)$ in \cref{eq_Pfilm_pbc_box_eq} with $\int_{-\infty}^\infty \d n\, e^{-y n^2} = (\pi/y)^{1/2}$ \footnote{Equivalently, one could apply the Poisson resummation formula [\cref{eq_Poisson_summation}] directly to \cref{eq_Pfilm_pbc_box_eq} [i.e., \cref{eq_Jacobi_theta} therein] and identify the ensuing zero mode $\bv{m}=\bv0$ as the bulk contribution.}. 
After subtraction of all bulk contributions, the canonical equilibrium CCF for $\tauLG\geq 0$ follows from \cref{eq_Pfilm_box_pbc_eq} as
\beq\begin{split} 
\bra\Kcal\pbc\ket\eq &= - \frac{1}{2A} \frac{\d}{\d L} \int_0^\infty \frac{\d s}{s}  \exp\left(-\frac{L^2\tauLG s}{4\pi^2}\right) \left\{\varrho^{-d+1}\pfrac{\pi}{s}^{d/2} - \vartheta(s) [\vartheta(\varrho^2 s)]^{d-1}  \right\} -  \onehalf L^{-d} \varrho^{d-1}  \\
&= \bra\Kcal\pbc\ket\eqGC -  \onehalf L^{-d} \varrho^{d-1} .
\end{split}\label{eq_CCF_eq_box_pbc_int}\eeq 
Here, we have used \cref{eq_aspectratio}, which implies
\beq \frac{1}{AL} = L^{-d} \varrho^{d-1},
\label{eq_aspectratio_vol}\eeq 
and have identified the grand canonical CCF as 
\beq \bra\Kcal\pbc\ket\eqGC = - \frac{\d}{\d L} \left[ \onehalf L^{-d+1} \int_0^\infty \frac{\d s}{s}  \exp\left(-\frac{L^2\tauLG s}{4\pi^2}\right) \left\{ \pfrac{\pi}{s}^{d/2} - \vartheta(s) [\varrho \vartheta(\varrho^2 s)]^{d-1}  \right\} \right].
\label{eq_CCF_eq_box_pbc_gc_int}\eeq 
The expression in \cref{eq_CCF_eq_box_pbc_int} fulfills the scaling form given in \cref{eq_scalform_CCF} (with $\initvar=0$; see Ref.\ \cite{gross_statistical_2017}).

Notably, \cref{eq_Pfilm_pbc_box_eq} can be written as $\bra \Rcal\pbc\ket\eq = -\d \Fcal\pbc\gc  / \d L$, where $\Fcal\pbc\gc = (1/2) \sum_{\nv_\parallel,n_d} \ln (\pv_{\nv_\parallel}^2 + (\lambda_{n_d}\pbc)^2)$ is the (unregularized) grand canonical free energy of the system (for $\Phi=0$) \footnote{The (finite) excess contribution of the free energy can be obtained by subtracting the associated bulk free energy from $\Fcal\pbc\gc$ \cite{gross_statistical_2017}. This procedure is analogous to determining the (finite) CCF by subtracting the bulk pressure from the the film pressure (the latter being infinite in a continuum theory as well) [see \cref{eq_CCF_def}]}. The expression in the square brackets in \cref{eq_CCF_eq_box_pbc_gc_int} thus represents the associated grand canonical residual finite-size free energy per area, $\Fcal\res/A$ \cite{gross_statistical_2017,dohm_critical_2011} \footnote{$\Fcal\res$ diverges logarithmically $\sim (1/2)AL^{-d+1} \varrho^{d-1} \ln(L^2\tauLG) = (1/2)\ln(L^2\tauLG)$ for $\tauLG\to 0$, which is due to the contribution of the zero mode (see Ref.\ \cite{gross_statistical_2017} for further discussions). However, this divergence drops out from \cref{eq_CCF_eq_box_pbc_gc_int} after taking the derivative with respect to $L$.}.
These results demonstrate that the stress tensor formalism used here leads to the same expression for the equilibrium CCF as the approach based on the free energy.

\begin{figure}[t]\centering
    \includegraphics[width=0.42\linewidth]{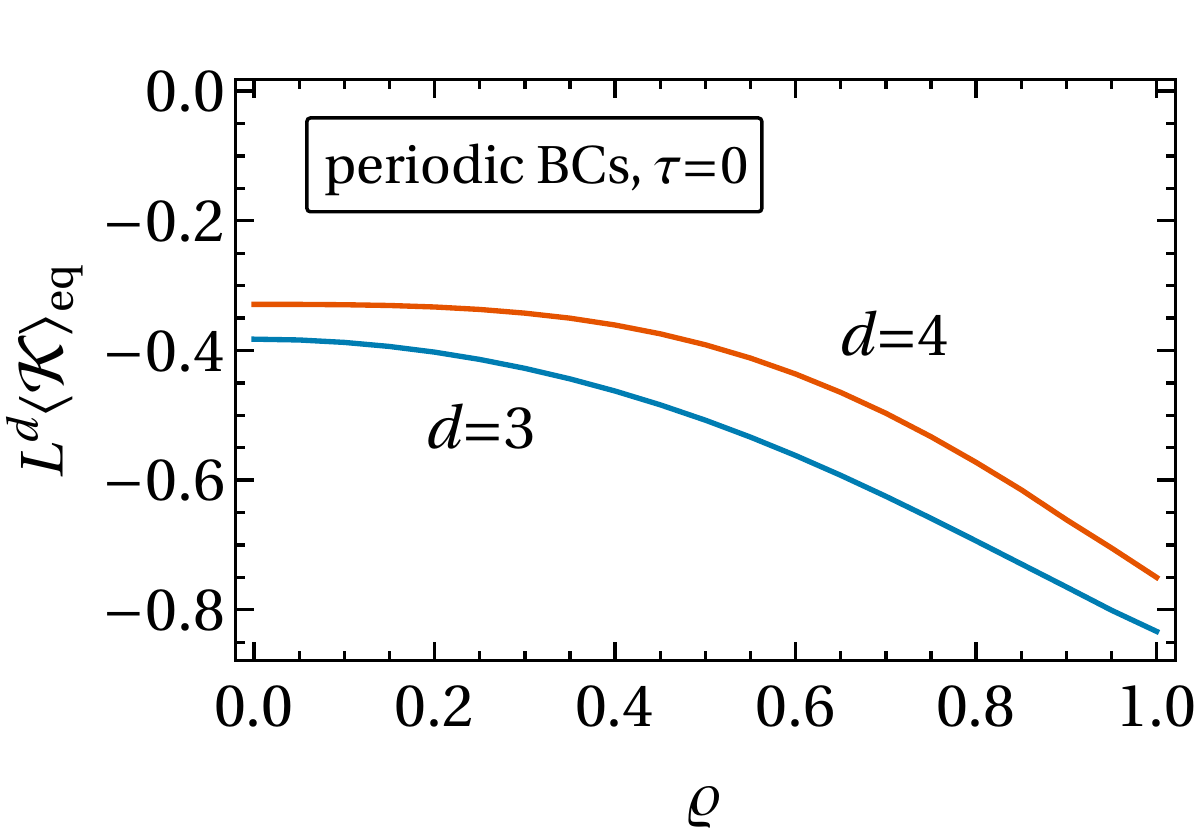} 
    \caption{Canonical CCF in thermal equilibrium [\cref{eq_CCF_eq_box_pbc_int}] for a box with periodic \bcs in all directions at bulk criticality ($\tauLG=0$) as a function of the aspect ratio $\varrho$, for $d=3$ and $d=4$ dimensions. $\Kcal$ is taken in units of $k_B T_c$ so that it has the units of an inverse volume. 
    }
    \label{fig_Keq_pbc_rho}
\end{figure}

The limit $\tauLG\to 0$ of \cref{eq_CCF_eq_box_pbc_gc_int} is singular, and must be performed after computing the integral and the derivative (see \cref{app_CCF_box_reg} for further discussions).
The canonical equilibrium CCF obtained from \cref{eq_CCF_eq_box_pbc_int} at bulk criticality, i.e., $\tauLG\to 0$, is shown in \cref{fig_Keq_pbc_rho} as a function of the aspect ratio $\varrho$.
An asymptotic analysis (see \cref{app_CCF_box_reg}) reveals that, for a cubic geometry ($\varrho=1$) and at bulk criticality, the grand canonical CCF attains the remarkably simple value
\beq \bra\Kcal\pbc\ket\eqGC\big|_{\tauLG\to 0, \varrho=1} \simeq -\frac{1}{L^d} \frac{1}{d}.
\label{eq_CCF_pbc_box_rho}
\eeq

\begin{figure}[t]\centering
    \subfigure[]{\includegraphics[width=0.44\linewidth]{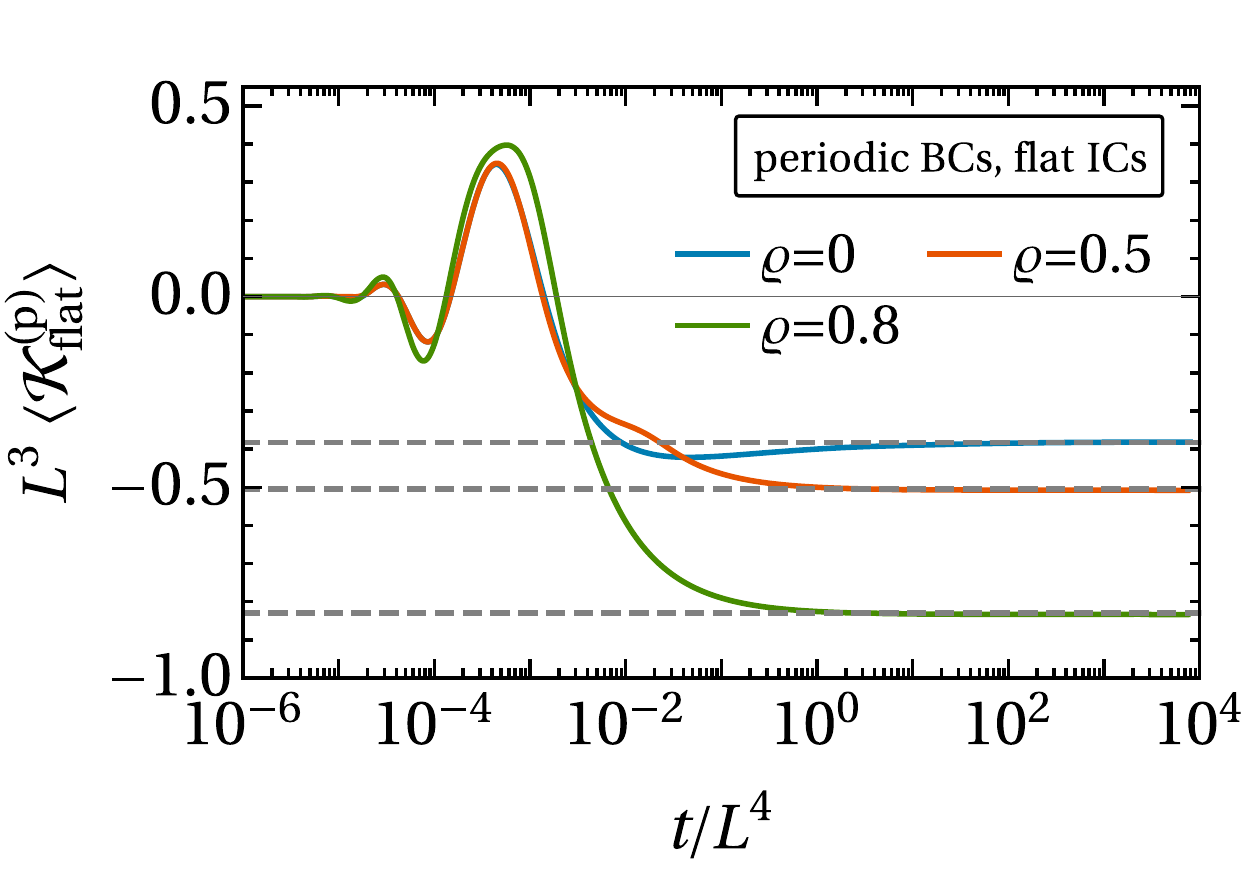} } \qquad
    \subfigure[]{\includegraphics[width=0.42\linewidth]{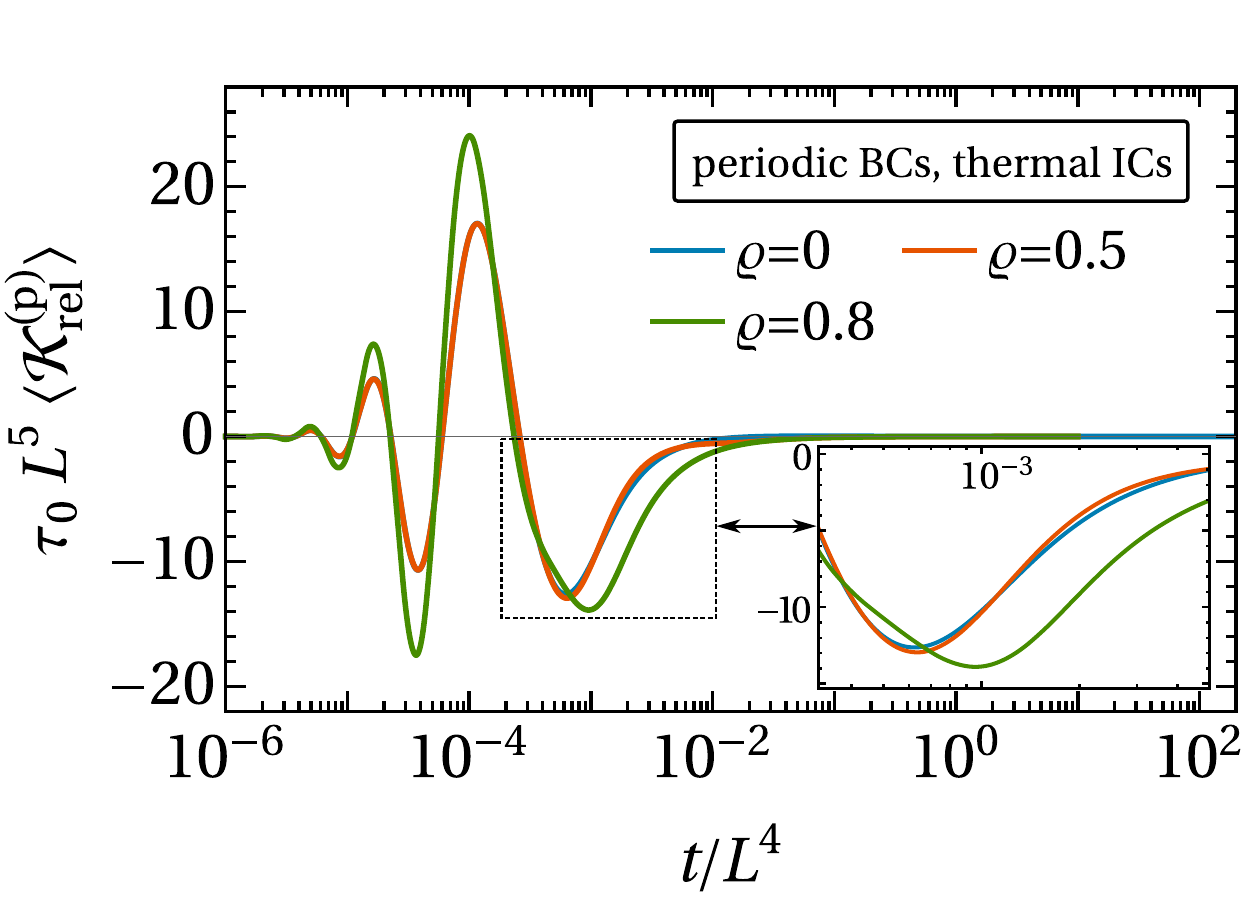} }
    \caption{Time evolution of the dynamic CCF $\bra\Kcal\pbc(t)\ket$ [in $d=3$ and per $k_B T_c$, \cref{eq_CCF_pbc_box}] for a box with periodic \bcs for various aspect ratios $\varrho$ ($\varrho = 0$ is the thin film limit). 
    At time $t=0$, the system is instantaneously quenched to the bulk critical temperature ($\tauLG=0$). 
    Panel (a) shows the CCF for flat ICs. The dashed lines indicate the equilibrium CCF at $T_c$ [see \cref{eq_CCF_eq_box_pbc_int} and \cref{fig_Keq_pbc_rho}].
    Panel (b) shows the contribution $\tauInit \bra\Kcal\pbc\rel(t)\ket$ (see Subsec.\ \ref{sec_CCF_pbc_box_rel}) to the CCF for thermal ICs [see \cref{eq_IC_correl}], where $\tauInit=1/\initvar$ is a measure for the initial temperature [see \cref{eq_static_OZ_correl}]. In both panels, the CCF and the time are scaled according to \cref{eq_scalform_CCF}, such that the universal behavior of the CCF scaling function is exhibited in the plot.
    }
    \label{fig_CCF_box_per}
\end{figure}

\subsubsection{Dynamic CCF for flat ICs}
\label{sec_CCF_dyn_box_pbc}

The time-dependent CCF $\bra\Kcal\flt\pbc(t) \ket$ for flat ICs is obtained by evaluating \cref{eq_CCF_pbc_box} using \cref{eq_Sflat}, i.e., $\Ccal_b(r,t) = \Scal\flt(r,t) = \Scal\stat(r) - \Scal\dyn(r,t)$.
According to \cref{eq_Sflat_asympt}, $\Scal\flt$ vanishes exponentially in the limit $t\to 0$, which implies  
\beq \bra\Kcal\flt\pbc(t=0)\ket=0.
\eeq 
At late times $t\gg L^\zdyn$ (which corresponds to $\tphys/\amplTrelax \gg (L/\amplXip)^z$, $z=4$), $\bra\Kcal\flt\pbc(t)\ket$ approaches the equilibrium value of the CCF given in \cref{eq_CCF_eq_box_pbc_int}.
Obtaining this result from the dynamics is, however, intricate because not only $\Scal\stat$, but also $\Scal\dyn$ contributes to the late-time limit of $\bra\Kcal\flt\pbc(t)\ket$, despite the fact that $\Scal\dyn(r,t)$ and its derivatives vanish algebraically for large $t$ [see \cref{eq_Sdyn_limt}]:
\beq \pd_r^2 \Scal\dyn(r,t\to \infty) \sim \frac{1}{r} \pd_r \Scal\dyn(r,t\to \infty)  \sim -t^{-d/4}.
\label{eq_Sdyn_deriv_asympt}\eeq 
The reason for the nonzero contribution of $\Scal\dyn$ to $\bra\Kcal\pbc\flt(t\gg L^\zdyn)\ket$ is the $d$-fold summation in \cref{eq_CCF_pbc_box}, which, as shown in \cref{app_dyn_constr}, balances the rather mild algebraic decay in \cref{eq_Sdyn_deriv_asympt} \footnote{This effect does not arise for a thin film (see \cref{sec_CCF_pbc_film}), because in this case the one-fold sum is insufficient to balance the algebraic decay in \cref{eq_Cb_Dz2_lim_tinf}.}.
The summation in \cref{eq_CCF_pbc_box} therefore does not commute with the limit $t\to\infty$.

The contribution to the equilibrium CCF stemming from $\Scal\dyn$ can be calculated exactly in $d=1$ spatial dimensions (see \cref{app_dyn_constr}), while it has to be determined numerically for $d>1$.
The universal behavior of $\bra\Kcal\pbc\flt(t)\ket$ is illustrated in \cref{fig_CCF_box_per}(a) as a function of time for $d=3$ and various aspect ratios $\varrho$. 
For an accurate estimate it suffices to take into account only terms with $|m_{1,\ldots,d}|\lesssim \Ocal(100)$  in the numerical evaluation of \cref{eq_CCF_pbc_box}, the precise number for truncation depending somewhat on the value of $t/L^z$ being considered. The computation can be further sped up by exploiting the isotropy of the expression in the lateral directions.
Notably, for $\varrho=1$, the contribution to $\bra\Kcal\flt\pbc(t)\ket$ stemming from $\Scal\stat$ vanishes identically (see \cref{app_static_CCF}), implying, in particular, that the equilibrium value $\bra\Kcal\flt\pbc\ket\eq<0$ (see \cref{fig_Keq_pbc_rho}) arises within the dynamics solely due to $\Scal\dyn$.
In \cref{fig_CCF_box_per} one observes that changing the aspect ratio mainly affects the late-time equilibrium value of the CCF.

\subsubsection{Dynamic CCF for thermal ICs}
\label{sec_CCF_pbc_box_rel}

Analogously to \cref{eq_CCF_pbc_thIC}, the dynamic CCF $\bra\Kcal\pbc\th(t)\ket$ for thermal ICs, obtained by inserting \cref{eq_Stherm} into \cref{eq_CCF_pbc_box}, can be split up into the contributions $\bra\Kcal\pbc\flt(t)\ket$ and $\bra\Kcal\pbc\rel(t)\ket$. In contrast to $\bra\Kcal\pbc\flt(t)\ket$ discussed in the preceding subsection, $\bra\Kcal\pbc\rel(t)\ket$ does not contribute to the late-time limit of the CCF, because according to \cref{eq_Srel_late_time} the relevant derivatives of $\Scal\rel$ decay more rapidly than those in \cref{eq_Sdyn_deriv_asympt}:
\beq \pd_r^2 \Scal\rel(r,t\to\infty) \sim \frac{1}{r} \pd_r \Scal\dyn(r,t\to\infty) \sim -t^{-(d+2)/4}.
\label{eq_Srel_deriv_asympt}\eeq 
Instead, $\bra\Kcal\pbc\rel(t)\ket$, which is illustrated in \cref{fig_CCF_box_per}(b), contributes to $\bra\Kcal\pbc\th(t)\ket$ only at intermediate times with a magnitude $\propto 1/\tauInit$ [see \cref{eq_IC_correl}].

\subsection{Cubical box with Neumann \bcs}
\label{sec_CCF_Nbc_box}

The calculation of the CCF for Neumann \bcs in a cuboidal box proceeds analogously to \cref{sec_CCF_Nbc_film,sec_CCF_pbc_box}. Therefore we summarize here only the main steps.
In the box geometry, the film pressure is given by the same expression as for the thin film [\cref{eq_Pfilm_Nbc}], i.e.,
\beq \bra\Pcal_f\Nbc(t)\ket = \bra \Rcal\Nbc(t) \ket - \frac{1}{2} \bra \Qcal\Nbc(t) \ket ,
\label{eq_Pfilm_Nbc_box}\eeq 
but with [see  \cref{eq_mL_notation} for the notation]
\begin{subequations}
\begin{align}
\bra \Rcal\Nbc(t) \ket &= -2 \sum_{\{m_\alpha=-\infty\}}^{\{\infty\}} \pd_z^2 \Ccal_b(\{\rvp, z\}, t)\big|_{\rv = \Lcaltildev} =  2\bra \Rcal\pbc(t) \ket\big|_{2L}\, ,
\label{eq_CCF_Nbc_box_Poisson_1} \\
\bra \Qcal\Nbc(t) \ket &= 2 \sum_{\{m_\alpha=-\infty\}}^{\{\infty\}} \nabla^2_\parallel \Ccal_b(\{\rvp, z\}, t)\big|_{\rv = \Lcaltildev} = 2\bra \Qcal\pbc(t) \ket\big|_{2L}\, , \label{eq_CCF_Nbc_box_Poisson_2}\end{align} \label{eq_CCF_Nbc_box_Poisson}
\end{subequations}
\hspace{-0.15cm}where $\bra\Rcal\pbc\ket$ and $\bra\Qcal\pbc\ket$ refer to the expressions for a box with periodic \bcs given in \cref{eq_CCF_pbc_box_correlators}, which are to be evaluated here for the film thickness $2 L$.
Using \cref{eq_Cb_derivs}, the CCF follows from \cref{eq_Pfilm_Nbc_box} as
\beq \bra \Kcal\Nbc(t)\ket = - \sum_{m_x=-\infty}^\infty \cdots \sum_{m_z=-\infty}^\infty \Theta(\bv{m}) \left[ \left(1+ \frac{z^2}{r^2}\right) \pd_r^2 \Ccal_b + \left(\frac{d}{r} - \frac{z^2}{r^3}\right) \pd_r\Ccal_b \right]_{\rv = \Lcaltildev}, 
\label{eq_CCF_Nbc_box}\eeq
where, analogously to \cref{eq_CCF_pbc_box}, the term pertaining to $\bv{m}=\bv0$ [see \cref{eq_Theta_m}] is identified with the \emph{bulk} pressure and is therefore absent in the sum.

\subsubsection{Equilibrium CCF}

\begin{figure}[t]\centering
    \includegraphics[width=0.42\linewidth]{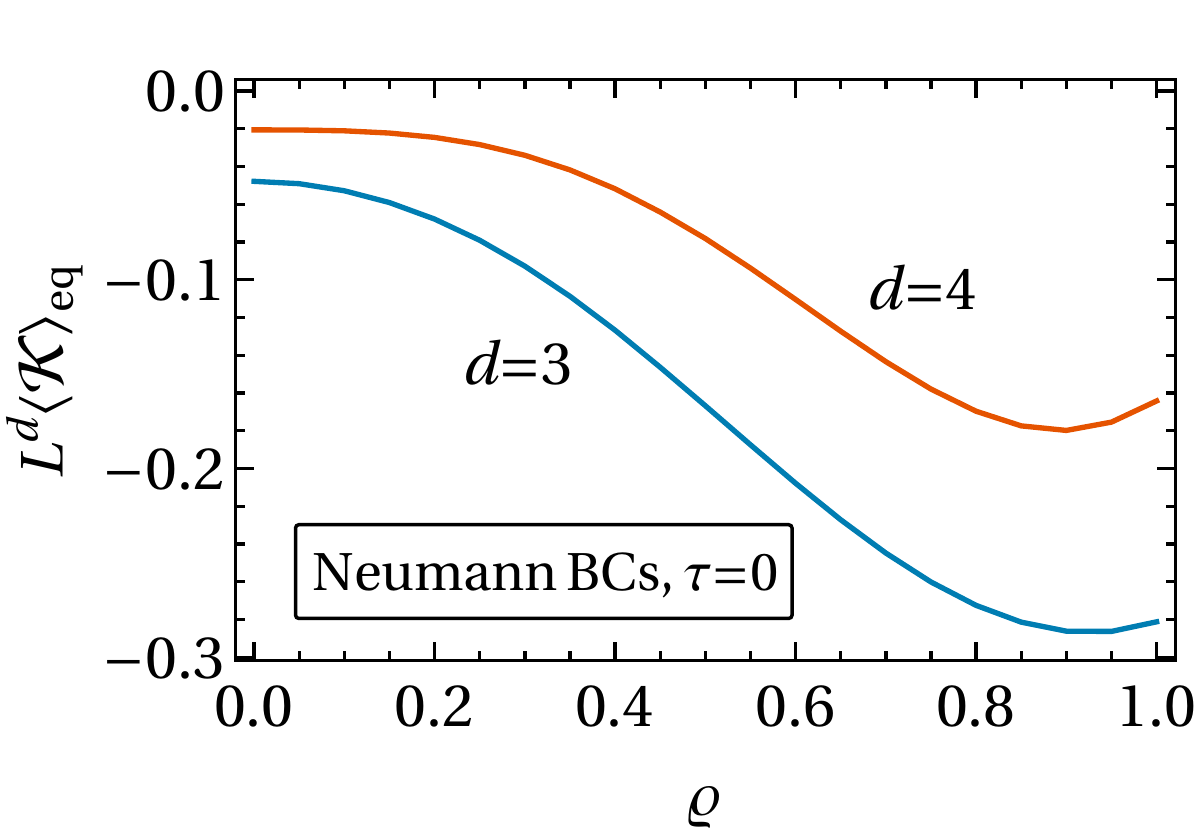} 
    \caption{Canonical CCF in thermal equilibrium  [\cref{eq_CCF_eq_box_pbc_int}] for a box with Neumann \bcs in the transverse direction and periodic \bcs in the lateral directions at bulk criticality ($\tauLG=0$) as a function of the aspect ratio $\varrho$, for $d=3$ and $d=4$ dimensions. $\Kcal$ is taken in units of $k_B T_c$ so that it has the units of an inverse volume.
    }
    \label{fig_Keq_Nbc_rho}
\end{figure}

In order to obtain the equilibrium CCF, we proceed as in \cref{sec_CCF_eq_box} and consider \cref{eq_Tzz_dyn} in equilibrium for arbitrary $\tauLG\geq 0$.
This renders the film pressure  
\beq\begin{split} \bra\Pcal_f\Nbc\ket\eq &= \bra \Rcal\Nbc \ket\eq - \frac{1}{2}\bra \Qcal\Nbc \ket\eq +  \onehalf \tauLG \Ccal\Nbc\eq(\bv0_\parallel,z\in\{0,L\})  \\
&= \bra \Rcal\Nbc \ket\eq - \frac{1}{2}\bra \Qcal\Nbc \ket\eq + \onehalf \tauLG \Ccal\Nbc\eqGC(\bv0_\parallel,z\in\{0,L\})  -  \onehalf L^{-d} \varrho^{d-1}, 
\end{split}\label{eq_Pfilm_Nbc_can_eq}\eeq 
where we have used \cref{eq_OPcorrel_box_eq}.
Since the relations between periodic and Neumann \bcs provided by \cref{eq_CCF_Nbc_box_Poisson} hold also in the equilibrium case, the analogous form of \cref{eq_Qcal_pbc_Schwinger} for Neumann \bcs follows by using \cref{eq_OPcorrel_box_Nbc_zcoinc,eq_CCF_Nbc_box_Poisson,eq_Qcal_pbc_Schwinger} as $\bra\Qcal\Nbc \ket\eq = \bra\Rcal\Nbc \ket\eq + \tauLG \Ccal\Nbc\eqGC(\bv0_\parallel,z\in \{0,L\}) - 2\delta(\bv0)$.
Inserting this into \cref{eq_Pfilm_Nbc_can_eq} renders
\beq \bra\Pcal_f\Nbc\ket\eq = \onehalf \bra \Rcal\Nbc \ket\eq + \delta(\bv0) - \onehalf L^{-d} \varrho^{d-1} .
\label{eq_Pfilm_box_Nbc_eq}\eeq 
The quantity $\delta(\bv0)$ is defined as in \cref{eq_Pfilm_box_pbc_eq} and represents a bulk contribution, while the term $-L^{-d}\varrho^{-d}/2$ is due to the global OP conservation.
The equilibrium CCF for a box with Neumann \bcs follows from \cref{eq_Pfilm_box_Nbc_eq} as \footnote{An alternative form for the CCF, which is equivalent to \cref{eq_CCF_eq_box_Nbc}, is given in Ref.\ \cite{gross_statistical_2017}, where the CCF is obtained from the residual finite-size contribution to the free energy.}
\beq \bra \Kcal\Nbc(\tauLG, \varrho, L)\ket\eq =  \bra\Kcal\Nbc(\tauLG, \varrho, L)\ket\eqGC - \onehalf L^{-d} \varrho^{d-1} ,
\label{eq_CCF_eq_box_Nbc}\eeq 
where, using \cref{eq_Pfilm_box_pbc_eq,eq_CCF_Nbc_box_Poisson_1}, the grand canonical CCF $\bra\Kcal\Nbc\ket\eqGC$ can be expressed in terms of the corresponding one for periodic \bcs [\cref{eq_CCF_eq_box_pbc_gc_int}] as 
\beq \bra\Kcal\Nbc(\tauLG, \varrho, L)\ket\eqGC = \bra\Kcal\pbc(\tauLG, 2\varrho, 2L)\ket\eqGC = 2^{-d}\bra\Kcal\pbc(4\tauLG, 2\varrho, L)\ket\eqGC.
\label{eq_CCF_eq_box_Nbc_gc}\eeq 
In the last relation, we have used the fact that $\tauLG$ enters the underlying expression only in the combination $L^2\tauLG$.
The CCF in \cref{eq_CCF_eq_box_Nbc} fulfils the scaling form given in \cref{eq_scalform_CCF} (with $\initvar=0$) \cite{gross_statistical_2017} and is illustrated in \cref{fig_Keq_Nbc_rho} at bulk criticality, i.e., $\tauLG\to 0$, as a function of $\varrho$.

\subsubsection{Dynamic CCF}
\label{sec_CCF_Nbc_box_dyn}

\begin{figure}[t]\centering
    \subfigure[]{\includegraphics[width=0.435\linewidth]{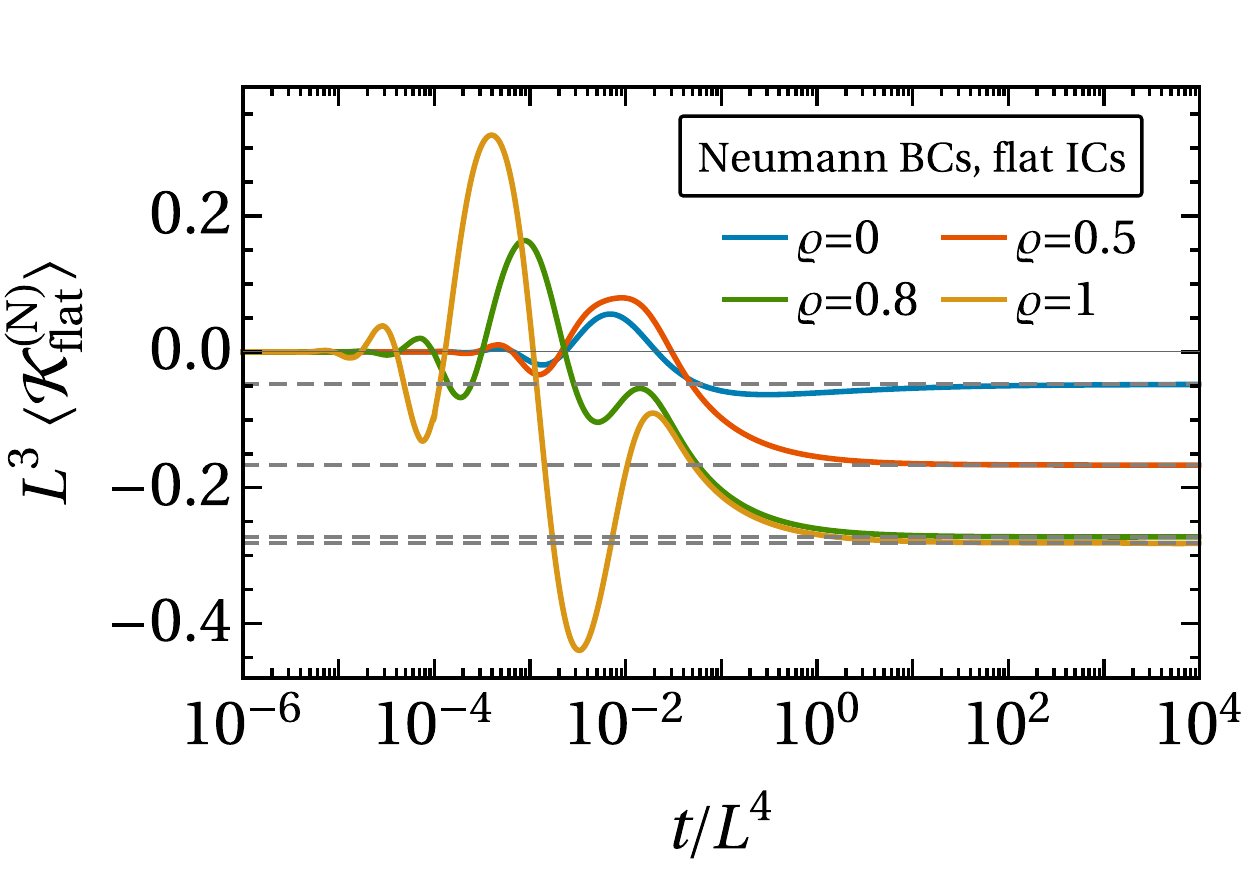} } \qquad
    \subfigure[]{\includegraphics[width=0.42\linewidth]{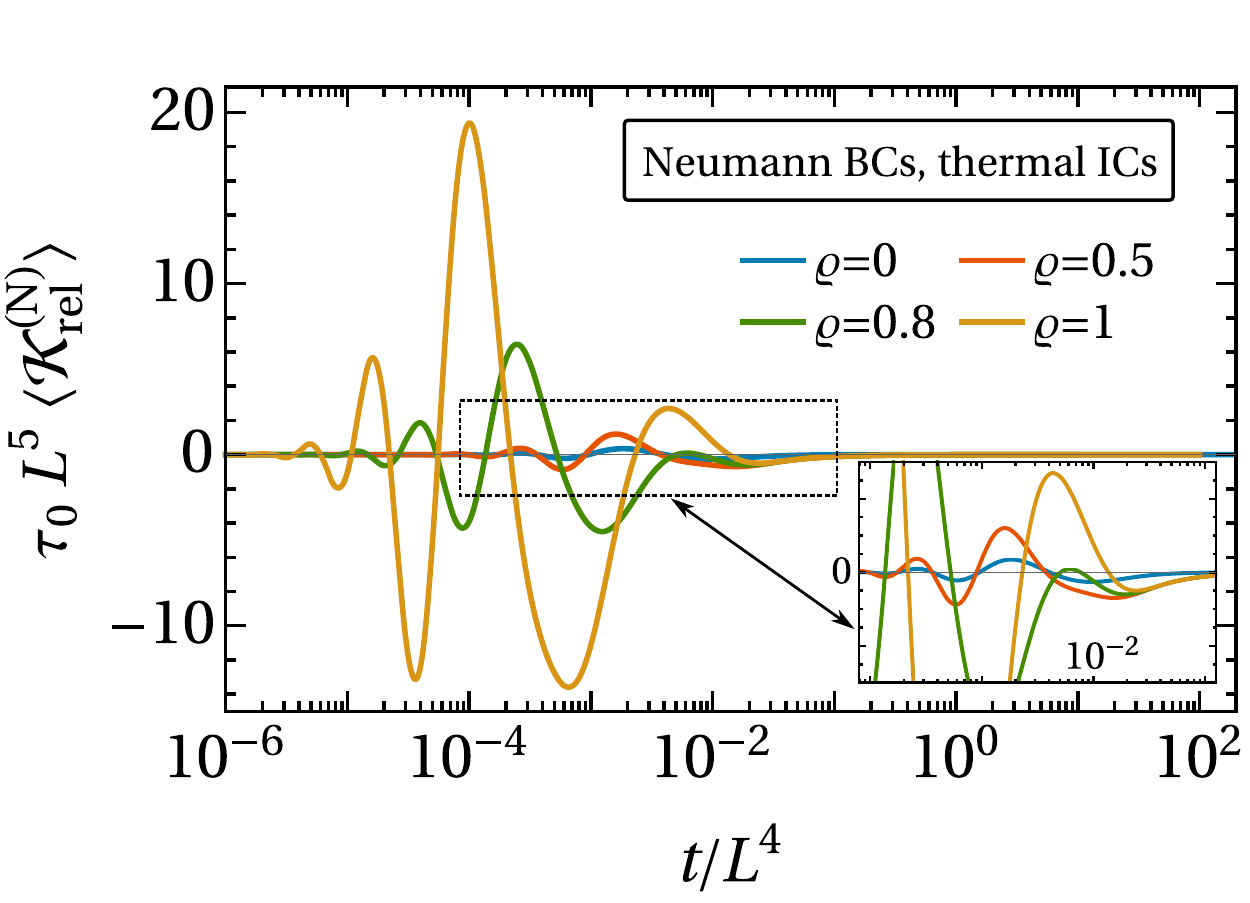} }
    \caption{Time evolution of the dynamic CCF [in $d=3$ and per $k_B T_c$, \cref{eq_CCF_Nbc_box}] for a box with Neumann \bcs in the transverse direction (and periodic \bcs in the lateral directions) for various aspect ratios $\varrho$ ($\varrho = 0$ is the thin film limit). 
    At time $t=0$, the system is instantaneously quenched to the bulk critical temperature ($\tauLG=0$). 
    Panel (a) shows the CCF for flat ICs. The dashed lines indicate the equilibrium CCF at $T_c$ [see \cref{eq_CCF_eq_box_Nbc} and \cref{fig_Keq_Nbc_rho}].
    Panel (b) shows the contribution $\tauInit \bra\Kcal\Nbc\rel(t)\ket$ [see Subsec.\ \ref{sec_CCF_Nbc_box_dyn}] to the CCF for thermal ICs [see \cref{eq_IC_correl}], where $\tauInit=1/\initvar$ is a measure of the initial temperature [see \cref{eq_static_OZ_correl}]. In both panels, CCF and time are scaled according to \cref{eq_scalform_CCF}, such that the universal behavior of the CCF scaling function is exhibited in the plot.
    }
    \label{fig_CCF_box_Neu}
\end{figure}

The time-dependent CCF for \emph{flat} and \emph{thermal} ICs follows from \cref{eq_CCF_Nbc_box} upon inserting Eq.~\eqref{eq_Sflat} or \eqref{eq_Stherm} for the bulk correlation function $\Ccal_b$. 
Analogously to the case of periodic \bcs (see \cref{sec_CCF_dyn_box_pbc}), at late times $t\gg L^z$, the dynamic CCF receives contributions both from the static and the dynamic correlators, $\Scal\stat$ and $\Scal\dyn$.
Accordingly, in \cref{eq_CCF_Nbc_box} the limit $t\to\infty$ must be performed after the summation in order to obtain the correct equilibrium value [\cref{eq_CCF_eq_box_Nbc}] of the CCF.

As illustrated in \cref{fig_CCF_box_Neu}(a), the dynamic CCF for Neumann \bcs and flat ICs initially vanishes and approaches its late-time, equilibrium value [see \cref{eq_CCF_eq_box_Nbc,fig_Keq_Nbc_rho}] in a non-monotonous fashion.
\Cref{fig_CCF_box_Neu}(b) shows the contribution $\bra\Kcal\rel\Nbc(t)\ket$ to the CCF for thermal ICs, which is obtained by inserting $\Ccal_b=\Scal\rel$ [\cref{eq_Srel_def}] into \cref{eq_CCF_Nbc_box} [see also \cref{eq_CCF_pbc_thIC}].
In contrast to the case of periodic \bcs [see \cref{fig_CCF_box_per}], increasing the aspect ratio $\varrho$ from 0 to 1 has a pronounced effect on the time-dependent CCF, inducing, in particular, a shift of the location of the major maximum of the CCF towards shorter times.
A heuristic understanding of this behavior can be gained by noting that, according to \cref{eq_scalform_Cb_init}, the term in the square brackets in \cref{eq_CCF_Nbc_box} has the scaling form $r^{-d}K(\psi)$, $\psi=r^4/(512 t) = L^4 (r_\parallel^2/(2\varrho)^2 + z^2)^2/t$, where the scaling function $K(\psi)$ attains its maximum for $\psi\sim\Ocal(1)$ and vanishes for $\psi\to\infty$ with exponentially decaying oscillations  [see \cref{eq_Srel_largeR_asympt,eq_Sflat_asympt}].
We note the factor 2 multiplying $\varrho$ in the expression for $\psi$, which stems from the fact that the expression in the square brackets in \cref{eq_CCF_Nbc_box} has to be evaluated for $z=2m_z L$.
Accordingly, in \cref{eq_CCF_Nbc_box} for $\varrho\lesssim 1/2$, the transverse mode $\mv=\{0,\ldots,0,1\}$ is dominant and the temporal shape of the CCF depends only weakly on $\varrho$.
In contrast, for $\varrho\gtrsim 1/2$ and a lateral mode $\mv\in \{ \{1,0,\ldots ,0\}, \{0,1,0,\ldots ,0\},\ldots\}$, one reaches $\psi\sim\Ocal(1)$ already at small $t/L^4$. 
Thus these modes increasingly contribute to the CCF and are responsible for a shift of the maximum towards earlier times.
The same reasoning applies also to the CCF for periodic \bcs [see \cref{fig_CCF_box_per}], except that here the crossover between the two regimes occurs for $\varrho\simeq 1$ (which is thus not apparent in the plots).

\section{Summary and Outlook}
\label{sec_sum}

We have studied the non-equilibrium dynamics of a confined fluid quenched to its critical point.
The confinement is realized by a $d$-dimensional cuboid box of volume $V = L L_\parallel^{d-1}$, thickness $L$, and lateral size $L_\parallel$ (see \cref{fig_boxsketch}). 
Included in this setup is the limit of a thin film, for which $L_\parallel\to\infty$.
Our analysis is based on the (linearized) equations of model B, which describe a near-critical fluid with conserved OP, neglecting heat and momentum transport \cite{hohenberg_theory_1977}. 
In the case of a box, periodic or Neumann \bcs are imposed in the transverse ($z$) direction and, for technical reasons, periodic \bcs in the remaining, lateral directions. These \bcs ensure that the total integrated OP $\Phi$ is conserved [see \cref{eq_total_OP}]. Accordingly, in equilibrium, which is achieved at late times ($t\to\infty$), the canonical ensemble is realized.
We take the initial ($t=0$) OP to have a vanishing mean value and short-ranged correlations [see \cref{eq_IC_correl}], amounting to a constant amplitude $\propto \initvar$ of the initial OP variance (``thermal'' initial conditions, ICs). This includes the case of a vanishing initial OP $\phi(\rv,t=0)=0$, which we call ``flat'' ICs [see \cref{eq_IC_correl_flat}]. Physically, thermal ICs correspond to starting the quench from equilibrium at a super-critical temperature which is sufficiently high to ensure a short initial correlation length.

Following Refs.\ \cite{dean_out--equilibrium_2010, kruger_stresses_2018, gross_surface-induced_2018}, the present study focuses on the non-equilibrium critical Casimir force (CCF), which we define in terms of a generalized force acting on the confining boundaries of the system.
The generalized force and the associated dynamical stress tensor \cite{kruger_stresses_2018, gross_surface-induced_2018} are determined by the finite-size OP correlation function and derivatives thereof, evaluated at coinciding spatial points.

Our main results can be summarized as follows:
\begin{enumerate}
 \item We have presented analytical expressions for the static and dynamic correlation functions of an OP governed by model B at bulk criticality subject to periodic or Neumann \bcs. 
 These finite-size correlation functions are expressed, via the Poisson resummation formula, in terms of an infinite summation over ``image points'' of the bulk correlation function (see also Refs.\ \cite{diehl_field-theoretical_1981,diehl_dynamical_1993}). The latter is given in closed form in terms of generalized hypergeometric functions. In the Poisson representation of a finite-size quantity the bulk contribution can be explicitly identified, which facilitates the determination of the CCF.

 \item The formalism, which is based on the recently introduced dynamic stress tensor \cite{kruger_stresses_2018,gross_surface-induced_2018}, is shown to lead, in the late-time limit, to the same equilibrium CCF obtained previously within statistical field theory \cite{krech_free_1992,gross_statistical_2017}. In the case of a box geometry the total OP conservation gives rise to the canonical CCF \cite{gross_statistical_2017}. In contrast, for a thin film the OP conservation is immaterial and  the standard grand canonical CCF \cite{krech_free_1992} is recovered. For periodic and Neumann \bcs and a total OP $\Phi=0$, the value of the canonical CCF is (within the Gaussian approximation) reduced relative to the grand canonical one by the amount $1/(2 V)$ [see \cref{eq_CCF_eq_box_pbc_int,eq_CCF_eq_box_Nbc} and Ref.\ \cite{gross_statistical_2017}]. 

 \item For all geometries and \bcs considered here, the dynamic CCF vanishes initially. Physically, this can be understood as a consequence of the short-ranged correlations of the ICs [see \cref{eq_IC_correl}] and the symmetry-preserving character of the \bcs (i.e., there are no surface fields). The (non-zero) late-time, equilibrium value of the CCF is approached in an oscillatory growing fashion. This oscillatory behavior is in contrast to the more gentle growth of the dynamic CCF for quenches within model A dynamics (i.e., for non-conserved OP) in a critical film  \cite{dean_out--equilibrium_2010}, as well as to the purely transient, non-oscillatory forces reported for model B in systems far from criticality~\cite{rohwer_transient_2017,rohwer2018forces}.
 
 \item Thermal ICs give rise to an additional transient at intermediate times, superimposed on the dynamics of the CCF pertaining to flat ICs. Thermal ICs do not affect the late-time behavior of the CCF. Their influence diminishes upon decreasing the amplitude of the initial OP correlations [see \cref{eq_IC_correl}]. 
 
\end{enumerate}

Within the framework of boundary critical phenomena \cite{diehl_field-theoretical_1986,diehl_theory_1997}, imposing Neumann \bcs for the OP [\cref{eq_eigenspec}] at the Gaussian level corresponds to the so-called ``special'' surface universality class (SUC) \footnote{This correspondence between Neumann \bcs and the special SUC breaks down beyond the Gaussian approximation, see, e.g., Ref.\ \cite{diehl_critical_2009} for further discussion and references.}.
Besides the so-called ``normal'' SUC, the quench dynamics of which has been studied in Ref.~\cite{gross_surface-induced_2018}, another relevant SUC for fluids is the so-called ``ordinary'' SUC, corresponding (within the Gaussian approximation) to Dirichlet \bcs. Each static SUC splits up into several dynamic SUCs depending on the dynamic model considered.
In contrast to Neumann \bcs, standard Dirichlet \bcs do not entail a vanishing flux at the boundaries of the system. In fact, for semi-infinite geometries, in Refs.\ \cite{diehl_universality_1994, wichmann_dynamic_1995} non-conservative dynamics at a surface has been shown  to lead to a new dynamic SUC (called ``model $\mathrm{B_A}$'' in Ref.\ \cite{diehl_universality_1994}), which is distinct from the fully conservative model B \cite{diehl_boundary_1992} (denoted as ``model $\mathrm{B_B}$'' in this context). A modification of Dirichlet \bcs in order to implement the no-flux condition is possible but leads to transcendental eigenvalues \cite{gross_first-passage_2018,gross_first-passage_2018-1} and is left for future investigations.

Another direction into which the present study can be extended, concerns the dynamics following a quench to a slightly super-critical temperature.
In fact, in the high-temperature limit of model B, which in thermal equilibrium is free of long-ranged correlations, one observes a transient post-quench Casimir-like force induced solely by the dynamic conservation law \cite{rohwer_transient_2017,rohwer2018forces}, while the equilibrium CCF vanishes. 

It should also be rewarding to go beyond the Gaussian dynamics considered here, retaining the $\phi^4$ nonlinearity in the Landau-Ginzburg free energy functional. The nonlinear term is expected to lead, \emph{inter alia}, to corrections of the early-time behavior after the quench \cite{ritschel_dynamical_1996}. 
Furthermore, the effect of a nonzero initial mean OP, i.e., $\bra\phi(\rv,t=0)\ket=\Phi/V \neq 0$, could be investigated.
In the case of non-conserved dynamics, it has been shown that a non-zero initial OP leads to a universal initial growth behavior of the OP, the so-called ``critical initial slip'' \cite{janssen_new_1989, janssen_renormalized_1992, diehl_dynamical_1993, ritschel_universal_1995, ritschel_dynamical_1996}, which is described by a new dynamic critical exponent \footnote{It is known, however, that in model B no genuinely new initial-slip exponent appears \cite{janssen_new_1989}.}. 
Additionally, it would be insightful to explore the relationship between the present model and active matter systems in more detail \cite{caballero_bulk_2018}.

Regarding the experimental realization of our findings, quenches of binary liquid mixtures to their critical demixing point appear to be promising candidates.
Indeed, extending established experimental techniques for equilibrium CCFs (see, e.g., Ref.~\cite{hertlein_direct_2008}) to non-equilibrium scenarios is a timely issue. 
In this regard, the influence of thermal diffusion and momentum transport on the quench dynamics could be elucidated \cite{oerding_nonequilibrium_1993,roy_solvent_2018}. 
Ultimately these studies should aim at the ``normal'' SUC as the one to which actual fluids belong due to the generic presence of symmetry breaking surface fields.
Notably, while the ``ordinary'' SUC is experimentally realizable for fluids via a suitable tuning of the surface fields \cite{desai_critical_1995,nellen_tunability_2009,mohry_crossover_2010}, achieving conditions appropriate for the ``special'' SUC is still an open issue.

A more direct test of our predictions can be achieved via simulations, for which non-symmetry breaking \bcs are easily realizable. Recently, progress has been made, e.g., within molecular dynamics \cite{das_critical_2006, roy_structure_2016} and within the lattice Boltzmann method \cite{gross_simulation_2012, gross_critical_2012, belardinelli_fluctuating_2015} towards simulation of critical dynamics and determining CCFs \cite{puosi_direct_2016}. Moreover, these simulation methods typically realize the canonical ensemble and thus allow one to test the ensemble differences of the CCF \cite{gross_critical_2016, gross_statistical_2017}.

\begin{acknowledgments}The authors would like to thank H.\ W.\ Diehl for useful correspondence.\end{acknowledgments}

\appendix

\section{Poisson resummation formula}
\label{app_Poisson}

The Poisson resummation formula in its general form (see, e.g., Refs.\ \cite{dohm_diversity_2008, gruneberg_thermodynamic_2008, diehl_dynamical_1993}) states that, for a given function $\hat f(\bv{n})$ depending on $\bv{n}\in \mathbb{Z}^D$, 
\beq \sum_{\bv{n}} \hat f(\bv{n}) = \sum_\bv{m} F(\bv{m}),
\label{eq_Poisson_summation}\eeq 
where $\bv{m}\in \mathbb{Z}^D$,
\beq  F(\rv) \equiv \int_{\reals^D} \d^D k\, \hat f(\kv) e^{2\pi \im \kv\cdot\rv} = (2\pi)^D f(2\pi \rv),
\eeq  
and
\beq  f(\rv) = \int_{\reals^D} \frac{\d^D k}{(2\pi)^D}\, \hat f(\kv) e^{\im \kv\cdot\rv}
\label{eq_Fourier}\eeq
represents the standard ($D$-dimensional) inverse Fourier transform of $\hat f$.

Specifically, if $\hat f$ depends on a $D<d$ dimensional vector $\bv{n}\in \mathbb{Z}^D$ via the form $\hat f\left(\Big\{\frac{2\pi n_\alpha}{L_\alpha} \Big\}\right) $ with $L_\alpha \in \mathbb{R}$, $\alpha=1,\ldots, D$, \cref{eq_Poisson_summation} renders
\beq \sum_{\bv{n}} \hat f\left(\Big\{\frac{2\pi n_\alpha}{L_\alpha} \Big\}\right) = \frac{\prod_\alpha L_\alpha}{(2\pi)^D} \sum_{\bv{m}} \int \d^D k\, \hat f(\kv) e^{\im k_\alpha m_\alpha L_\alpha} = \prod_\alpha L_\alpha \sum_{\bv{m}} f(\rv= \Lcalv),
\label{eq_Poisson_periodic}\eeq 
where $\Lcalv$ is a shorthand notation for the $D$-dimensional vector $\{m_1 L_1, \ldots, m_D L_D \}$ [see \cref{eq_mL_notation}].
Another relevant case for the present study is obtained if $\hat f$ depends on $n=0,1,2,\ldots$ in the form $\hat f(n) = (2-\delta_{n,0}) \doublehat{f}\left( \frac{\pi |n|}{L} \right)$, with $L\in\mathbb{R}$ and some function $\doublehat{f}$. Here, \cref{eq_Poisson_summation} renders
\beq\begin{split} \sum_{n=0,1,\ldots} (2-\delta_{n,0}) \doublehat f\left( \frac{\pi |n|}{L} \right) &= 2 \sum_{n=0,1,\ldots} \doublehat f\left( \frac{\pi |n|}{L} \right) - \doublehat f(0) = \sum_{{n=-\infty,}\atop{n\neq 0}}^\infty \doublehat f\left( \frac{\pi |n|}{L} \right) + \doublehat f(0) \\
&= \sum_{n=-\infty}^\infty \doublehat f\left( \frac{\pi |n|}{L} \right) 
=  2L \sum_{m=-\infty}^\infty f(r= 2L m) ,
\end{split}\label{eq_Poisson_Neumann}\eeq 
where now, accordingly, $f$ in \cref{eq_Fourier} is to be evaluated by replacing $\hat f$ by $\doublehat{f}$ on the right-hand side.

\section{Static equilibrium correlation functions in a thin film}
\label{sec_film_correl}

Here, we present analytic expressions for the static equilibrium correlation function $\Ccal\eq(\rvp,z,z')$ [\cref{eq_OPcorrel_box_eq}] in a thin film for periodic as well as Neumann \bcs. 
In this case, according to \cref{eq_OPcorrel_thinfilm_eq}, $\Ccal\eq$ is not affected by the choice of the ensemble.

\subsection{Thin film with periodic \bcs}
\label{sec_film_correl_pbc}

Upon inserting $\Ccal_b=\Scal\stat$ [\cref{eq_Sstat}] into \cref{eq_OPcorrel_thinfilm_per}, the static correlation function for a thin film with periodic \bcs follows as 
\beq \Ccal\pbc\eq(r_\parallel, z) = \frac{ \Gamma(d/2-1)}{4\pi^{d/2}} \sum_{m=-\infty}^\infty \frac{1}{\left|(z+ mL)^2 + r_\parallel^2\right|^{d/2-1}} = \frac{ \Gamma(d/2-1)}{4\pi^{d/2} L^{d-2}} \sum_{m=-\infty}^\infty \frac{1}{\left|(\zeta+ m)^2 +  \hat r^2\right|^{d/2-1}},
\label{eq_OPcorrel_stat_film}\eeq 
where $\zeta\equiv z/L$ and $\hat r\equiv r_\parallel/L$.
Due to the periodicity in the $z$-direction, $\Ccal\pbc\eq$ depends only on one $z$ coordinate.
For $r_\parallel\neq 0$, the sum represents an inhomogeneous Epstein zeta function which can be determined according to  Eq.~(4.13) in Ref. \cite{elizalde_ten_2012}. This renders 
\beq \Ccal\pbc\eq(r_\parallel > 0, z) = \frac{1 }{L^{d-2}} \Big[ \frac{\Gamma(d/2-3/2)}{4\pi^{d/2-1/2} } \hat r^{3-d} + \frac{1}{\pi} \hat r^{(3-d)/2} \sum_{n=1}^\infty n^{d/2-3/2} \cos(2\pi n \zeta) K_{d/2-3/2}(2\pi n \hat r) \Big],
\label{eq_OPcorrel_stat_film_zeta}\eeq 
with the modified Bessel function $K_\nu$ of the second kind.
For $r_\parallel=0$, \cref{eq_OPcorrel_stat_film} can be expressed in terms of the Hurwitz zeta function $\zeta_H(s,q)$ \cite{olver_nist_2010} as
\beq \Ccal\pbc\eq(r_\parallel = 0, z) = \frac{ \Gamma(d/2-1)}{4\pi^{d/2}L^{d-2}} \left[ \frac{1}{\zeta^{d-2}} + \zeta_H(d-2, 1+\zeta) +  \zeta_H(d-2,1-\zeta) \right].
\label{eq_OPcorrel_stat_film_Hurwitz}\eeq 
For $d\to 3$ one has $\zeta_H(d-2, u) \simeq \frac{1}{d-3} - \Psi(u) + \Ocal(d-3)$, where $\Psi$ denotes the dilogarithm.
The logarithmic divergence of the sum in \cref{eq_OPcorrel_stat_film} for $d=3$ is reflected by the simple poles $\sim 2/(d-3)$ in \cref{eq_OPcorrel_stat_film_zeta,eq_OPcorrel_stat_film_Hurwitz}, stemming from the contributions $\Gamma(d/2-3/2)$ and $\zeta_H$, respectively.  
\Cref{eq_OPcorrel_stat_film,eq_OPcorrel_stat_film_zeta} are therefore valid for $d>3$.
For $d=4$, the sum in \cref{eq_OPcorrel_stat_film_zeta} can be carried out in closed form, yielding
\beq \Ccal\pbc\eq(r_\parallel,z)\big|_{d=4}  = \frac{1 }{4\pi L r_\parallel} \Big[ 1 +  \frac{e^{-2 \pi  \hat r}-\cos (2 \pi  \zeta )}{\cos (2 \pi  \zeta )-\cosh (2 \pi  \hat r)} \Big].
\label{eq_OPcorrel_stat_film_d4}\eeq 
For large lateral distances $\hat r \gg 1$, \cref{eq_OPcorrel_stat_film_d4} reduces to $\Ccal\pbc\eq(r_\parallel\gg L,z)\simeq (\pi/L) \Scal\stat(r_\parallel)$, i.e., $\Ccal\pbc\eq$ shows the same $r_\parallel$-dependence as the correlation function in a critical $d=3$ dimensional bulk system [\cref{eq_Sstat}].
On the other hand, in the limit $L\to \infty$ one obtains $\Ccal\pbc\eq(r_\parallel,z) \simeq \sfrac{1}{[4\pi^2 (r_\parallel^2 + z^2)]}$, i.e., the static bulk correlator [\cref{eq_Sstat}] for $d=4$. 

\subsection{Thin film with Neumann \bcs}
\label{sec_film_correl_nbc}

Upon using \cref{eq_OPcorrel_stat_film_zeta} and the last line of \cref{eq_OPcorrel_box_Nbc}, the following form of the static equilibrium correlation function for a thin film with Neumann \bcs is obtained:
\begin{multline}
\Ccal\Nbc\eq(r_\parallel > 0, z,z') =  \frac{1 }{L^{d-2}} \Bigg[\frac{\Gamma(d/2-3/2)}{4\pi^{d/2-1/2} } \hat r^{3-d}  + \frac{1}{\pi} \hat r^{(3-d)/2} \sum_{n=1}^\infty \left(\frac n 2 \right)^{(d-3)/2} \cos(\pi n \zeta) \cos(\pi n \zeta') K_{(d-3)/2}(\pi n \hat r) \Bigg], 
\label{eq_OPcorrel_stat_film_zetaN}
\end{multline}
with $\zeta\equiv z/L$ and $\hat r\equiv r_\parallel/L$.
In contrast to the periodic case, $\Ccal\Nbc\eq$ depends explicitly on $z'$. For $z' = 0$, \cref{eq_OPcorrel_stat_film_zetaN} differs from the periodic one [\cref{eq_OPcorrel_stat_film_zeta}] only by the replacement $n\to n/2$ in the summation. 
For $r_\parallel=z'=0$, one has 
\beq \Ccal\Nbc\eq(r_\parallel = 0, z) = 2^{3-d} \Ccal\pbc\eq(r_\parallel = 0, z/2).
\label{eq_OPcorrel_stat_film_HurwitzN}\eeq 
In the case of $d=4$ dimensions, \cref{eq_OPcorrel_stat_film_zetaN} can be determined explicitly, rendering [in analogy to \cref{eq_OPcorrel_stat_film_d4}]
\al{
\Ccal\Nbc\eq(r_\parallel,z,z')\big|_{d=4} &= \frac{1 }{8\pi L r_\parallel} 
\frac{\sinh (\pi  \hat r) \big[\cos (\pi  (\zeta-\zeta') )+\cos (\pi  (\zeta+\zeta') )-2 \cosh (\pi  \hat r)\big]}
{ \big[\cos (\pi  (\zeta-\zeta') )-\cosh (\pi  \hat r)\big]\big[\cosh (\pi  \hat r)-\cos (\pi  (\zeta+\zeta') )\big]}. 
\label{eq_OPcorrel_stat_film_d4N}
}

\section{Late-time asymptotics of the CCF for a thin film}
\label{app_CCF_film_asympt}

Here we derive the late time asymptotic behavior, reported in \cref{eq_CCF_asymp_larget}, of the dynamic CCF for periodic \bcs in the thin film geometry.
Using \cref{eq_Sflat,eq_CCF_pbc_eqampl}, one can write \cref{eq_CCF_pbc_final} as 
\beq \bra\Kcal(t)\ket -\bra\Kcal\ket\eq = \sum_{m=1}^\infty \mathpzc{h}(Lm,t),
\label{eq_Kasymp_sum}\eeq 
with
\beq \mathpzc{h}(z,t) \equiv 3 \pd_z^2 \Scal\dyn(z,t) + \frac{d-1}{z}\pd_z \Scal\dyn(z,t) = t^{-d/4} h(\psi), \qquad \psi=\frac{z^4}{512 t},
\label{eq_Kasymp_scalf}\eeq 
where we have expressed $\mathpzc{h}(z,t)$ in terms of a scaling function $h(\psi)$. 
The explicit form of $h(\psi)$ can be obtained from \cref{eq_Sdyn_def}, but it is rather lengthy and is thus not stated here. However, we note its limiting behaviors:
\beq\begin{split} 
h(\psi\to 0) &\simeq \frac{(2+d)\pi^{(1-d)/2}}{2^{7d/4} d\Gamma(1/2+d/4)}, \\
h(\psi\to \infty) &\simeq \frac{(d-1)\pi^{1-d/2} }{2^{9d/4} \Gamma(1-d/2) \sin(d\pi/2)} \psi^{-d/4}.
\end{split}\label{eq_Kasymp_limits}\eeq 
In order to proceed, we recall the Euler-Maclaurin formula: 
\beq \sum_{m=1}^\infty f(m) = \int_1^\infty \d m\, f(m) + \frac{f(1) + f(\infty)}{2} + \mathfrak{R},\qquad |\mathfrak{R}| \leq \frac{1}{12} \int_1^\infty \d m\, |f''(m)| ,
\eeq
where $\mathfrak{R}$ denotes the remainder.
Applying this formula to \cref{eq_Kasymp_sum}, and using \cref{eq_Kasymp_limits} as well as the scaling property expressed in \cref{eq_Kasymp_scalf}, we obtain, in the limit of late times $t \gg L^4$:
\beq \sum_{m=1}^\infty \mathpzc{h}(Lm,t) \simeq t^{1/4-d/4} \frac{2^{1/4} }{L} \int_0^\infty \d \psi \, \psi^{-3/4}  h(\psi) + \onehalf t^{-d/4} h(0) + \mathfrak{R}.
\label{eq_Kasymp_intform}\eeq 
We note that the lower integration boundary and the argument of $h$ in the second term vanish in this limit of late $t$. 
Moreover, owing to \cref{eq_time_resc}, the r.h.s.\ in \cref{eq_Kasymp_intform} indeed has the correct dimension $L^{-d}$.
The remainder $\mathfrak{R}$ is estimated as $|\mathfrak{R}| \leq t^{-1/2 - d/4} L^2 \int_0^\infty \d \psi\, \psi^{-3/4} |h''(\psi)|$.
According to \cref{eq_Kasymp_limits} [and the fact that $h''(\psi\to 0)$ is finite], the integrals over $h(\psi)$ and $h''(\psi)$ are finite and we thus conclude that the dominant behavior for large times in \cref{eq_Kasymp_intform} is provided by the first term on the r.h.s.
An analysis of this contribution yields the late-time asymptotic behavior
\beq  \bra\Kcal(t)\ket -\bra\Kcal\ket\eq \simeq -\frac{2^{3/4-7d/4} \pi^{1-d/2}}{\Gamma(1/4+d/4)} t^{1/4-d/4}.
\eeq

\section{Static contribution to the CCF}
\label{app_static_CCF}

Here, we discuss the contribution to the CCF which results from evaluating \cref{eq_CCF_pbc_box} with $\Ccal_b = \Scal\stat$ [\cref{eq_Sstat}].
This so-called ``static'' contribution leads to
\beq\begin{split} 
\bra \Kcal\pbc \ket\stat &= - \frac{\Gamma(d/2)}{2\pi^{d/2}} \sum_{\{m_\alpha=-\infty\}}^{\{\infty\}}  \Theta(\bv{m})\left[ \frac{d\, z^2}{r^{d+2}} -\frac{1}{ r^d}  \right]_{r_\alpha= L_\parallel m_\alpha, r_z = L m_z} \\
&= -  \sum_{\{m_\alpha=-\infty\}}^{\{\infty\}}  \Theta(\bv{m})\, \pd_z^2 \Scal\stat(r)\Big|_{r_\alpha= L_\parallel m_\alpha, r_z = L m_z} ,
\end{split}\label{eq_CCF_eq_box_pbc_gc}\eeq 
where the aspect ratio $\varrho$ is given in \cref{eq_aspectratio} and $\Theta$ in \cref{eq_Theta_m}.
Interestingly, for $\varrho=1$ one finds 
\beq \bra\Kcal\pbc\ket\stat\big|_{\varrho=1}=0,
\label{eq_CCF_pbc_box_rho1}\eeq 
which can be readily proven by writing the term in the square brackets of the first line of \cref{eq_CCF_eq_box_pbc_gc} as $\big[(d-1)r_z^2 - \sum_{\alpha=1}^{d-1} r_\alpha^2 \big]/r^{d+2} = r^{-d-2} \sum_{\alpha=1}^{d-1} (r_z^2-r_\alpha^2)$ and using the fact that, for $\varrho=1$, $r_z$ and $r_\alpha$ run over the same set of values.
Analogously, for Neumann \bcs, \cref{eq_CCF_eq_box_Nbc} implies 
\beq \bra\Kcal\Nbc \ket\stat\big|_{\varrho=1}=0.
\label{eq_CCF_Nbc_box_rho1}\eeq 

We emphasize that, in a box geometry and for both periodic and Neumann \bcs, $\bra\Kcal\ket\stat$ does in general \emph{not} represent the \emph{equilibrium} CCF, i.e., 
\beq \bra \Kcal \ket\stat \neq \bra\Kcal\ket\eq,\qquad \varrho>0.
\label{eq_Kstat_Keq_diff}\eeq 
The reason is that the equilibrium CCF acquires a contribution from $\Scal\dyn$ which does not vanish in the limit $t\to\infty$ (see the discussion in \cref{sec_CCF_dyn_box_pbc}).
This contribution vanishes only in the case of a thin film, for which one thus obtains  $\bra\Kcal\ket\stat= \bra\Kcal\ket\eq$ [see \cref{eq_CCF_pbc_eqampl}].

\section{CCF for a box with periodic \bcs near bulk criticality}
\label{app_CCF_box_reg}

Here, we analyze the behavior of the (grand canonical) equilibrium CCF for periodic \bcs as reported in \cref{eq_CCF_eq_box_pbc_gc_int} upon approaching the bulk critical point $\tauLG\to 0$ and discuss its relation to $\bra\Kcal\stat\ket$ defined in \cref{eq_CCF_eq_box_pbc_gc}.
Upon expressing the Jacobi theta function [\cref{eq_Jacobi_theta}] by means of the Poisson resummation formula [\cref{eq_Poisson_summation}] as 
\beq \vartheta(y) = \sqrt{\frac{\pi}{y}} \sum_{m=-\infty}^\infty e^{-\pi^2 m^2/y},
\eeq 
\cref{eq_CCF_eq_box_pbc_gc_int} takes on the form
\beq\begin{split} 
\bra\Kcal\pbc\ket\eqGC &= \frac{\d}{\d L} \left[ \frac{1}{(2\pi)^{d/2} L^{d-1}}  \sum_{\{m_\alpha=-\infty\}}^{\{\infty\}} \Theta(\bv{m}) \sqrt{\frac{L^2\tauLG}{\varrho^{-2}\sum_{\alpha=1}^{d-1} m_\alpha^2 + m_z^2}}\, K_{d/2}\left( \sqrt{L^2\tauLG\left( \varrho^{-2}\sum_{\alpha=1}^{d-1} m_\alpha^2 + m_z^2 \right)}\right) \right] \\
&= \frac{1}{ L^d }  \sum_{\{m_\alpha=-\infty\}}^{\{\infty\}} \Theta(\bv{m}) \Bigg\{  \kappa^{d/4} \hat r^{-2-d/2} K_{d/2}\left(2\pi \hat r\sqrt{\kappa}\right) \left[ \varrho^{-2} \sum_{\alpha=1}^{d-1} m_\alpha^2 - (d-1) m_z^2  \right] \\
&\qquad \qquad - 2\pi  \kappa^{1/2+d/4} \hat r^{-1-d/2} K_{1-d/2}\left(2\pi\hat r \sqrt{\kappa}\right)  m_z^2 \Bigg\},
\end{split}\label{eq_CCF_eq_box_pbc_gc_Bessel}\eeq  
where $\kappa\equiv  L^2\tauLG/(4\pi^2)$, $\hat r \equiv \varrho^{-2}\sum_{\alpha=1}^{d-1} m_\alpha^2 + m_z^2$, and $K_n(z)$ denotes the modified Bessel function of the second kind. The function $\Theta(\mv)$ is defined in \cref{eq_Theta_m} and accounts for the last term in the curly brackets in \cref{eq_CCF_eq_box_pbc_gc_int}. We note that here the total derivative is required, because $\varrho$ depends on $L$ [see \cref{eq_aspectratio}].

We henceforth focus on a cubic box geometry, i.e., $\varrho=1$.
In this case, the first term in the curly brackets in \cref{eq_CCF_eq_box_pbc_gc_Bessel} vanishes after the summation over $\mv$ \footnote{This can be readily seen by writing the term in the square brackets in that line as $\sum_{\alpha=1}^{d-1} ( m_\alpha^2 -  m_z^2 )$ and noting that $m_\alpha$ and $m_z$ run over the same set of integer values [compare \cref{eq_CCF_pbc_box_rho1}].}.
Next, we analyze the behavior of the last term in the curly brackets in \cref{eq_CCF_eq_box_pbc_gc_Bessel}, i.e.,
\beq \mathcal{J} \equiv -\frac{2\pi \kappa^{1/2+d/4}}{  L^d }  \sum_{\{m_\alpha=-\infty\}}^{\{\infty\}} \Theta(\bv{m})\,   \hat r^{-1-d/2} K_{1-d/2}\left(2\pi\hat r \sqrt{\kappa}\right)  m_z^2.
\label{eq_Knz_aux}\eeq 
For nonzero $\tauLG>0$, one has $\mathcal{J}<0$. In the limit $\tauLG\to 0$, if taken \emph{before} the sum over $\bv{m}$, one obtains $\kappa^{1/2+d/4}\allowbreak K_{1-d/2}(2\pi \hat r \sqrt{\kappa}) \to 0$, which is consistent with \cref{eq_CCF_pbc_box_rho1}.
However, accepting \cref{eq_CCF_eq_box_pbc_gc_Bessel} as the definition of the CCF, the actual value of $\bra\Kcal\pbc\ket\eqGC$ at bulk criticality must be obtained by taking the limit $\tauLG\to 0$ \emph{after} the summation over $\bv{m}$. 

In order to determine the limit of $\mathcal{J}$ for $\kappa\to 0$, we approximate the sum in \cref{eq_Knz_aux} by an integral.
Upon introducing $d$-dimensional spherical coordinates, one obtains, in fact, a $\tauLG$-independent result:
\beq \mathcal{J} \underset{\kappa\to0}{\longrightarrow}  \frac{2\pi \kappa^{1/2+d/4}}{L^d}  \int_0^\infty \d r \int \d\Omega \, r^{d/2-2} r_z^2\, K_{1-d/2}\left(2\pi r \sqrt{\kappa}\right)  = -\frac{1}{L^d} \frac{1}{d}.
\label{eq_Knz_aux2}\eeq 
In the calculation, we rotated the coordinate system such that $r_z=r \cos \theta$ and used the fact that the surface area of the $(d-1)$-dimensional sphere is given by $\Omega_{d-1} \equiv \int\d\Omega = 2\pi^{d/2}/\Gamma(d/2) = \Omega_{d-1} \int_0^\pi \d \theta\, \sin^{d-2}\theta$.
In summary, \cref{eq_CCF_eq_box_pbc_gc_Bessel,eq_Knz_aux2} render the value
\beq \bra\Kcal\pbc\ket\eqGC\big|_{\tauLG\to 0, \varrho=1} \to -\frac{1}{L^d} \frac{1}{d},
\label{eq_CCF_pbc_box_rho1_est}\eeq 
which is different from \cref{eq_CCF_pbc_box_rho1}.
Notably, \cref{eq_CCF_pbc_box_rho1_est} agrees accurately with a numerical evaluation of \cref{eq_CCF_eq_box_pbc_gc_int}.

\section{Dynamic contribution to the equilibrium CCF}
\label{app_dyn_constr}

In this appendix we analyze the dynamic contribution to the equilibrium CCF for a box with periodic \bcs, as expressed in \cref{eq_CCF_pbc_box}, stemming from the non-fluctuating property of the zero mode at late but finite times ($t\gg L^4$).
An analogous analysis for Neumann \bcs leads to essentially the same result, but is, due to the approximative character of the calculation, not included here.
We note that, while we consider flat ICs [see \cref{eq_Sflat}], the obtained asymptotic results are valid also for thermal ICs, because their asymptotic contribution to the CCF is subdominant at late times [see \cref{eq_Cb_Dz2_rel_lim_tinf,eq_Cb_Dz2_lim_tinf}].

\subsection{Exact calculation in dimension $d=1$}
\label{app_asympt_1d}

In spatial dimension $d=1$, the exact asymptotic behavior of the CCF at late times ($t\to \infty$) can be determined analytically.
By using \cref{eq_Sflat}, in $d=1$ \cref{eq_CCF_pbc_box} reduces to
\beq \bra\Kcal\pbc(t)\ket\big|_{d=1} = \frac{3}{2}  \sump_{m = -\infty}^{\infty} \pd_r^2 \left[\Scal\stat(m L) - \Scal\dyn(m L, t)  \right],
\label{eq_Kpbc_1d}\eeq 
where the prime indicates the absence of the term $m=0$.
We remark that, in $d=1$ \cref{eq_Sstat} renders $\Scal\stat(r)=-r/2$, which does not contribute to $\bra\Kcal\pbc(t)\ket$.
The sum in \cref{eq_Kpbc_1d} can be determined via the Abel-Plana summation formula \cite{saharian_generalized_2007},
which, for an even function $f(m)=f(-m)$, states that
\beq \sump_{m=-\infty}^\infty f(m)= 2\int_0^\infty \d x\, f(x)  - f(0) + 2\im \int_0^\infty \d t \frac{f(\im t)-f(-\im t)}{e^{2\pi t}-1}  .
\label{eq_AbelPlana2}\eeq 
For $f\equiv \pd_r^2 \Scal\dyn$, the last term on the r.h.s.\ of \cref{eq_AbelPlana2} vanishes, while the second term decays $\propto t^{-1/4}$ and can be neglected for large $t$.
Using the expression for $\Scal\dyn$ in Fourier space given in \cref{eq_Sdyn_def}, one is then left with
\beq\begin{split} \bra\Kcal\pbc(t\gg L^4)\ket\big|_{d=1} &\simeq -\frac{3}{2 L} \int_{-\infty}^\infty \d r\, \int_{-\infty}^\infty \frac{\d k}{2\pi} e^{\im k r} e^{-2k^4 t}  = -\frac{3}{2L}  \int_{-\infty}^\infty \d k\, \delta(k) e^{-2k^4 t} \\
&= -\frac{3}{2L}.
\end{split}\label{eq_dynCCF_asympt_R}\eeq 

A direct calculation analogously to \cref{eq_Pfilm_pbc_box_eq,eq_CCF_eq_box_pbc_int} of the CCF in dimension $d=1$ \emph{at} equilibrium yields
\beq\begin{split} 
\bra\Kcal\pbc\ket\eq\big|_{d=1} &= \frac{L \tauLG}{4\pi^2} \int_0^\infty \d s\, \exp\left(-\frac{L^2\tauLG}{4\pi^2}s\right) \left[  \pfrac{\pi}{s}^{1/2} - \vartheta(s) \right] \\
&= \frac{1}{L} \left\{ \frac{\sqrt{L^2\tauLG}}{2} \left[ 1 - \coth\left(\frac{\sqrt{
L^2 \tauLG} }{2}\right)  \right] - \onehalf  \right\} ,
\end{split}\label{eq_CCF_pbc_1d}\eeq 
where, in order to obtain the last expression, we have used the expansion of $\coth$ in terms of simple fractions (see \S 1.421 in Ref.\ \cite{gradshteyn_table_2014}).
At the bulk critical point ($\tauLG=0$), \cref{eq_CCF_pbc_1d} reduces to 
\beq \bra\Kcal\pbc\ket\eq\big|_{d=1,\tauLG\to 0} = -\frac{3}{2 L},
\eeq 
in agreement with the asymptotic estimate in \cref{eq_dynCCF_asympt_R}.

\subsection{Asymptotic estimate in dimensions $d>1$}

In spatial dimensions $d>1$, the late time behavior of the CCF can be estimated asymptotically.
To this end, we define $\kcal$ as the term in the square brackets in \cref{eq_CCF_pbc_box} and note that, according to \cref{eq_Sflat,eq_Sdyn_def}, $\kcal$ can generally be expressed as
\beq \kcal(\{r_\parallel,z\},t) = \kcal\stat(\{r_\parallel,z\}) - t^{-d/4} \kcal\dyn(\psi, \psi_z),\qquad \psi \equiv \frac{r^4}{512 t}, \psi_z \equiv \frac{z^4}{512 t},
\label{eq_kcal_scalf}\eeq 
with $r^2=r_\parallel^2 + z^2$. The functions $\kcal\stat$ and $\kcal\dyn$ represent the contributions stemming from $\Scal\stat$ and $\Scal\dyn$, respectively. 
The explicit form of these functions does not matter for the following discussion; but in principle it can be obtained from \cref{eq_CCF_pbc_box}. 
We henceforth consider the case $\psi_z \sim \psi$ and do not indicate the dependence on $\psi_z$ separately. 

The leading behavior of $\kcal\dyn$ for $\psi\ll 1$, i.e., for short distances $r \ll r^*\equiv (512 t)^{1/4}$, follows from the asymptotic relations in \cref{eq_Cb_Dz2_lim_tinf}:
\beq \kcal(\{r_\parallel,z\},t)\big|_{\psi\ll 1} \simeq \kcal\stat(\{r_\parallel,z\},t) -  C t^{-d/4},\qquad C = \frac{2+d}{2^{1+7d/4}\pi^{(d-1)/2} d\, \Gamma(1/2 + d/4)}.
\label{eq_kcal_small_psi}\eeq 
In the opposite limit $\psi\gg 1$, i.e., for large distances $r \gg r^*\equiv (512 t)^{1/4}$, $\kcal\dyn$ approaches $\kcal\stat$ via exponentially damped oscillations [see \cref{eq_Sflat_asympt}], implying
\beq \kcal(\{r_\parallel,z\},t)\big|_{\psi\gg 1} \simeq \frac{2^{d/2}(r^2+2 z^2) \psi^{d/6}}{\sqrt{3}\pi^{d/2} r^{d+2}} e^{-3\psi^{1/3}/2} \cos\left[(d\pi - 9\sqrt{3}\psi^{1/3})/6\right].
\label{eq_kcal_large_psi}\eeq 

For large but finite times $t\gg L^4$, we use Eqs.\ \eqref{eq_kcal_scalf}--\eqref{eq_kcal_large_psi} in order to estimate the CCF in \cref{eq_CCF_pbc_box} as
\begin{multline} \bra \Kcal\pbc(t\to\infty)\ket \sim \bra\Kcal\pbc\ket\stat \\ - \sum_{m_x=-\infty}^\infty \cdots \sum_{m_z=-\infty}^\infty \Theta(\bv{m}) \Big\{ \theta(\varepsilon r^*-r) \kcal\dyn(\{r_\parallel,z\},t)\big|_{\psi\ll 1} + \theta(r-r^*/\varepsilon) \kcal(\{r_\parallel,z\},t)\big|_{\psi\gg 1} \Big\}_{\rv=\Lcalv}, 
\label{eq_CCF_pbc_box_est0}\end{multline}
where $\bra\Kcal\pbc\ket\stat$ accounts for the contribution stemming from $\kcal\stat$ in \cref{eq_kcal_scalf} (which is discussed separately in \cref{app_static_CCF}); $\varepsilon\ll 1$ is a small positive, but otherwise arbitrary real number which ensures that $r\ll r^*$ or $r\gg r^*$, respectively. 
In \cref{eq_CCF_pbc_box_est0} we have neglected the contribution from $\kcal\dyn$ for $\psi\sim\Ocal(1)$, which, according to \cref{eq_kcal_scalf}, is suppressed for large $t$ and contributes only within the limited range where $r\sim r^*$.
For $t\gg L^4$ and $r\gg r^*$, $\psi$ varies strongly between neighboring values of $\mv$ which occur in the sum in \cref{eq_CCF_pbc_box_est0}. This implies that in \cref{eq_kcal_large_psi} a steep exponential decay  occurs within a short range in $r$. 
Accordingly, also the last term in \cref{eq_CCF_pbc_box_est0} can be neglected. Thus the sum in \cref{eq_CCF_pbc_box_est0} is dominated by the contributions from the limit $\psi\ll 1$.
Since for $t\gg L^4$ these contributions vary weakly between neighboring values of $\mv$, a reasonable approximation of the asymptotic behavior can be obtained by replacing the sums by integrals. This renders
\beq \bra \Kcal\pbc(t\to\infty)\ket 
\sim  \bra\Kcal\pbc\ket\stat - \frac{C}{AL} t^{-d/4} \int \d^d r\,\theta(\varepsilon r^* - r).
\label{eq_CCF_pbc_box_est1}\eeq
In the second term on the r.h.s.\ of \cref{eq_CCF_pbc_box_est1}, the integral amounts to the volume $(\varepsilon r^*)^d \pi^{d/2}/\Gamma(d/2+1)$ of the $d$-dimensional sphere of radius $\varepsilon r^*$. Upon inserting $r^* = (512 t)^{1/4}$ one obtains 
\beq \bra \Kcal\pbc(t\to\infty)\ket \sim \bra\Kcal\pbc\ket\stat - L^{-d} \rho^{d-1} \varepsilon^d \tilde C, \qquad \tilde C \simeq \frac{2^{d-3} (2+d) \Gamma(d/4) }{\Gamma(1+d/2)^2} ,
\label{eq_CCF_pbc_box_est}\eeq 
which is time independent.
The last term in \cref{eq_CCF_pbc_box_est} provides an estimate of the dynamical contribution to the equilibrium CCF, induced by the late time-behavior of model B.
We emphasize that, due to \cref{eq_Kstat_Keq_diff}, this term does not correspond to the last term in \cref{eq_CCF_eq_box_pbc_int}.
Furthermore, due to the arbitrariness of $\varepsilon$ and the involved approximations, the numerical value of this term carries a significant uncertainty.
Nevertheless, for $d=3$ and $d=4$, one has $\tilde C\sim \Ocal(1)$, such that, for a reasonable value of $\varepsilon\sim \Ocal(0.1)$, \cref{eq_CCF_pbc_box_est} predicts a value for $\bra \Kcal\pbc(t\to\infty) \ket$ which is close to the exact one obtained from \cref{eq_CCF_eq_box_pbc_int} for $\tau\to 0$ [see also \cref{fig_Keq_pbc_rho}].



%

\end{document}